\documentclass[sn-basic]{sn-jnl}

\jyear{2021}%

\theoremstyle{thmstyleone}%
\theoremstyle{thmstyletwo}%

\theoremstyle{thmstylethree}%

\usepackage{graphicx}
\usepackage{subfig}
\usepackage{multirow}
\usepackage{multicol}
\usepackage{algorithm}
\usepackage{amsmath}
\usepackage{arabtex}
\usepackage{utf8}
\usepackage{caption}
\usepackage{graphicx}
\usepackage{arydshln}

\usepackage{xcolor}
\usepackage{soul}

\begin{document}

\title[Experimental Evaluation of  Baselines for Forecasting Social Media Timeseries]{Experimental Evaluation of  Baselines for Forecasting Social Media Timeseries}


\author[1]{\fnm{Kin Wai} \sur{NG}}\email{kinwaing@usf.edu}

\author[1]{\fnm{Frederick} \sur{Mubang}}\email{fmubang@usf.edu}

\author[1]{\fnm{Lawrence O.} \sur{Hall}}\email{lohall@usf.edu}

\author[2]{\fnm{John} \sur{Skvoretz}}\email{skvoretz@usf.edu}

\author[3]{\fnm{Adriana} \sur{Iamnitchi}}\email{a.iamnitchi@maastrichtuniversity.nl}

\affil[1]{\orgdiv{Department of Computer Science and Engineering}, \orgname{University of South Florida}, \orgaddress{\street{4202 E Fowler Ave}, \postcode{33620}, \city{Tampa}, \state{FL}, \country{USA}}}

\affil[2]{\orgdiv{Department of Sociology}, \orgname{University of South Florida}, \orgaddress{\street{4202 E Fowler Ave}, \postcode{33620}, \city{Tampa}, \state{FL}, \country{USA}}}

\affil[3]{\orgdiv{Institute of Data Science}, \orgname{Maastricht University}, \orgaddress{\street{Minderbroedersberg 4-6}, \postcode{6211 LK}, \city{Maastricht}, \country{The Netherlands}}}


\abstract{
Forecasting social media activity can be of practical use in many scenarios, from understanding trends, such as which topics are likely to engage more users in the coming week, to identifying unusual behavior, such as coordinated information operations or PumpNDump efforts. To evaluate a new approach to forecasting, it is important to have baselines against which to assess performance gains.
We experimentally evaluate the performance of four baselines for forecasting activity in several social media datasets that record discussions related to three different geo-political contexts synchronously taking place on two different platforms, Twitter and YouTube. Experiments are done over hourly time periods.
Our evaluation identifies the baselines which are most accurate for particular metrics and thus provide guidance for future work in social media modeling.  
}




\maketitle

\section{Introduction}

The importance of baselines for scientific progress cannot be overestimated.  In their absence, the evaluation of substantive theoretical models is compromised:  exact and accurate fit is rare, if not impossible, and so whether the fit is ``good enough'' is left to the eye of the beholder.  It is much better to have an understanding of how much can be accomplished by relatively simple and widely utilized models of the process of interest and then compare the value added by one's substantive theoretical model.  These simple and widely used models are ``baselines.''  In certain circumstances, they can themselves be successful at capturing key regularities at relatively low cost in terms of parameters estimated \citep{mayhew84}.

In a new domain of phenomena, like that of social media timeseries, baselines are even more important.  And in such a new domain, it is worth exploring multiple baseline candidates since it is not \textit{a priori} known which baseline does the best job at capturing key regularities.  Indeed, it is possible, as we show, that which baseline does best depends on which key regularities one believes it is essential to capture.  In the next section, we introduce three different baselines for social media time series:  ARIMA \citep{zhang1998forecasting, box2015time, hipel1994time, siami2018comparison, prasha-naam}, Hawkes process ~\cite{hawkes1971spectra, bacry2015hawkes,masuda2013self,rizoiu2017expecting, zhao2015seismic, lukasik2016hawkes, pinto2015trend,bacry2020sparse}, and the Shifted Baseline \citep{prasha-naam, Hernandez2020DeepLearning, socialcube,pb-medic, pb-27-solarwind,perez-ortiz-weather}. They have each been widely used in the field of time series analysis.

Our objective is to use data from multiple social media platforms streams to assess which of these baselines has better predictive power against the empirical data, so called ground truth (GT) data streams for capturing key properties (e.g., volume of activities, temporal patterns, distribution characteristics).  
Via experimental evaluation from six independent datasets (three geo-political contexts over two different social media platforms) and a total of 30 time series of topics in social media, we show that each baseline has advantages and limitations based on the end-task and particular time series characteristics that we aim to capture.  Two prediction time periods, daily and weekly, are examined.
We observed that, for example, ARIMA better estimates the overall volume of activities over a long forecasting horizon, but it fails to produce accurate temporal patterns/signals present in the GT.
When the detection of anomalous or irregular periods of activity is of interest, comparing our predictions against ARIMA could be problematic and might lead to misleading conclusions.
For these scenarios, Hawkes processes can be a more appropriate benchmark.
Simple baseline approaches that consider shifting past observations as the forecast proved to be highly competitive and a strong benchmark for many scenarios, especially, for next day forecasts.
However, when topics are heavily influenced by external variables (e.g., real-world events, economic fluctuations, emergency responses, etc.), the shifted baseline is likely to be less useful over longer prediction windows.
Finally, we show that ensemble approaches could provide guidance for better benchmarks. 
\section{Traditional Baselines for Time Series}

\emph{What is a competitive baseline against which to compare approaches for predicting time series for social media activity?} 
This question was motivated by our work on the DARPA SocialSim project with a focus on forecasting social media activity in selected contexts at micro-level granularity~\citep{renhao-github-atp, Hernandez2020DeepLearning, github-renhao-cve2vec, socialcube, prasha-naam, ng2021forecasting, bollenbacher2021challenges, blythe2019darpa}. 
In the DARPA challenges, our forecasts were evaluated against what we call below the Shifted Baseline, while reviewers of our work submitted for publication consistently suggested a comparison with ARIMA models.  
This paper addresses this disjuncture systematically.  
We selected three well-known time-series generators used in different applications, and apply them to three datasets from two social media platforms each.
The three baselines typically used for time series forecasting are briefly presented below. 
In addition, we combine two of these baselines, ARIMA and Hawkes, into an ensemble that simply averages their outputs in an attempt to obtain a  better baseline.

\subsection{ARIMA} 

ARIMA is a traditional approach to modeling dynamical systems.
This model is derived from a composition of autoregressive and moving average modeling processes, which are commonly represented by $p$ and $q$, with the addition of a differencing term (I) represented by $d$. 
In short, ARIMA assumes that data are linear and follow a particular probability distribution.
That is, for any observed variable $X$, it can be decomposed into  fixed trend, seasonal, and irregular components over time.
The output of the ARIMA model can be expressed as a linear combination of past values of $X_t$ and lagged forecast errors $\epsilon_t$:

\begin{equation}
    X_t = \mu + \phi_1 X_{t-1} + \dots + \phi_p X_{t-p} - \theta_1 \epsilon_{t-1} - \dots - \theta_q \epsilon_{t-q}
\end{equation}

where $\phi_1, \dots, \phi_p$ and $\theta_1, \dots, \theta_q$ are the regression weights to be estimated, $\mu$ represents the constant trend component, and $\epsilon_{t-1}, \dots, \epsilon_{t-q}$ are the random errors.   

Despite suffering from limitations when presented with non-linear tasks due to its linearity assumption~\citep{zhang1998forecasting}, ARIMA is widely accepted and used as a base model in most time series research~\citep{box2015time, hipel1994time, siami2018comparison, prasha-naam}.

\subsection{Hawkes Processes}

Hawkes processes have been applied in a variety of settings in order to describe or predict  univariate or multivariate data.
Unlike discrete-time models (in which temporal data is aggregated into uniform time intervals), Hawkes processes rely on discrete events in continuous time (that is, with exact timestamps) to model the likelihood of an event occurring in the future.
They are based on the idea that events that are observed in time frequently cluster together. 
For example, a post published by an influential social media user is likely to be followed by a significant number of interactions from other social media users.
Hawkes processes model the inter-arrival timing of these events/interactions by using a conditional intensity function, which can be influenced by past/historical events. 
Hawkes processes can be defined as follows:

\begin{equation}
    \lambda (t) = \mu + \sum_{t_i < t} \varphi (t - t_i)
\end{equation}

where $\mu$ is the conditional intensity rate, and $\varphi$ is a triggering kernel function that takes as input the delay between the current and previous timestamps of events. In this paper, we used the exponential decay kernel function defined by \begin{math} \varphi(x) = \alpha \beta \exp(-\beta x) \end{math}, where $\alpha$ and $\beta$ regulate the growth and decay rate of events based on observed data. 

Hawkes processes were originally defined to describe earthquake dynamics~\citep{hawkes1971spectra}, but previous work has used them for forecasting tasks such as estimating price fluctuations in stocks~\citep{bacry2015hawkes} or reproducing conversation dynamics as event sequences~\citep{masuda2013self}.
In social media, Hawkes processes have been adopted to model information diffusion and to predict the popularity of content~\citep{rizoiu2017expecting, zhao2015seismic, lukasik2016hawkes, pinto2015trend}.
Despite some success, they often struggle to capture characteristic patterns usually observed on social media discussions such as topics with low volume of activities~\citep{bacry2020sparse}.

\subsection{Shifted Baseline}

The Shifted (or Persistence) Baseline model is the third  common baseline approach we investigate. It is defined as follows. Let \(T\) be the current timestep of interest, and let \(S\) is the length of the desired predicted time series. The Shifted Model predicts the time series for period \(T +1\) to \(T + S\) by moving forward, or ``shifting forward'' the time series that spans period \(T -S +1\) to \(T\). The underlying assumption of this model is that the immediate future of the time series will approximately replicate its immediate past.

The Shifted baseline has been used in many domains:  in social media time series prediction \citep{prasha-naam, Hernandez2020DeepLearning, socialcube}, in healthcare time series prediction (patients' weekly average expenditures on certain pain medications) \citep{pb-medic}, and in weather time series forecasting \citep{pb-27-solarwind,perez-ortiz-weather}.

\subsection{Ensemble Hawkes and ARIMA}

The quantitative evaluations presented later suggested a combination of the Hawkes processes approach and the ARIMA approach in an ensemble could be useful. As shown in the next section, Hawkes processes offer better approximations of fluctuations in activity over time, while ARIMA offers better approximations of the amplitude of activity.  Therefore, these results suggest an ensemble of the two approaches might prove valuable.  Our ensemble is simple:  it averages the two source predictions over time, assuming that each approach contributes equally to the final predicted value.

\section{Datasets}

We use data from three different geo-political contexts, each with their own different reactions on social media.
The datasets contain a record of user activity on two social media platforms, Twitter and YouTube. 
The focus on multiple datasets from different contexts and platforms attempts to capture regularities attributable to a baseline rather than features related to data or a social media platform. 

\subsection{Three Geo-Political Contexts}
Our datasets include a period of political unrest in Venezuela in early 2019, discussions around the Chinese-Pakistan Economic Corridor, and  discussions around the Chinese-East Africa Development Project.  

\textbf{Venezuelan Political Crisis of 2019 (Vz19)}: In 2019, a humanitarian, economic and political crisis engulfed Venezuela as the presidency was disputed between Juan Guaid\'{o} and Nicol\'{a}s Maduro, each claiming to be the country's rightful president. 
The political turmoil had its roots in the controversial re-election of Nicol\'{a}s Maduro as the head of the state on January 10th, which was boycotted by opposition politicians and condemned by the international community as fraudulent.
Consequently, on January 23rd, the opposition-led National Assembly ratified Juan Guaid\'{o} as the country's interim president in an effort to restore democracy and constitutional rights. 
This event contributed to unprecedented conflicts and high political tension which resulted in nationwide protests, militarized responses, international humanitarian aid intervention attempts, and incidents of mass violence and arrests. 
Our dataset covers two major events unfolding in the country. 
First, the declaration of Juan Guaid\'{o} as Venezuela's interim president which triggered massive anti-government protests across the country.
Second, the arrival of humanitarian aid for the country which was met with violent standoffs between protesters and military forces. 
Our models are trained on data from the first event and evaluated on data from the second.

\textbf{Chinese-Pakistan Economic Corridor (CPEC)}: 
CPEC is a strategic economic project under the Belt and Road Initiative (BRI) launched by China aimed at strengthening and modernizing road, port and energy transportation systems in Pakistan. 
While China's investment in Pakistan has largely taken the form of infrastructure development, the project has been heavily criticized with claims of lack of transparency, self-benefit and imposing unsustainable debt on the Southeast Asian country~\citep{david_2021}.
For example, the Indian government firmly stands against CPEC as the project involves traversing parts of the disputed Jammu and Kashmir regions~\citep{adnan_2020}.
Conflict around this project plays out in discussions on social media, where it has been reported that both Pakistani and Chinese state actors strategically promote Chinese interests to facilitate building projects and garner further investment. 
On the other hand, factions opposed to CPEC perpetuate unfavorable narratives such as the project only being beneficial for China to boost its influence and trading power~\citep{ahmed_2021, afzal2020all}. 

\textbf{Belt and Road Initivative in East Africa (BRIA)}: 
As part of the same BRI flagship project, China has committed a substantial amount of resources and investment across Africa, especially in the east African nations.
Similar to the CPEC scenario, several concerns have arisen over China's strategic intentions in Africa.
Chinese officials claim BRIA is a purely economic construct to enhance international cooperation.
However, other parties suggest that the project is a geopolitical tool to boost China's global strategic influence~\citep{lokanathan2020china}. 
Due to the heavy influence of Chinese media in Africa~\citep{thussu2018china}, BRIA activities have been promoted extensively on the continent through public pronouncements from ambassadors and government officials. 
At the same time, many citizens have openly expressed resistance to China's BRIA as they believe it creates an era of external dependency, internal conflicts, and a rising debt trap~\citep{lokanathan2020china}.

\subsection{Twitter and YouTube Data Collection}

The datasets used in this work were collected by Leidos and are part of the DARPA-sponsored SocialSim program~\citep{DARPA}. 
Twitter data was collected using the GNIP API and YouTube data was collected through its public API query tool.
The collection was based on a list of keywords relevant to each context and was compiled by our SocialSim collaborators.
Table~\ref{tab:data_keywords} presents the search terms for data collection and the 66-day time period considered in this work. 
For Twitter, activities consist of tweets, retweets, replies and quotes. 
For YouTube, activities consist of videos and comments.
The resultant  social media datasets consist of 13,984,719 Twitter activities done by 1,052,526 users and 116,725 YouTube activities by 70,651 users in Vz19; 1,892,182 Twitter activities by 631,920 users and 133,931 YouTube activities by 73,268 users in CPEC; and lastly 1,046,827 Twitter activities by 472,663 users and 5,703 YouTube activities by 4,168 users in BRIA. 

\begin{table}
\centering
\caption{Search terms used for data collection and time periods considered.}
\label{tab:data_keywords}
\begin{tabular}{|c|c|c|} 
\hline
Domain & Period                                                                           & Search Terms                                                                                                                                                                                                                                                                                                                                                                                                                                                                                                                                                                                                           \\ 
\hline
Vz19   & \begin{tabular}[c]{@{}c@{}}December 24, 2018 - \\ February 28, 2019\end{tabular} & \begin{tabular}[c]{@{}c@{}}\#23Ene, \#23Feb, Aid Venezuela, \#BravoPueblo, \\Caracas, Maracaibo, Chacao, Maturin,\\ \#Chavismo, FANB, \#FreeVenezuela, \#FueraDictadura, \\Fuerza Venezuela, GNB,\\ \#Guaid\'{o}Presidente, Juan Guaid\'{o}, Nicol\'{a}s Maduro, \\\#LasCallesSonDelChavismo, \\ Libertad para Venezuela, Freedom for Venezuela, \\\#VamosBien, \#MaduroDictator,\\ \#SOSVenezuela, \#VenezolanosEnElMundo, \\Venezuela Aid Live, \#WeAreMaduro, \\ Yankee go Home, \#HandsOffVenezuela, \\\#FebreroRebelde, \#NoMasDictatura,\\ \#AbajoCadenas, Venezuela Crisis Humanitaria, \\Maduro Ilegitimo\end{tabular}  \\ 
\hline
CPEC   & \begin{tabular}[c]{@{}c@{}}March 30, 2020 - \\ June 4, 2020\end{tabular}         & \begin{tabular}[c]{@{}c@{}}gwadar, china-pak, china pakistan economic corridor, \\cpec, cathaypak, gwadar port,\\ cphcg, pakchina\_friendship, baltistan, baluchistan, \\balochistan, kashmir, Imran Khan,\\ Belt and Road Initiative, beltandroadpod,\\~OborChina, \#BRI, \#obor, one belt and road\end{tabular}                                                                                                               \\ 
\hline
BRIA   & \begin{tabular}[c]{@{}c@{}}February 1, 2020 - \\ April 7, 2020\end{tabular}      & \begin{tabular}[c]{@{}c@{}}ctgnafrica, cdafricanews, china belt and road, \\ukanda mmoja barabara moja,\\ bring kenyans in china home, evacuate ugandans in wuhan,\\~kenya under construction,\\ keep kenya moving, china confucius institute, uhuru kenyatta, \\Yoweri Museveni, John Magufuli, Paul Kagame, \\Emmerson Mnangagwa, \\Edgar Lungu,Karuma Hydropower, Isimba Hydroelectric, \\Entebbe-Kampala Expressway,\\ Mombasa-Nairobi Standard Gauge Railway, \\Standard Guage Railway\end{tabular}                                                                                                               \\
\hline
\end{tabular}
\end{table}

\subsection{Topic Assignment}

As an effort to investigate how activity/information diffuses for different topics of discussion in the social media ecosystem, Leidos employed annotators and subject matter experts (SME) to identify a set of relevant topics relating to each of the three contexts. 
After a thorough exploration of the datasets corpus, SME crafted a total of 18 topics for Vz19, 12 topics for CPEC and 12 topics for BRIA.
For each of the three contexts, the annotators created initial annotation sets comprised of a small subset of tweets and YouTube posts (i.e., videos and comments), which were then labeled with their corresponding topics.
The annotations were conducted manually and over a subsample of 11,218 distinct text documents for each of the three contexts.
The annotation process consisted of an 8 to 1 ratio of single annotator annotations to all annotator annotations. 
The 18 topics identified in Vz19 had a weighted average Cohen's Kappa score of 0.64 for inter-annotator agreement.
The 12 topics in CPEC had inter-annotator agreement of 0.74 for Cohen's Kappa.
Lastly, the 12 topics in BRIA had an average Cohen's Kappa score of 0.65.

Due to the substantial variation in agreement scores among individual topics, we decided to focus this work on the  5 topics with the highest inter-annotator agreement scores from each context. 
Table~\ref{tab:topic_annotators} presents the topics chosen alongside their agreement scores and a brief description.
Because it is not feasible to manually label millions of messages, BERT models~\citep{devlin2019bert} were trained and tested for topic annotation.
The BERT models for each context were trained on 10,097 distinct text documents and evaluated on a 10\% test set (1,121 texts).
Stratified sampling was used to ensure that the train and test sets had approximately the same percentage of samples of each topic class as in the original manually annotated corpus.
The F1 score of the BERT models on Vz19 and CPEC annotation test sets were 0.66 and 0.73, respectively. 
For BRIA, a XLM-Roberta-Large pre-trained model~\citep{conneau2019unsupervised} was used instead to label the texts. Its F1 score was 0.83 on the annotation test set.

After topic assignment and only considering the top 5 topics in inter-annotator agreement (see Table~\ref{tab:topic_annotators}), we ended up with 7,986,918 activities by 757,496 users in Twitter and 66,788 YouTube activities by 43,342 for Vz19; 407,621 Twitter activities by 184,961 users and 1,072 YouTube activities by 889 users for CPEC; and 905,018 Twitter activities by 437,941 users and 3,365 YouTube activities by 2,616 users for BRIA.
Table~\ref{tab:topic_annotators} also shows the number of activities broken down per topic in both Twitter and YouTube.
We observe that both platforms exhibit clear differences in terms of the magnitude of activities.
Topic activity on Twitter is orders of magnitude larger than YouTube activity in every context.
This is expected as YouTube is mainly used for audiovisual media consumption rather than micro-blogging, where large-scale user interactions are more common. 
We also find differences across contexts, for instance, the Venezuelan political crisis in 2019 (Vz19) spawned much larger discussions, compared to CPEC and BRIA, on both social media platforms as different external events unfolded in the country.

\begin{table}[htpb]
\centering
\caption{Topic Inter-Annotator Agreement scores (IAA) and corresponding descriptions in each context. We include the aggregated volume of activities per topic in both Twitter and YouTube over the periods we consider.}
\label{tab:topic_annotators}
\scalebox{0.70}{
\begin{tabular}{|c|l|r|rr|l|}
\hline
\multirow{2}{*}{Domain} & \multicolumn{1}{c|}{\multirow{2}{*}{Topic}} & \multicolumn{1}{c|}{\multirow{2}{*}{IAA}} & \multicolumn{2}{c|}{Volume} & \multicolumn{1}{c|}{\multirow{2}{*}{Description}} \\
 & \multicolumn{1}{c|}{} & \multicolumn{1}{c|}{} & \multicolumn{1}{c}{Twitter} & \multicolumn{1}{c|}{YouTube} & \multicolumn{1}{c|}{} \\ \hline
\multirow{8}{*}{Vz19} & maduro/dictator & 0.78 & \multicolumn{1}{r|}{1,281,021} & 25,776 & \begin{tabular}[c]{@{}l@{}}Posts that describe Maduro as a dictator, or when its\\ administration is represented as a dictatorship.\end{tabular} \\ 
 & arrests & 0.70 & \multicolumn{1}{r|}{842,546} & 1,944 & \begin{tabular}[c]{@{}l@{}}Describes arrests or people who have been taken\\ prisoners during on-the-ground events.\end{tabular} \\ 
 & other/ch\'{a}vez & 0.70 & \multicolumn{1}{r|}{958,450} & 18,372 & \begin{tabular}[c]{@{}l@{}}Posts with terms related to the late president of\\ Venezuela, Hugo Ch\'{a}vez.\end{tabular} \\  
 & military & 0.69 & \multicolumn{1}{r|}{3,502,074} & 12,895 & \begin{tabular}[c]{@{}l@{}}Posts about the Venezuelan military, security forces,\\ or other armed militarized organization.\end{tabular} \\ 
 & international/aid & 0.68 & \multicolumn{1}{r|}{2,336,919} & 12,168 & \begin{tabular}[c]{@{}l@{}}Descriptions of foreign humanitarian aid sent or\\ requested to be sent to Venezuela.\end{tabular} \\ \hline
\multirow{8}{*}{CPEC} & controversies/china/border & 0.91 & \multicolumn{1}{r|}{180,078} & 260 & \begin{tabular}[c]{@{}l@{}}Discussions related to the disputed Chinese-Indian\\ border.\end{tabular} \\
 & leadership/sharif & 0.89 & \multicolumn{1}{r|}{82,327} & 132 & \begin{tabular}[c]{@{}l@{}}Positively discusses leadership by Nawaz Sharif,\\ former prime minister of Pakistan.\end{tabular} \\ 
 & controversies/china/uighur & 0.85 & \multicolumn{1}{r|}{25,854} & 52 & \begin{tabular}[c]{@{}l@{}}Posts that bring attention to the Uighur \\ controversies in Xinjiang China.\end{tabular} \\
 & controversies/pakistan/baloch & 0.85 & \multicolumn{1}{r|}{84,621} & 410 & \begin{tabular}[c]{@{}l@{}}Statements that advocate for Baloch independence, or\\ emphasize exploitation of Baloch by Pakistan or China.\end{tabular} \\ 
 & benefits/development/roads & 0.83 & \multicolumn{1}{r|}{43,831} & 222 & \begin{tabular}[c]{@{}l@{}}Discussions on road or railway projects in Pakistan\\ related to CPEC.\end{tabular} \\ \hline
\multirow{8}{*}{BRIA} & covid & 0.82 & \multicolumn{1}{r|}{774,284} & 2,018 & \begin{tabular}[c]{@{}l@{}}Coronavirus and related topics such as vaccines in\\ Africa.\end{tabular} \\  
 & debt & 0.82 & \multicolumn{1}{r|}{23,803} & 205 & \begin{tabular}[c]{@{}l@{}}Debts or loans, where the lender is China and the\\ lendee is an African nation\end{tabular} \\ 
 & infrastructure & 0.78 & \multicolumn{1}{r|}{28,397} & 205 & \begin{tabular}[c]{@{}l@{}}Roads, railways, ports or other infrastructure project\\ as part of the BRI in Africa.\end{tabular} \\ 
 & travel & 0.78 & \multicolumn{1}{r|}{224,582} & 815 & \begin{tabular}[c]{@{}l@{}}Travel between countries, usually by flights. Often in\\ the context of COVID and related to Africa and China.\end{tabular} \\ 
 & mistreatment & 0.68 & \multicolumn{1}{r|}{58,306} & 670 & \begin{tabular}[c]{@{}l@{}}Description of mistreatment of particular groups of\\ people such as Kenyans, Africans, or Chinese people.\end{tabular} \\ \hline
\end{tabular}
}
\end{table}

\section{Evaluation}

Our objective is to assess how well different baselines can capture key properties of time series in the context of per-topic social media activity.
To this end, we evaluate the baselines described above against real data and across a set of relevant performance metrics. 

\subsection{Experimental Setup}

We evaluate performance in forecasting social media activity on two platforms, Twitter and YouTube, on a number of topics from three different geo-political contexts.
We split the time series samples into training, validation and testing sets for each context as shown in Table~\ref{tab:data_split}.
The baselines forecast one week of social media activity at hour granularity, thus, 168 datapoints corresponding to the 168 hours in a week. 

For Hawkes, we use the sequence of previous events timestamps in order to estimate an intensity function that approximates the likelihood of events happening in the future.
The model parameters are: (1) the intensity rate, $\mu$, (2) the infectivity factor which represents how many expected events are caused by a previous event, $\alpha$, and (3) the exponential delay which is the probability distribution of time between events, $\beta$.
These parameters are fit from the training data using an expectation-maximization algorithm that seeks to maximize the log-likelihood function. 
The output of the fit model results in timestamps of future events, which we aggregate into discrete-time intervals for evaluation (in our case, hour granularity).

For ARIMA, we use the validation set to select a combination of optimized parameters $(p, q, d)$ for each topic.
That is, each hour of predicted activity per topic is dependent on the previous $p$ hours of activities and the previous $q$ hours of estimation errors. 
 The differencing factor $d$ indicates the order of transforming a non-stationary time series into a stationary one.

We ran a grid search over different combinations of $(p, q, d)$.
At hourly granularity, we considered 24, 48, 72, and 96 previous hours for both $p$ and $q$ values. 
Values of 0, 1 and 2 were considered for $d$.
Table~\ref{tab:arima_params} shows the parameter combinations that obtained the lowest RMSE score over the validation period for each topic. 

\begin{table}[htpb]
\centering
\caption{ARIMA best parameter combinations over the validation set for each topic on two different platforms.}
\label{tab:arima_params}
\begin{tabular}{|c|l|cc|}
\hline
\multicolumn{1}{|c|}{\multirow{2}{*}{Domain}} & \multicolumn{1}{c|}{\multirow{2}{*}{Topic}} & \multicolumn{2}{c|}{(p, q, d)} \\  
\multicolumn{1}{|c|}{} & \multicolumn{1}{c|}{} & \multicolumn{1}{c}{Twitter} & YouTube \\ \hline
\multirow{5}{*}{Vz19} & arrests & \multicolumn{1}{c|}{(72, 1, 48)} & (48, 2, 72) \\ 
 & international/aid & \multicolumn{1}{c|}{(24, 2, 48)} & (72, 1, 96) \\ 
 & maduro/dictator & \multicolumn{1}{c|}{(96, 2, 72)} & (96, 1, 72) \\ 
 & military & \multicolumn{1}{c|}{(72, 1, 72)} & (72, 1, 96) \\  
 & other/chavez & \multicolumn{1}{c|}{(48, 2, 72)} & (48, 1, 24) \\ \hline
\multirow{5}{*}{CPEC} & benefits/development/roads & \multicolumn{1}{c|}{(24, 1, 72)} & (72, 2, 72) \\
 & controversies/china/border & \multicolumn{1}{c|}{(24, 2, 72)} & (48, 2, 72) \\ 
 & controversies/china/uighur & \multicolumn{1}{c|}{(24, 1, 48)} & (24, 2, 48) \\ 
 & controversies/pakistan/baloch & \multicolumn{1}{c|}{(48, 2, 96)} & (48, 2, 72) \\  
 & leadership/sharif & \multicolumn{1}{c|}{(24, 2, 48)} & (24, 1, 24) \\ \hline
\multirow{5}{*}{BRIA} & covid & \multicolumn{1}{c|}{(96, 1, 48)} & (72, 2, 96) \\  
 & debt & \multicolumn{1}{c|}{(96, 1, 96)} & (48, 1, 96) \\  
 & infrastructure & \multicolumn{1}{c|}{(24, 1, 24)} & (72, 1, 96) \\ 
 & mistreatment & \multicolumn{1}{c|}{(96, 1, 96)} & (48, 1, 48) \\ 
 & travel & \multicolumn{1}{c|}{(24, 1, 48)} & (24, 2, 72) \\ \hline
\end{tabular}
\end{table}

We report six performance metrics. Here, $y_t$ indicates the true value of the time series and $\hat{y}_t$ the forecast series. $Y_t$ and $\hat{Y}_t$ represent the aggregated volume of activities for the ground truth and the forecast, respectively:

\textbf{Absolute percentage error (APE)}, which evaluates the error over the aggregated volume of activities within the forecasting period, 
\begin{equation}
    APE = \frac{\mid Y_t - \hat{Y}_t \mid}{Y_t} \times 100
\end{equation}

\textbf{Root mean squared error (RMSE)}, which measures the differences between predicted and actual values over time, 
\begin{equation}
    RMSE = \sqrt{\frac{1}{n}\sum_{t=1}^{n} (y_t - \hat{y}_t)^2}
\end{equation}

\textbf{Symmetric mean absolute percentage error (sMAPE)}, which measures relative errors over time, 
\begin{equation}
    sMAPE = \frac{1}{n} \sum_{t=1}^{n} (\frac{200 \times \mid y_t - \hat{y}_t \mid}{\mid y_t \mid + \mid \hat{y}_t \mid})
\end{equation}

\textbf{Dynamic time warping (DTW)}, for measuring the similarity between two time series that may vary in timing. DTW takes as input two time sequences, X and Y, of length n, where $X = \{x_1, \dots, x_n\}$ and $Y = \{y_1, \dots, y_n\}$ and computes the matching cost based on the following formulation,
\begin{equation}
    D(i,j) = d(x_i, y_i) + \min\{D(i, j-1), D(i-1, j), D(i-1,j-1)\}
\end{equation}
where $d$ is a distance function (e.g., euclidean distance), and consequently,
\begin{equation}
    DTW(X,Y) = \sum_{i=1}^{n}\sum_{j=1}^{n}D(i, j)
\end{equation}

\textbf{Volatility error (VE)}, for measuring the absolute difference between the standard deviation ($\sigma$) of the ground truth and forecasted time series,
\begin{equation}
    VE = \mid \sigma(y_t) - \sigma(\hat{y}_t) \mid
\end{equation}

\textbf{Skewness error (SkE)}, for measuring asymmetry in the volume of activities over time. The SkE metric is measured by computing the adjusted Fisher–Pearson standardized moment coefficient (Eq. 9) of both the ground truth and forecast time series, 
\begin{equation}
    G_1 = \frac{\sqrt{n(n-1)}}{n-2} \times \frac{\sum_{i=1}^{n} (x_i - \overline{x})/n}{s^3}
\end{equation}
where $n$ is the length of the time series, $\overline{x}$ is the sample mean and $s$ is the sample standard deviation, and then calculating their absolute difference,
\begin{equation}
    SkE = \mid G_1 - \hat{G_1}\mid
\end{equation}
where $G_1$ and $\hat{G_1}$ are the standardized moment coefficients for the ground truth and forecast, respectively.

We selected these metrics for evaluation as they measure specific characteristics of time series that we believe are of relevance to social media forecasting solutions.
First, we use APE to assess how well a model can capture the total volume of activities per-topic.
This property is important for scenarios where the identification of viral/popular topics across social media platforms is of interest.
For temporal patterns and volume of activity over time, we use DTW, RMSE and sMAPE measures.
The temporal characteristics of time series are valuable for identifying anomalous periods of activity and making informed decisions based on predictions. 
Finally, we measure skewness and volatility errors, which are important characteristics that can describe social media time series distributions in terms of their shape and variability. The authors of \citep{vam-venez-xplore} noted the importance of these 2 particular metrics because of instances in which a model reported strong performance using more typical metrics such as RMSE, but were weak at capturing variability or spikiness upon visual representation.

\begin{table}[htpb]
\centering
\caption{Time periods chosen for data split into training, validation and testing sets. 
}
\label{tab:data_split}
\begin{tabular}{|l|l|l|l|}
\hline
\multicolumn{1}{|c|}{Domain} & \multicolumn{1}{c|}{Training (52 days)} & \multicolumn{1}{c|}{Validation (1 week)} & \multicolumn{1}{c|}{Testing (1 week)} \\ \hline
Vz19 & 2018-12-24 to 2019-02-14 & 2019-02-15 to 2019-02-21 & 2019-02-22 and 2019-02-28 \\ \hline
CPEC & 2020-03-30 to 2020-05-21 & 2020-05-22 to 2020-05-28 & 2020-05-29 and 2020-06-04 \\ \hline
BRIA & 2020-02-01 to 2020-03-24 & 2020-03-25 to 2020-03-31 & 2020-04-01 and 2020-04-07 \\ \hline
\end{tabular}
\end{table}

\subsection{Hardware and Runtime Costs}
All experiments were run on a computer with an Intel Xeon E5-2630 v4 CPU. Each CPU had two sockets, 10 cores, 2 threads, and 512 GB of memory.  
The ARIMA-based models took roughly 12 hours per topic.
We ran ARIMA experiments in parallel across 5 computers, and ended up with a total training time duration of 36 hours for all topics. 
Hawkes processes took on average 2 minutes for each topic and a total run time of 1.5 hours for all topics.
The shifted baseline is trivially created by moving historical predictions forward; thus, there is no training phase involved.

\subsection{Empirical Results} 

In comparing the performance of the baselines against the ground truth, we answer the following questions:
\begin{enumerate}
    \item Which baselines better capture the volume of activity? 
    \item Which baselines better capture the temporal patterns? 
    \item Which baselines better capture the skewness and volatility? 
    \item How does the duration of the forecast window affect the performance of the baselines?
\end{enumerate}

A high level overview of results is found in Table~\ref{tab:aggregate_counts_performance} which presents the performance of baselines for predicting time series of activities across social media topics over different metrics, contexts and platforms at hour granularity. 
Highlighted in bold are the scores for the baseline that performed best for a metric and topic for a platform. 
Before we discuss the results in detail, we want to emphasize that they vary significantly by topic and by platform. 
Thus, we try to identify consistent patterns, but stress the patterns hold in most but not all cases. 

We then utilized an aggregate metric, called the \textit{Overall Normalized Metric Error} (ONME) \citep{vam-venez-xplore} that better illustrates relative performance among models. Tables \ref{tab:onme-full-results} and \ref{tab:onme-summary} contain the results of this metric. Utilizing this new metric, we find that the Shifted Baseline seems to be the strongest performing baseline.

\begin{table}[htpb]
\centering
\caption{Forecasting performance for total activity predicted across topics in each domain and platform. Baselines were set to forecast one week of activity hour granularity (168 hours). To compute each metric, we aggregated the predicted activity across all topics and compare with the total number of activities in ground truth. Thus, the performance results are implicitly weighted by the number of activities in each topic. Bold entries are for the best aggregate metric.}
\begin{tabular}{|c|l|r|r|r|r|}
\hline
 Platform & \multicolumn{1}{c|}{Metric} & \multicolumn{1}{c|}{ARIMA} & \multicolumn{1}{c|}{Hawkes} & \multicolumn{1}{c|}{Shifted} & \multicolumn{1}{c|}{Hawkes+ARIMA} \\ \hline
 \multirow{6}{*}{Twitter (Vz19)} & APE & \textbf{35.54} & 74.61 & 67.88 & 55.07 \\
  & RMSE & \textbf{20216.11} & 24015.25 & 24312.27 & 21898.09 \\
   & SMAPE & 81.25 & 93.59 & 98.17 & \textbf{76.38} \\
   & DTW & \textbf{0.53} & 0.67 & 0.83 & 0.57 \\
   & Skewness & 1.37 & \textbf{0.05} & 1.13 & 0.90 \\
   & Volatility & 17562.68 & 17836.05 & \textbf{15222.80} & 18309.18 \\ \hline 
 \multirow{6}{*}{YouTube (Vz19)} & APE & \textbf{40.70} & 81.07 & 52.39 & 60.88 \\
   & RMSE & \textbf{128.06} & 173.08 & 148.72 & 148.04 \\
  & SMAPE & \textbf{45.13} & 119.75 & 73.88 & 66.81 \\
  & DTW & 0.55 & \textbf{0.40} & 0.68 & 0.49 \\
  & Skewness & 2.71 & 1.65 & \textbf{1.45} & 2.52 \\
  & Volatility & 92.85 & 104.28 & \textbf{77.50} & 106.69 \\ \hline
 \multirow{6}{*}{Twitter (CPEC)} & APE & 270.56 & 75.97 & \textbf{0.64} & 97.29 \\
  & RMSE & 2082.84 & \textbf{767.49} & 1055.06 & 1029.92 \\
  & SMAPE & 123.11 & 108.25 & \textbf{84.60} & 98.15 \\
  & DTW & 0.92 & \textbf{0.45} & 0.74 & 0.91 \\
  & Skewness & 3.04 & \textbf{0.29} & 0.75 & 3.11 \\
  & Volatility & 133.45 & 447.03 & 266.56 & \textbf{208.28} \\ \hline
  \multirow{6}{*}{YouTube (CPEC)} & APE & 46.11 & 61.14 & \textbf{5.70} & 29.53 \\
  & RMSE & 1.54 & 1.57 & 2.13 & \textbf{1.52} \\
  & SMAPE & \textbf{103.58} & 113.69 & 116.64 & 114.56 \\
  & DTW & 0.73 & \textbf{0.61} & 0.63 & 0.74 \\
  & Skewness & 2.75 & \textbf{0.83} & 1.50 & 0.93 \\
  & Volatility & 0.74 & 0.47 & \textbf{0.33} & 0.62 \\ \hline
\multirow{6}{*}{Twitter (BRIA)} & APE & \textbf{33.30} & 39.34 & 43.84 & 36.32 \\ 
  & RMSE & \textbf{430.54} & 733.79 & 731.99 & 533.80 \\
  & SMAPE & 60.34 & 69.60 & \textbf{54.31} & 61.53 \\
  & DTW & 0.48 & \textbf{0.45} & 0.44 & \textbf{0.45} \\
  & Skewness & 3.09 & 0.13 & \textbf{0.08} & 0.40 \\
  & Volatility & 294.21 & 142.47 & 213.30 & \textbf{120.38} \\ \hline
  \multirow{6}{*}{YouTube (BRIA)} & APE & 53.99 & 78.71 & \textbf{10.27} & 11.03 \\
  & RMSE & 2.80 & 3.42 & 3.55 & \textbf{2.64} \\
  & SMAPE & 127.34 & 116.49 & 117.78 & \textbf{113.58} \\
  & DTW & 0.87 & \textbf{0.44} & 0.66 & 0.49 \\
  & Skewness & 3.05 & \textbf{1.43} & 1.76 & 1.44 \\
  & Volatility & 2.04 & 0.17 & \textbf{0.14} & 0.97 \\ \hline
\end{tabular}
\label{tab:aggregate_counts_performance}
\end{table}

\subsubsection{Volume of Activity}

\begin{figure*}[htbp]
\centering
\begin{tabular}{cc}
	\subfloat[Vz19 (Twitter)]{
		\includegraphics[width=0.4\linewidth]{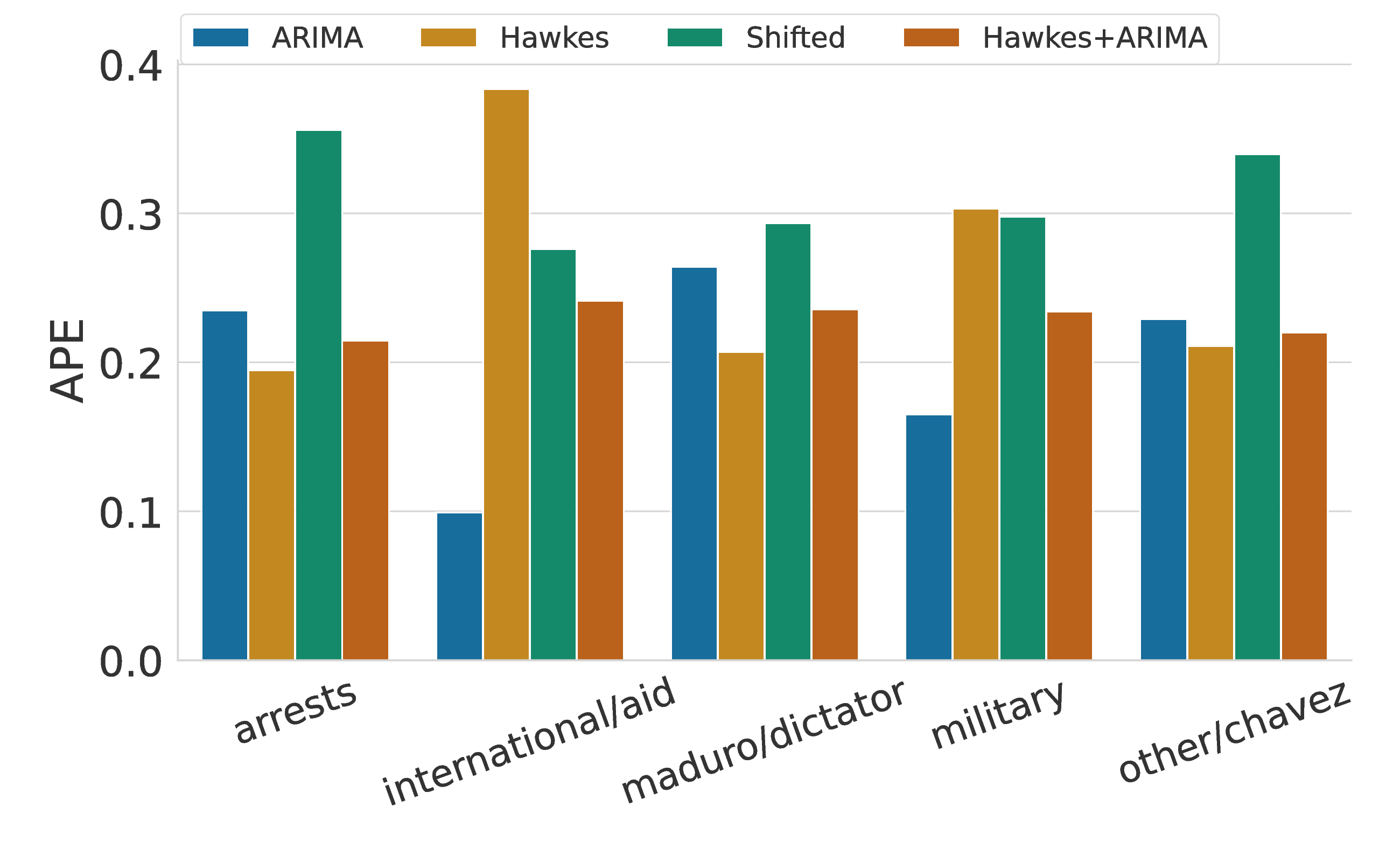}
			\label{fig:cp4_tw_ape_h}
	}
	&
	\subfloat[Vz19 (YouTube)]{
		\includegraphics[width=0.4\linewidth]{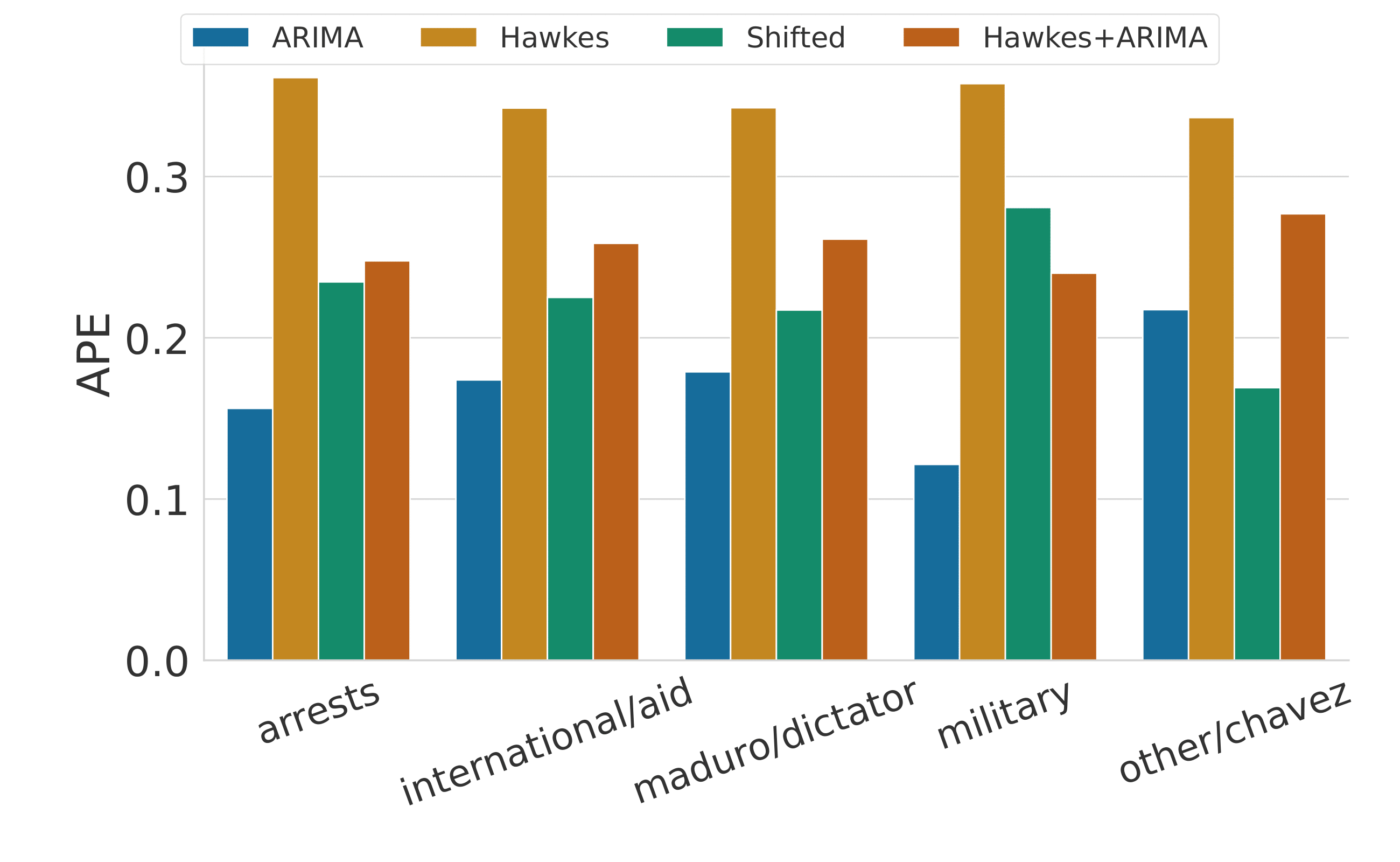}
			\label{fig:cp4_yt_ape_h}
	}
	
	\\
	\subfloat[CPEC (Twitter)]{
		\includegraphics[width=0.4\linewidth]{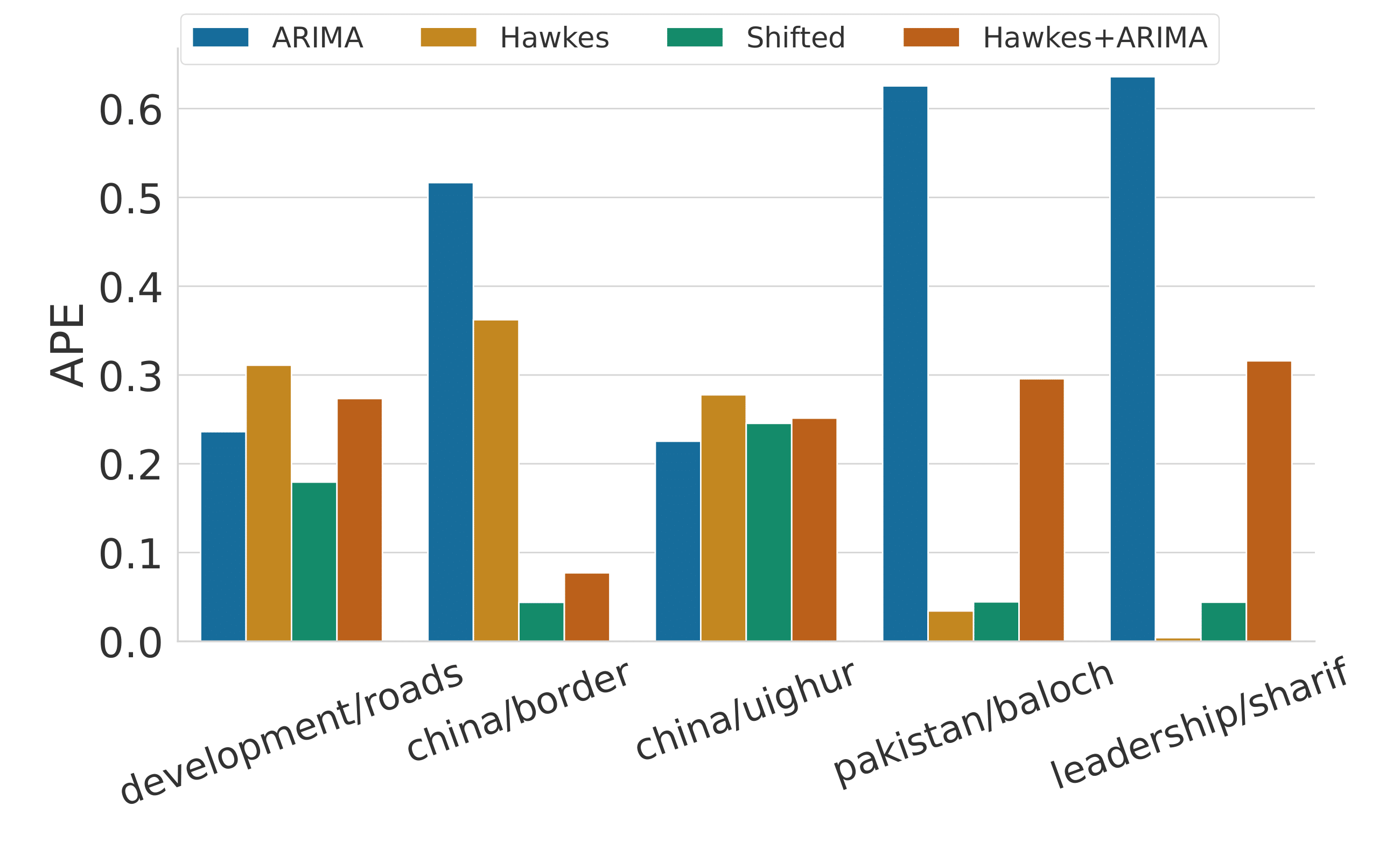}
			\label{fig:cp5_tw_ape_h}
	}
&
	\subfloat[CPEC (YouTube)]{
		\includegraphics[width=0.4\linewidth]{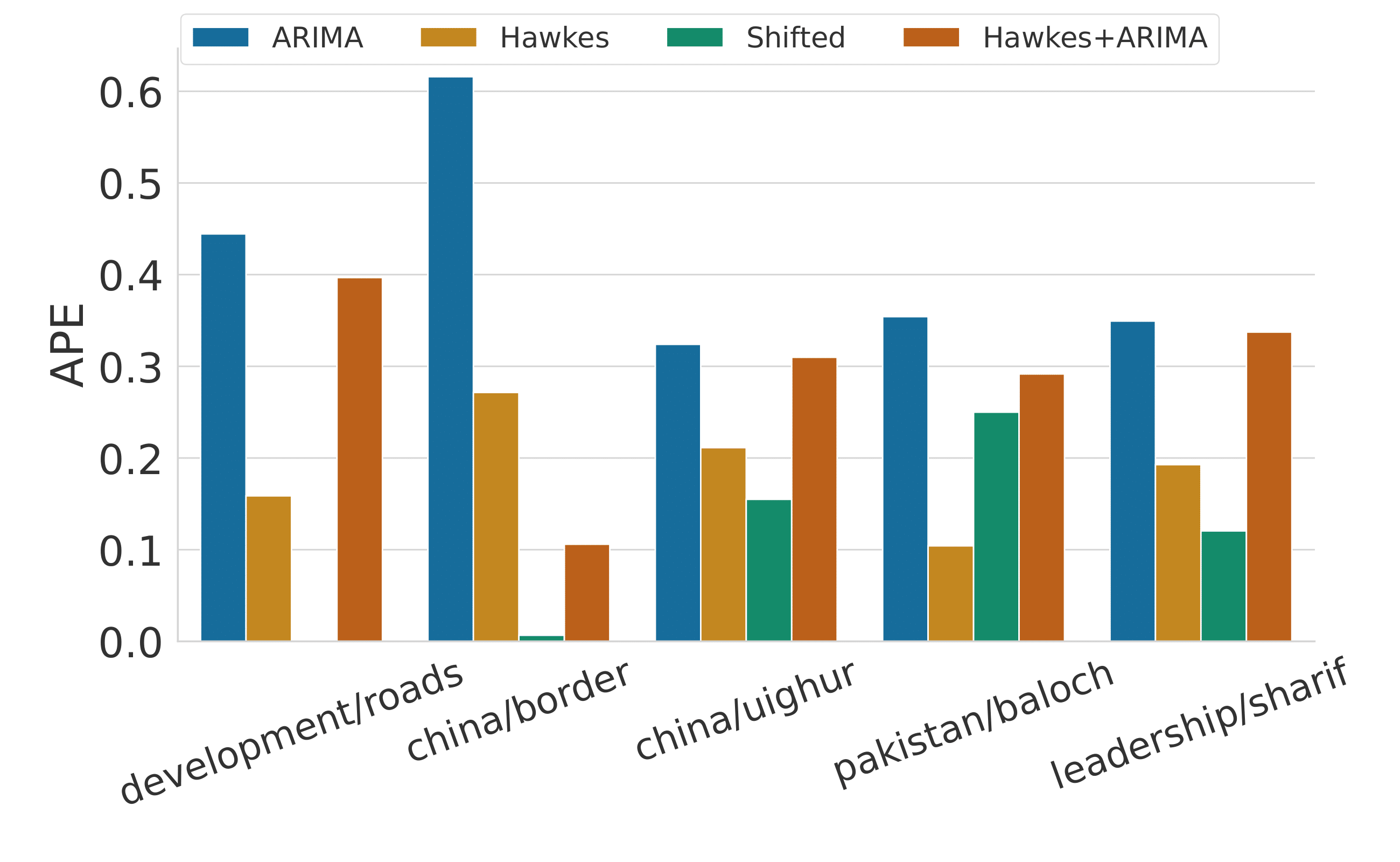}
			\label{fig:cp5_yt_ape_h}
	}
	\\
	\subfloat[BRIA (Twitter)]{
		\includegraphics[width=0.4\linewidth]{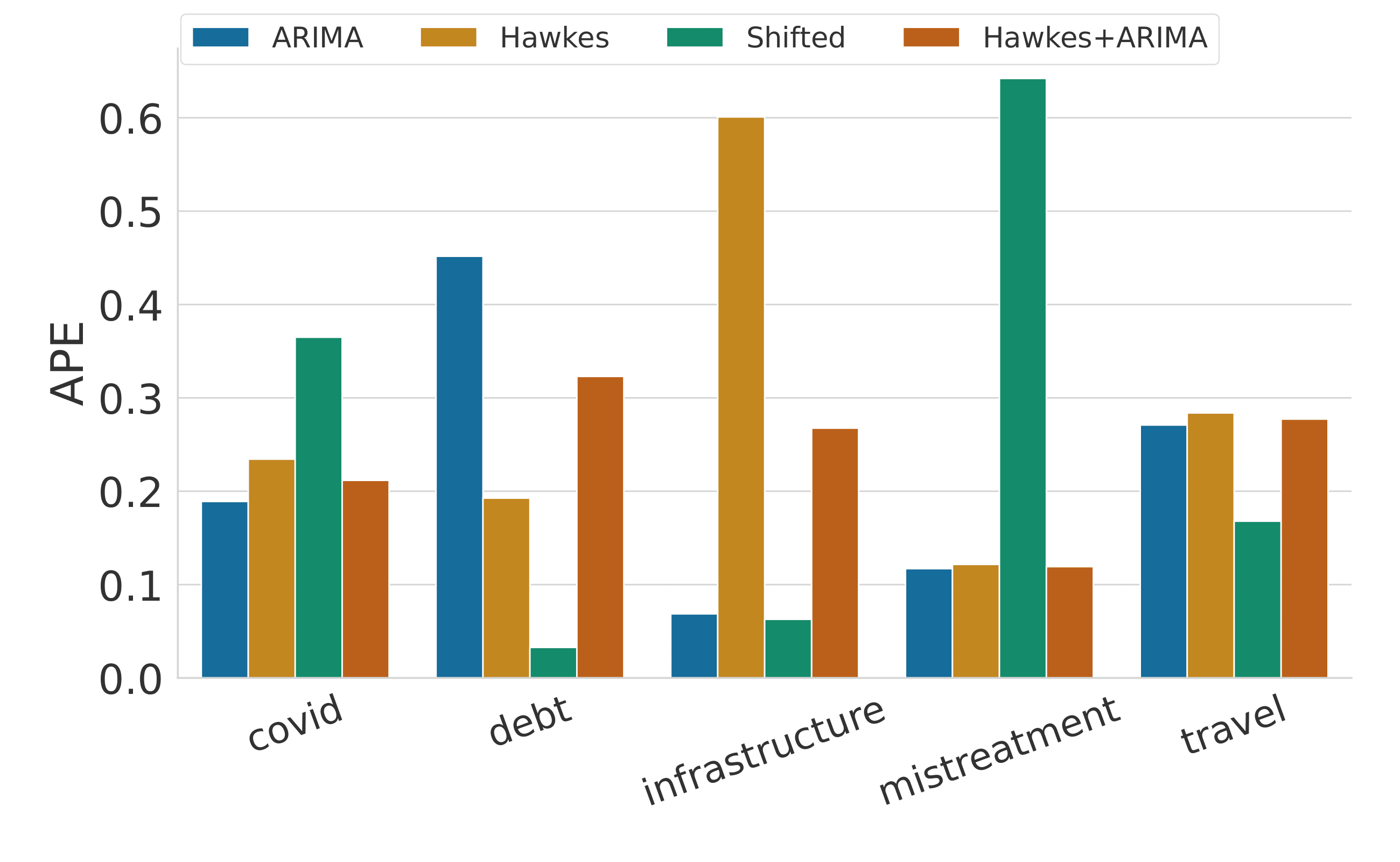}
			\label{fig:cp6_tw_ape_h}
	}
	&
	\subfloat[BRIA (YouTube)]{
		\includegraphics[width=0.4\linewidth]{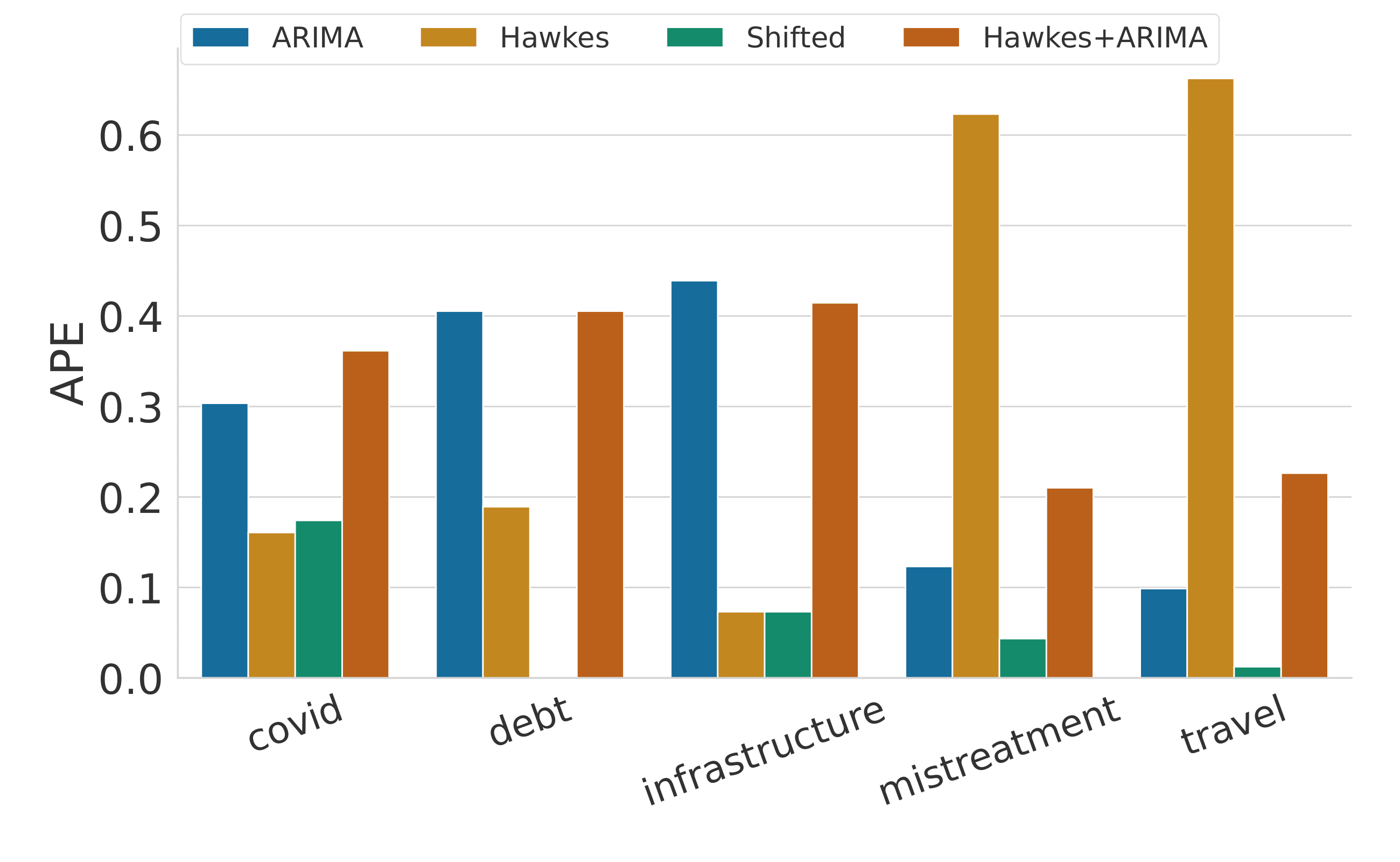}
			\label{fig:cp6_yt_ape_h}
	}
\end{tabular}
	 \caption{APE hourly performance across topics for each context and over two platforms (lower is better). APE scores for each topic are normalized between 0 to 1 relative to the sum of the baselines' errors. The results are for one week predictions at hourly granularity.}
	 \label{fig:ape_topic_performance_h}
\end{figure*}

\begin{figure*}[htbp]
\centering
\begin{tabular}{ccc}
	\subfloat[Vz19 (Twitter)]{
		\includegraphics[width=0.4\linewidth]{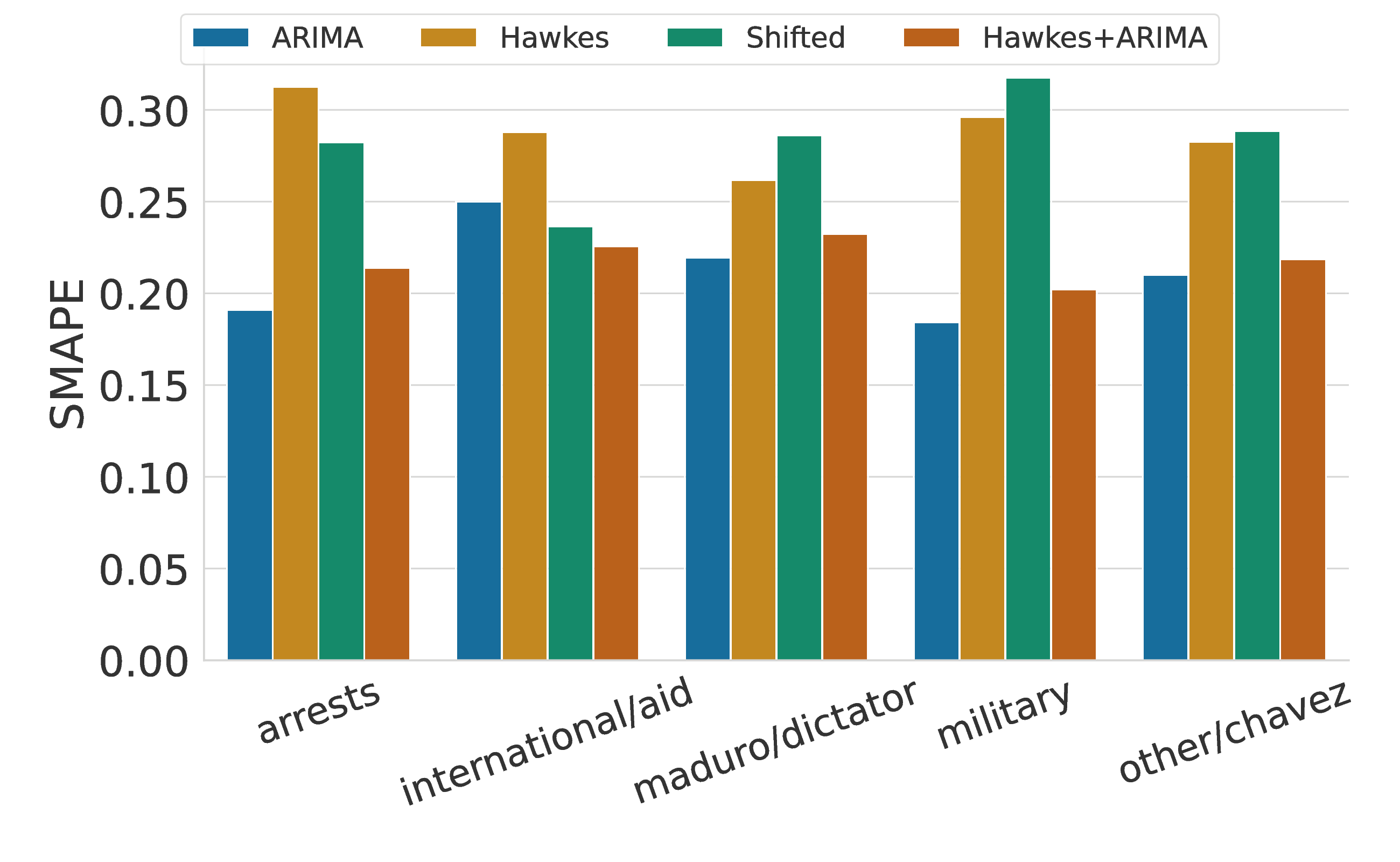}
			\label{fig:cp4_tw_smape_h}
	}
	&
	\subfloat[Vz19 (YouTube)]{
		\includegraphics[width=0.4\linewidth]{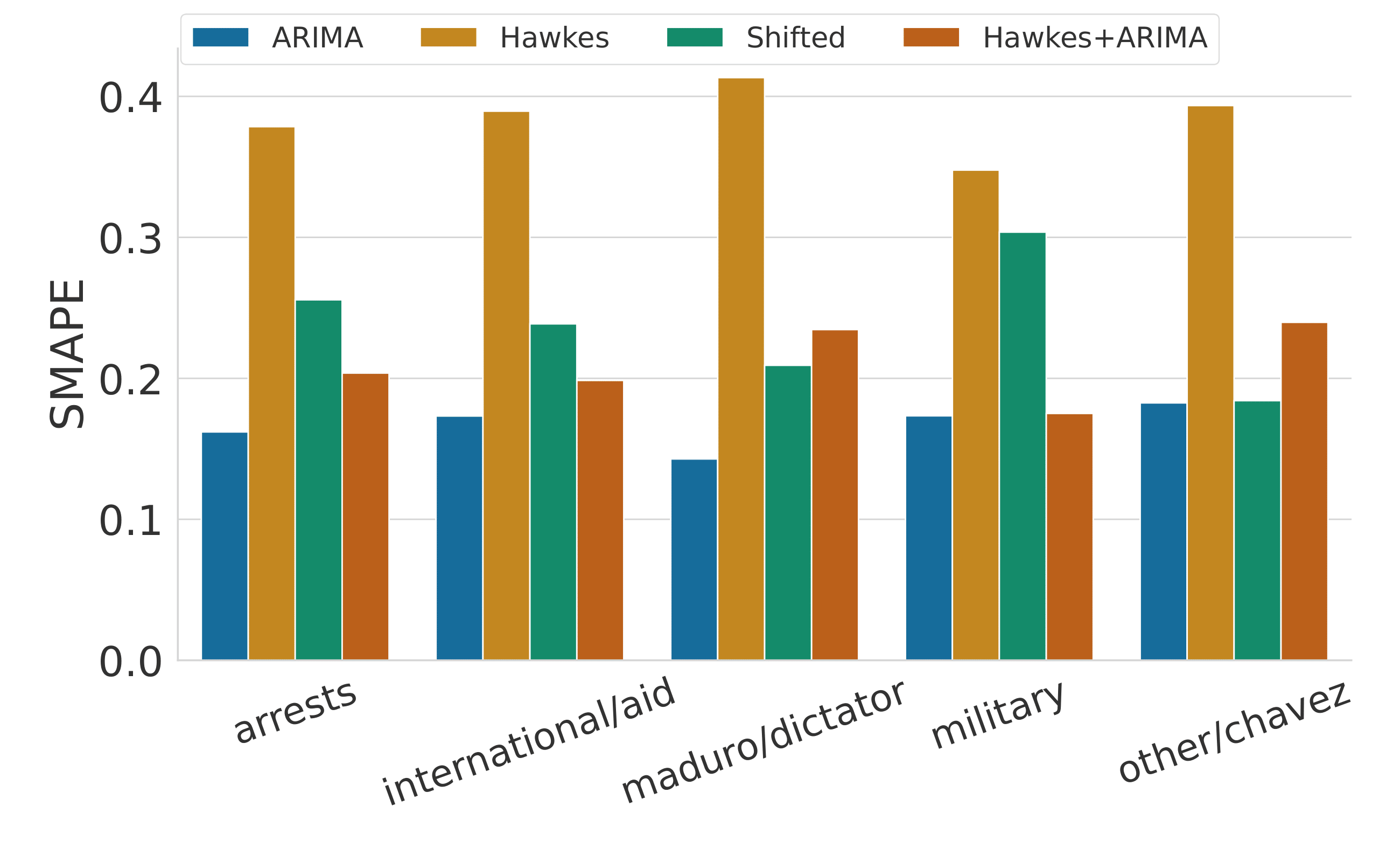}
			\label{fig:cp4_yt_smape_h}
	}
	\\
	\subfloat[CPEC (Twitter)]{
		\includegraphics[width=0.4\linewidth]{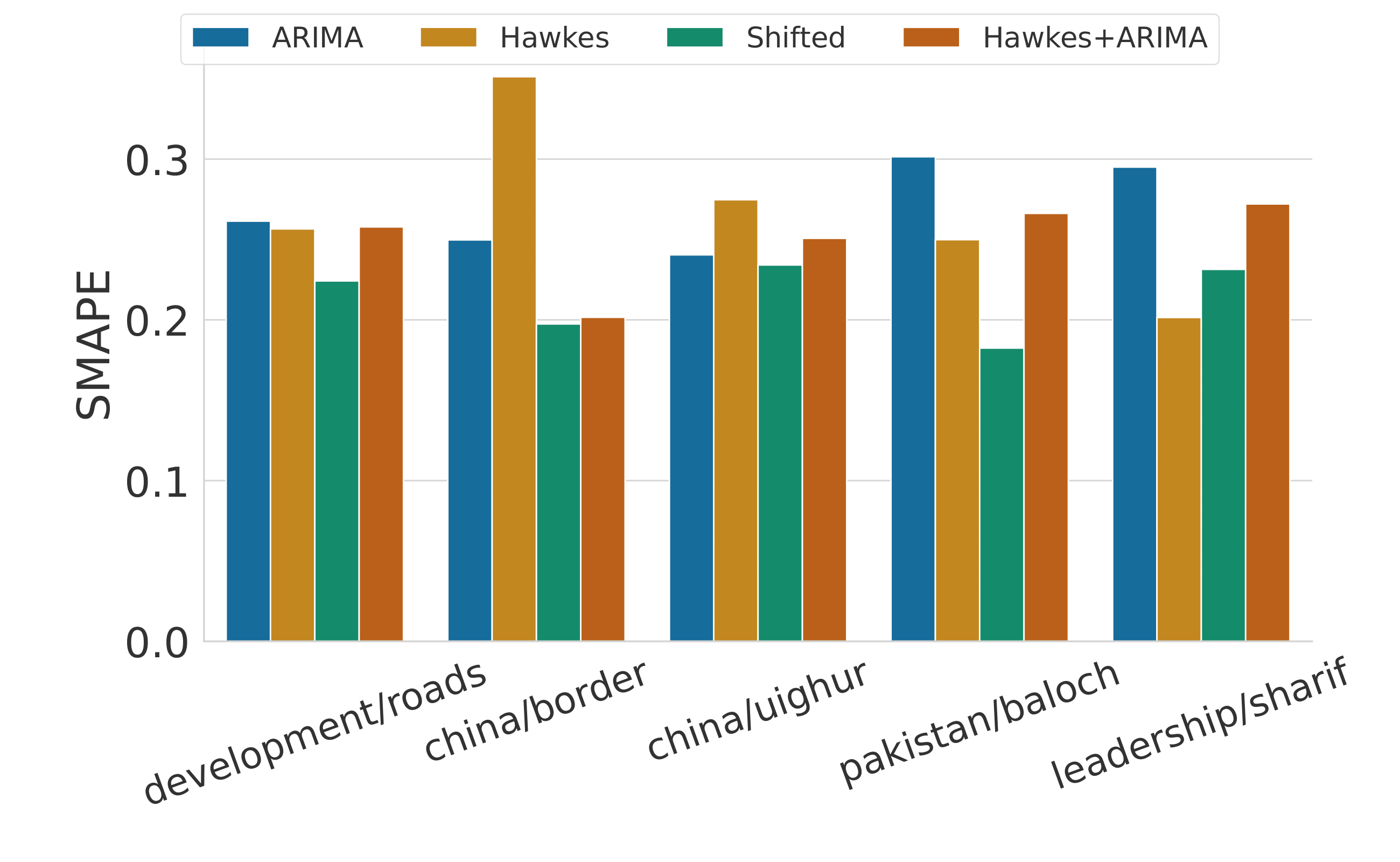}
			\label{fig:cp5_tw_smape_h}
	}
	&
	\subfloat[CPEC (YouTube)]{
		\includegraphics[width=0.4\linewidth]{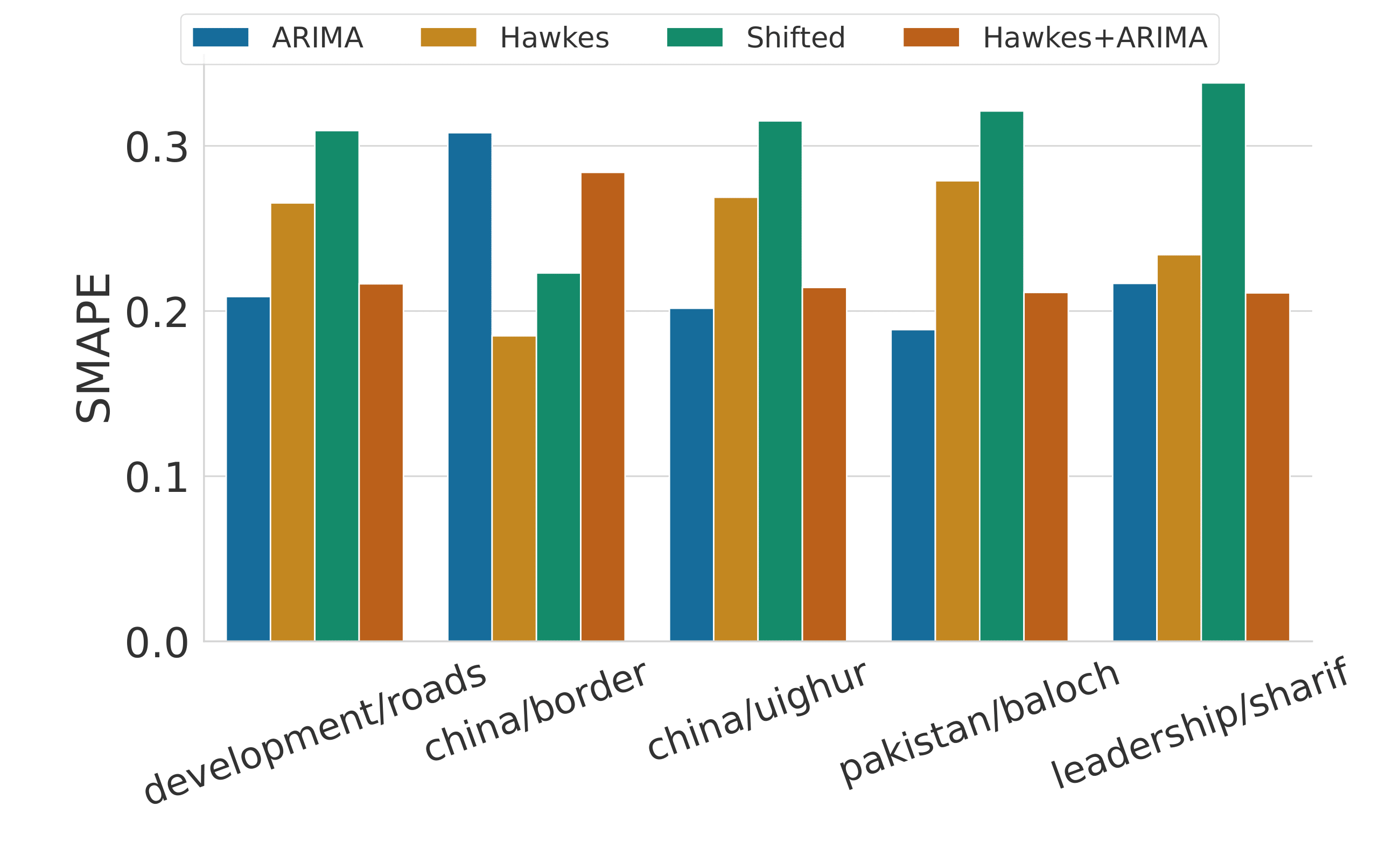}
			\label{fig:cp5_yt_smape_h}
	}
	\\
	\subfloat[BRIA (Twitter)]{
		\includegraphics[width=0.4\linewidth]{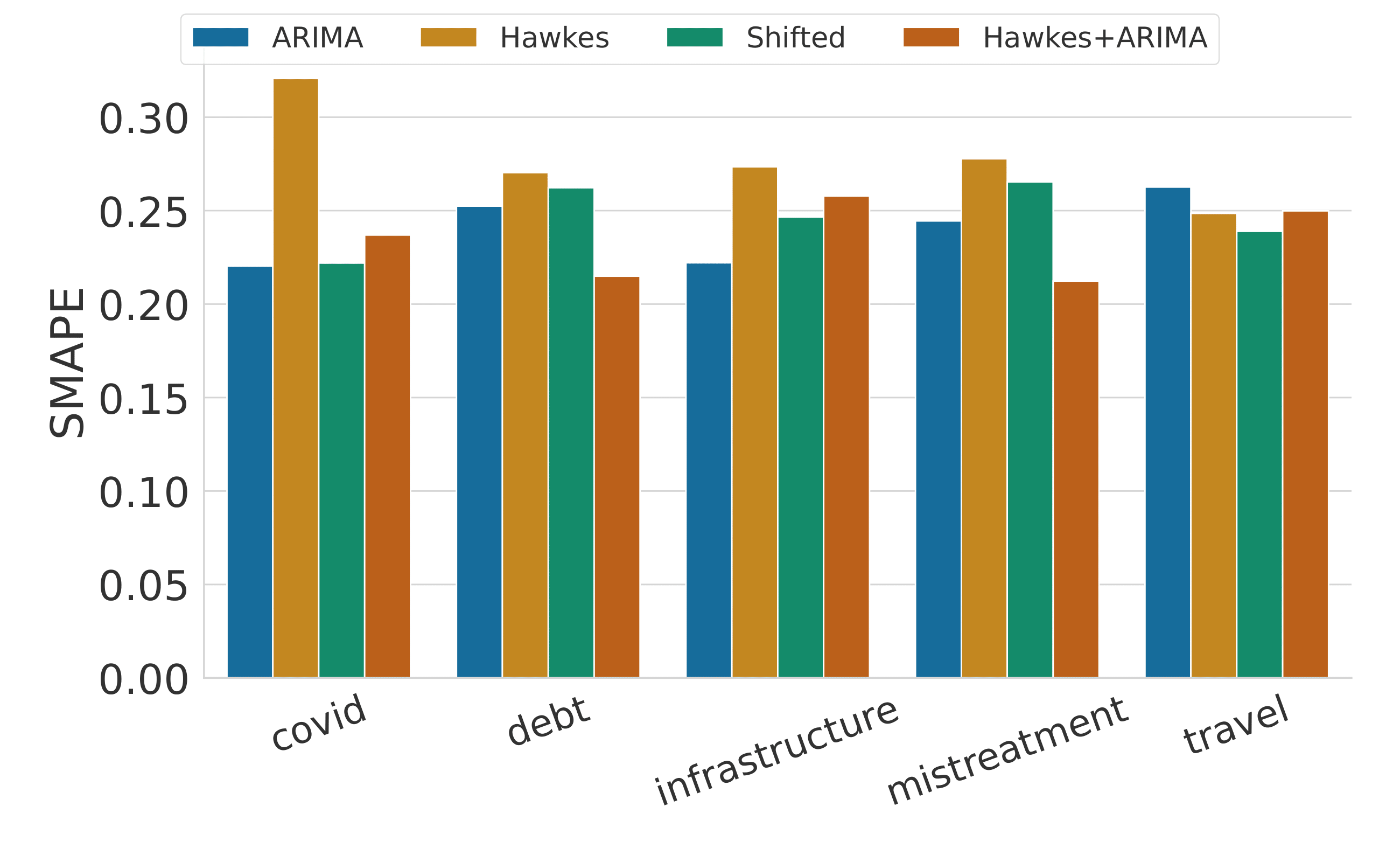}
			\label{fig:cp6_tw_smape_h}
	}
	&
	\subfloat[BRIA (YouTube)]{
		\includegraphics[width=0.4\linewidth]{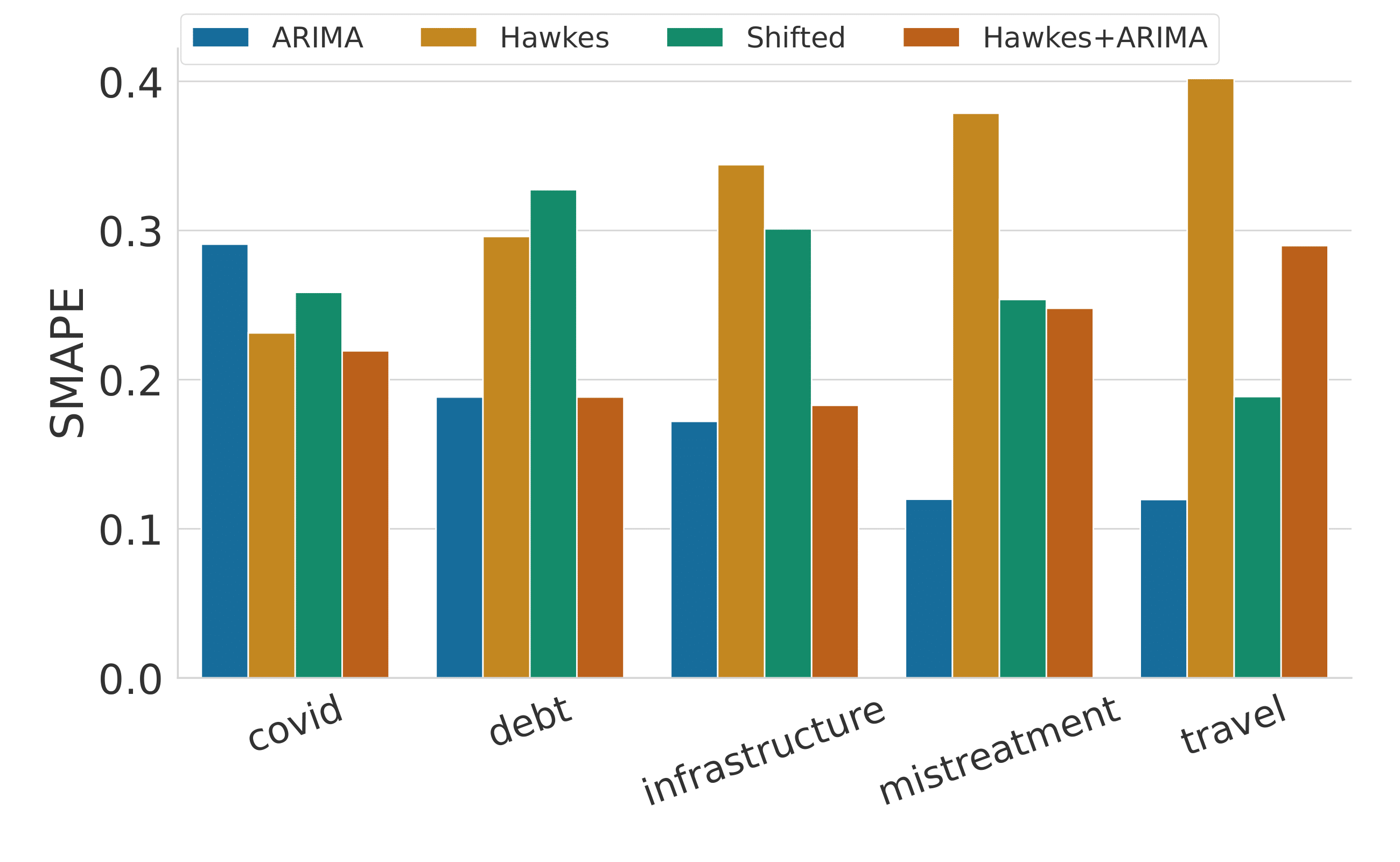}
			\label{fig:cp6_yt_smape_h}
	}

\end{tabular}
	 \caption{sMAPE hourly performance across topics for each context and over two platforms (lower is better). sMAPE scores for each topic are normalized between 0 to 1 relative to the sum of the baselines' errors. The results are for one week predictions at hourly granularity.}
	 \label{fig:smape_topic_performance_h}
\end{figure*}

APE and sMAPE are measures for the volume of activity. 
Figure~\ref{fig:ape_topic_performance_h} presents the APE and Figure~\ref{fig:smape_topic_performance_h} the sMAPE per topic for Twitter and YouTube. 
We make the following observations.  

First, there is consistency across topics within the same context, but not across contexts with regards to predicting the total volume of activities.
For example, in CPEC, ARIMA performs poorly across most topics on the two platforms (Figures~\ref{fig:cp5_tw_ape_h} and~\ref{fig:cp5_yt_ape_h}), yet in Vz19 it is best at predicting the total volume of activity overall (2 out of 5 topics in Twitter and 4 out of 5 topics in YouTube). 
Another example is that Hawkes processes perform the worst on all Vz19 topics in YouTube (Figure~\ref{fig:cp4_yt_ape_h}).  
A third example is the shifted baseline which performs the best overall in CPEC (6 out of 10 topics) and BRIA (7 out of 10 topics). 
One reason for the good performance of the shifted baseline in CPEC/BRIA (or, conversely, its poor performance for Vz19 topics) is the context of our datasets: social media activity in Vz19 (Venezuela political unrest) is strongly correlated to exogenous activity (such as street protests, power outages, political declarations), thus simply replaying the past does not capture potential reactions to current events. 
Moreover, the shifted baseline replays the actions from seven days before, while Hawkes processes and ARIMA at least take into account recent history. 

Second, we note that ARIMA performs consistently better than other baselines in sMAPE for 12 out of 15 topics on YouTube (as shown in Figures~\ref{fig:cp4_yt_smape_h}, \ref{fig:cp5_yt_smape_h}, \ref{fig:cp6_yt_smape_h}).
For low-activity topics, as is the case for most topics in YouTube, ARIMA can more accurately predict the volume of activities per hour while baselines such as Hawkes or the Shifted baseline often over predict.
On the other hand, for Twitter, there is not a persistent winner across most topics using sMAPE.
Yet ARIMA is rarely ranked as the worst performing baseline in this metric (only 3 out of 15 topics in total).

\subsubsection{Temporal Patterns}

\begin{figure*}[htbp]
\centering
\scalebox{0.85}{
\begin{tabular}{ccc}
	\subfloat[international/aid (Vz19)]{
		\includegraphics[width=0.32\linewidth]{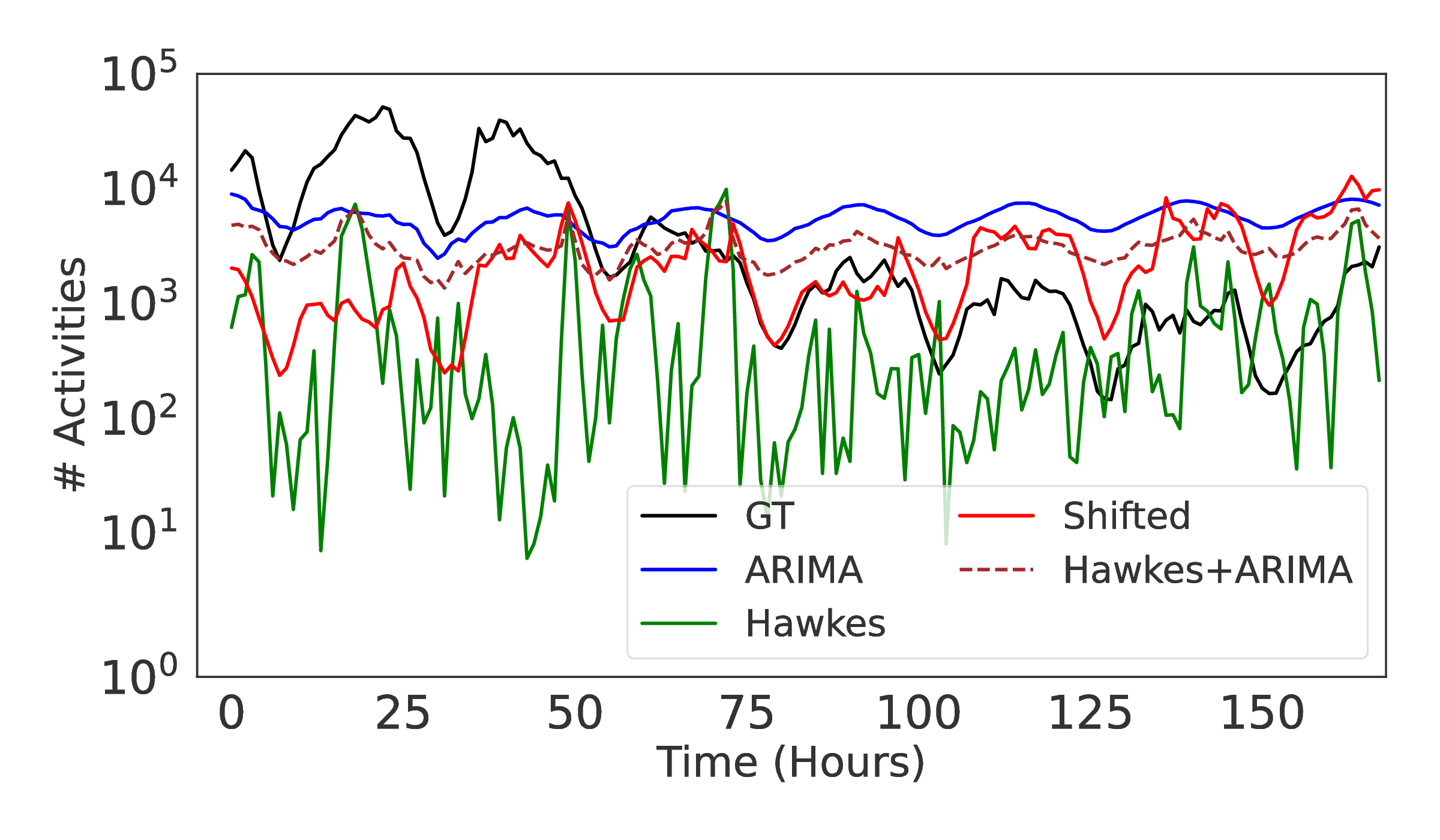}
			\label{fig:international_aid_tw_h}
	}
	&
	\subfloat[military (Vz19)]{
		\includegraphics[width=0.32\linewidth]{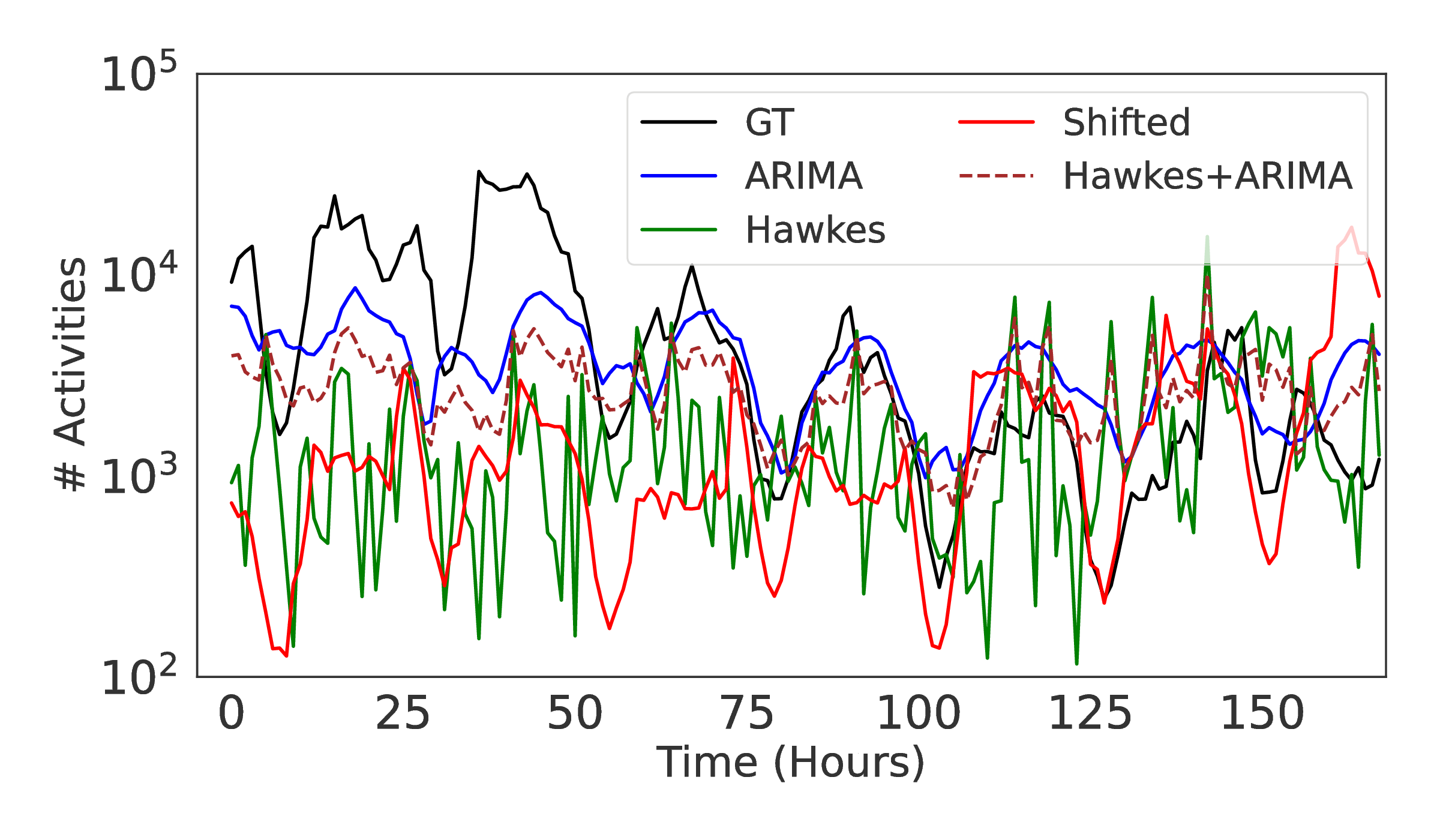}
			\label{fig:military_tw_h}
	}
&

	\subfloat[maduro/dictator (Vz19)]{
		\includegraphics[width=0.32\linewidth]{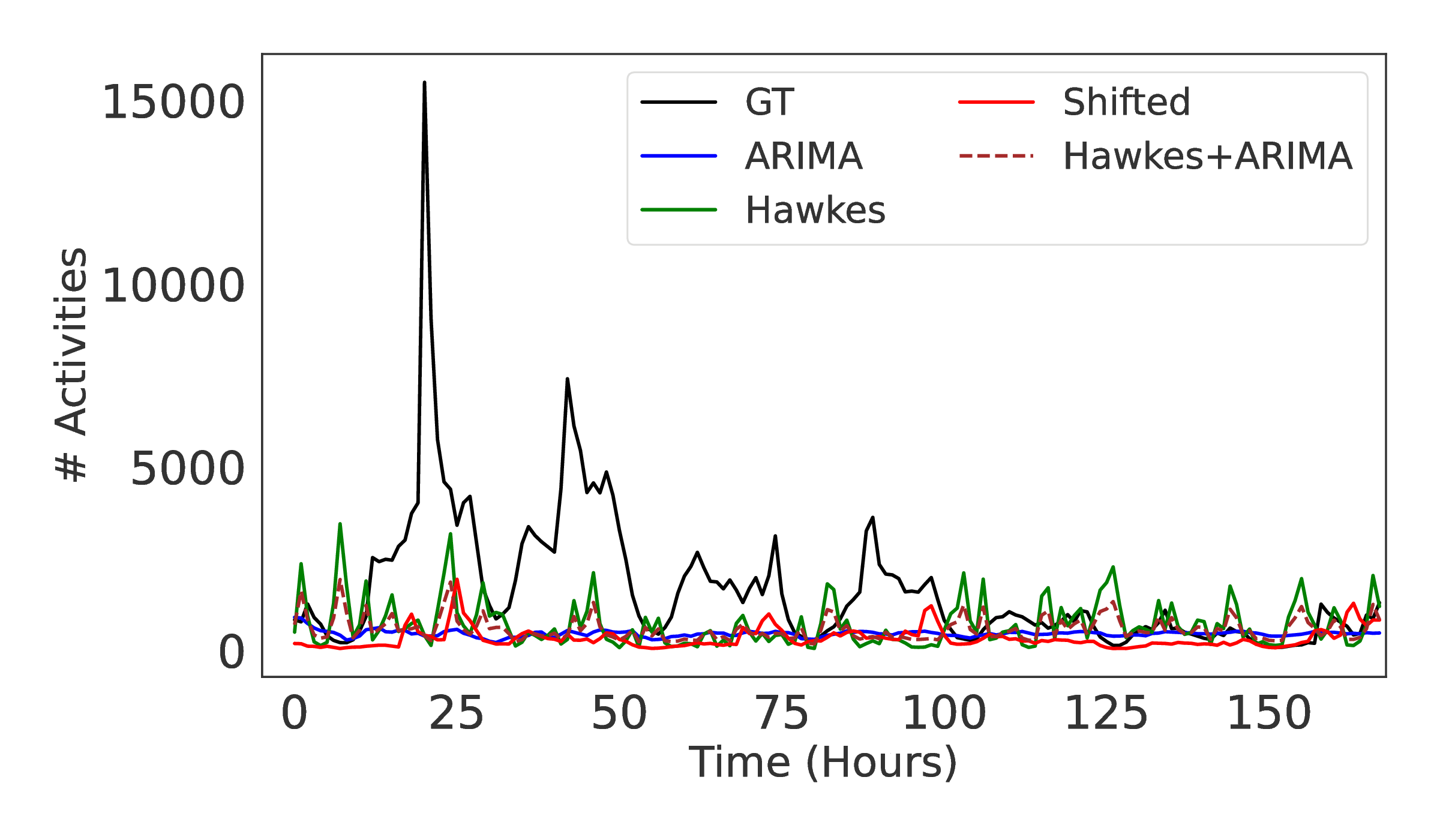}
			\label{fig:maduro_dictator_tw_h}
	}
\\
	\subfloat[controversies/china/border (CPEC)]{
		\includegraphics[width=0.32\linewidth]{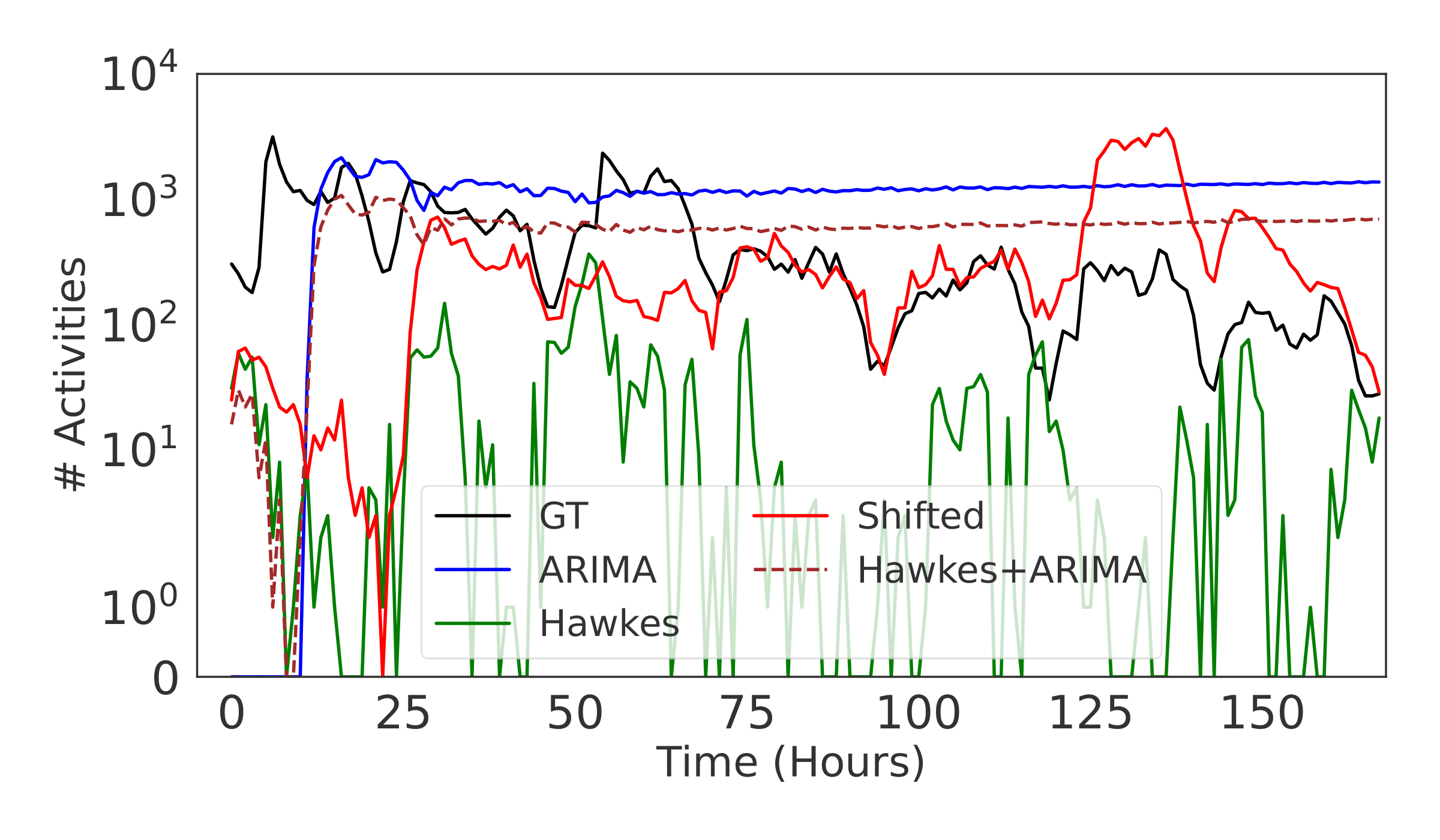}
			\label{fig:china_border_tw_h}
	}
	&
	\subfloat[leadership/sharif (CPEC)]{
		\includegraphics[width=0.32\linewidth]{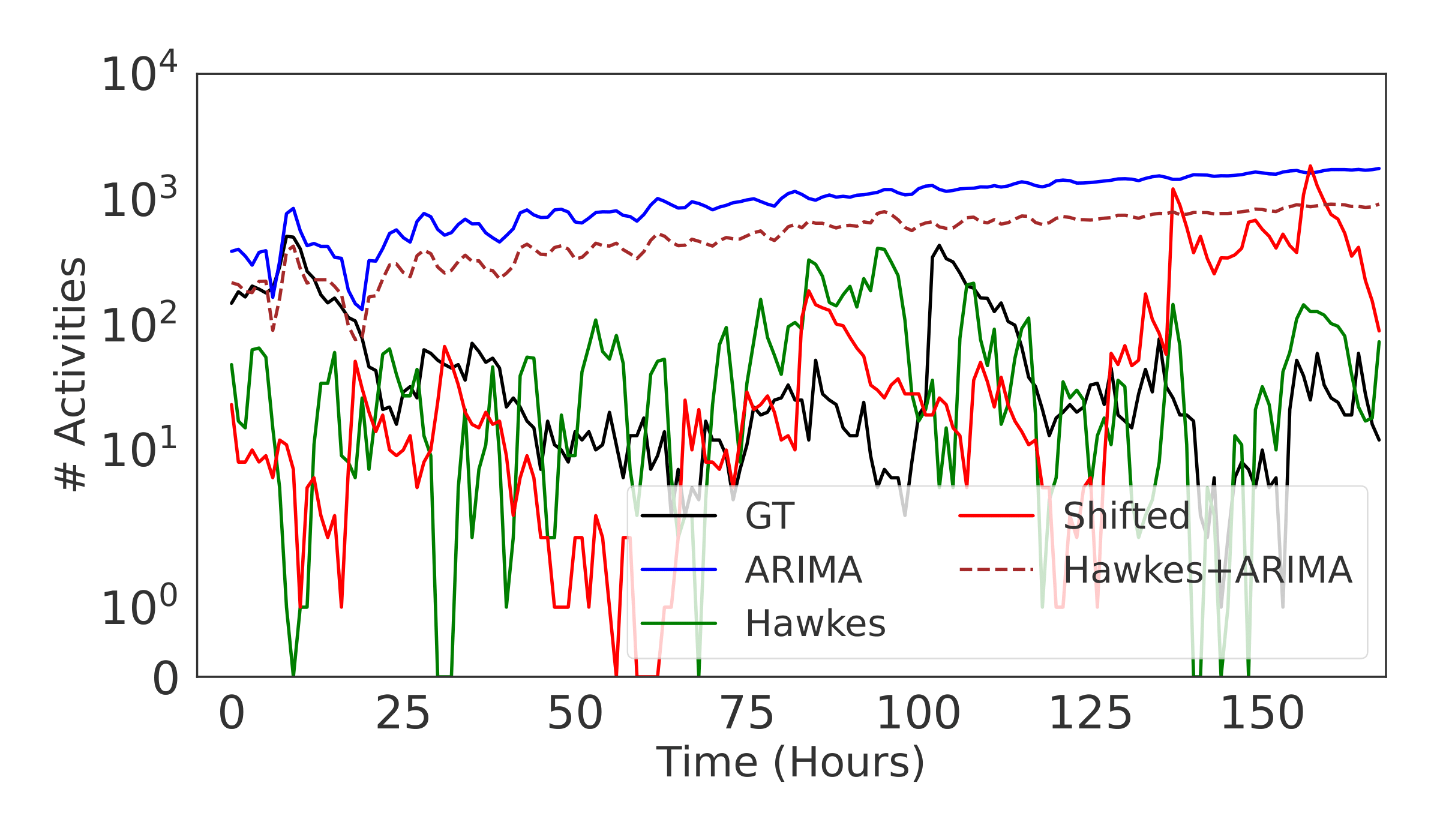}
			\label{fig:leadership_sharif_tw_h}
	}
	&
	\subfloat[controversies/pakistan/baloch (CPEC)]{
		\includegraphics[width=0.32\linewidth]{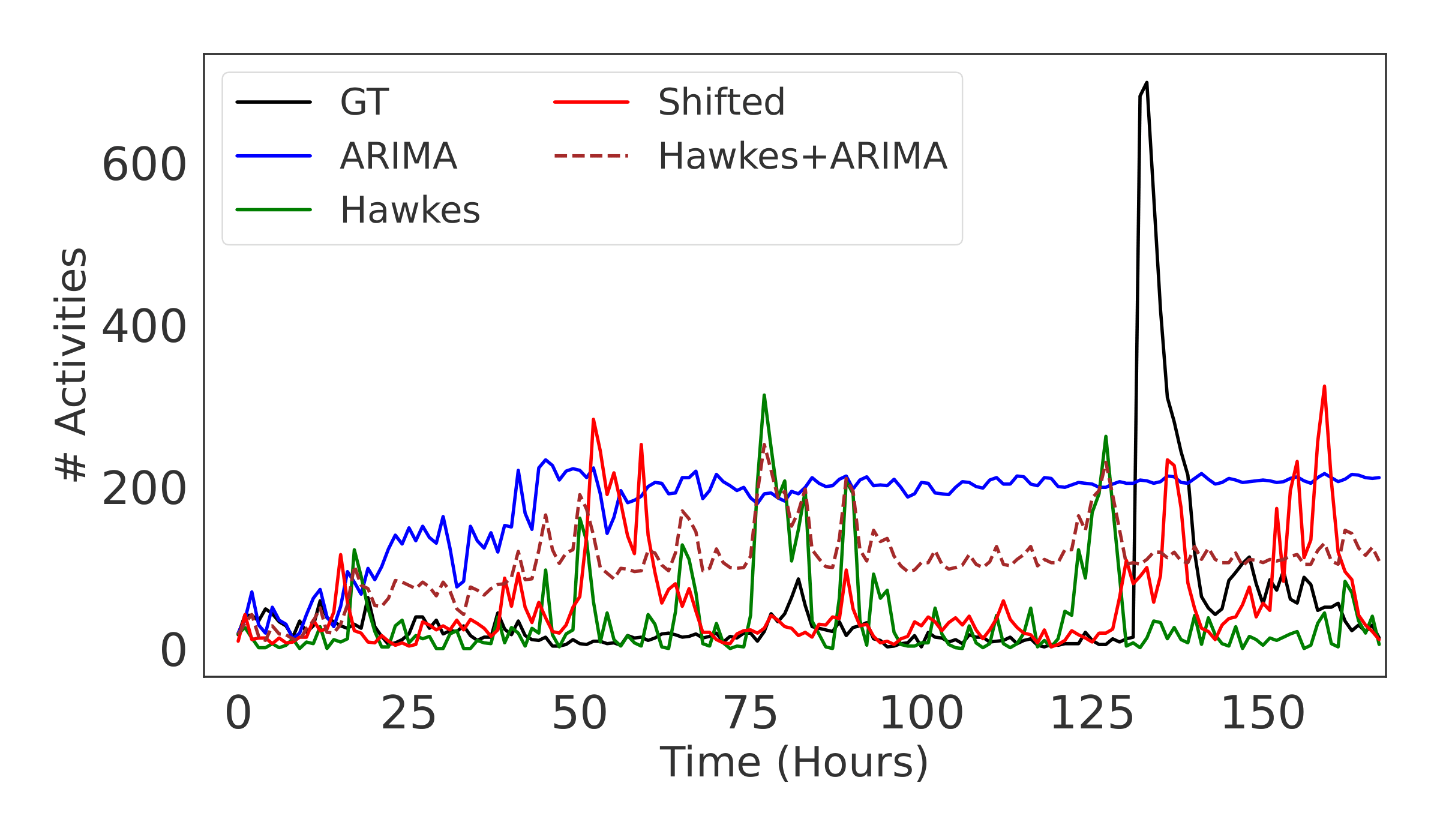}
			\label{fig:pakistan_baloch_tw_h}
	}
\\
	\subfloat[covid (BRIA)]{
		\includegraphics[width=0.32\linewidth]{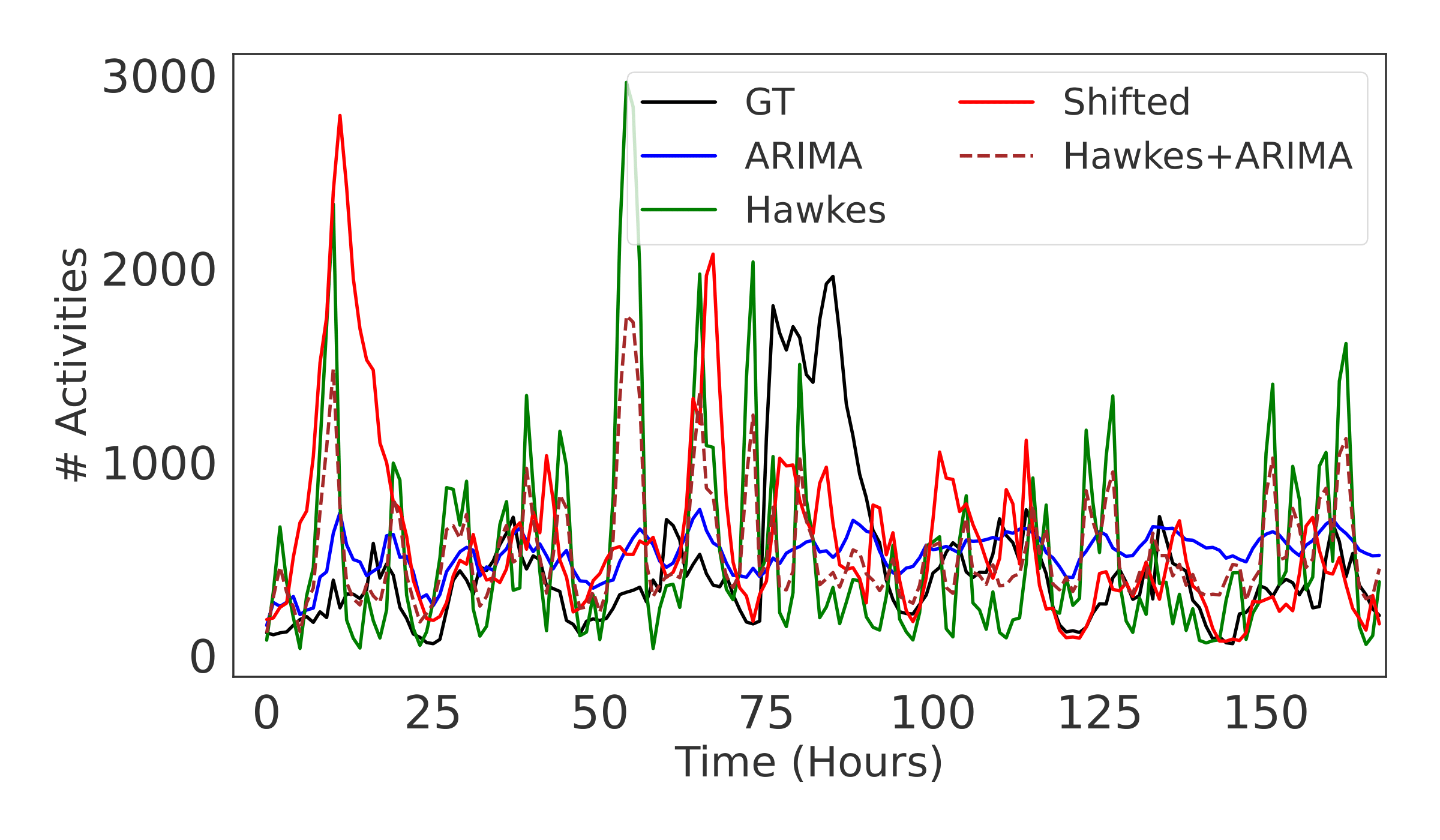}
			\label{fig:covid_tw_h}
	}
	&
	\subfloat[travel (BRIA)]{
		\includegraphics[width=0.32\linewidth]{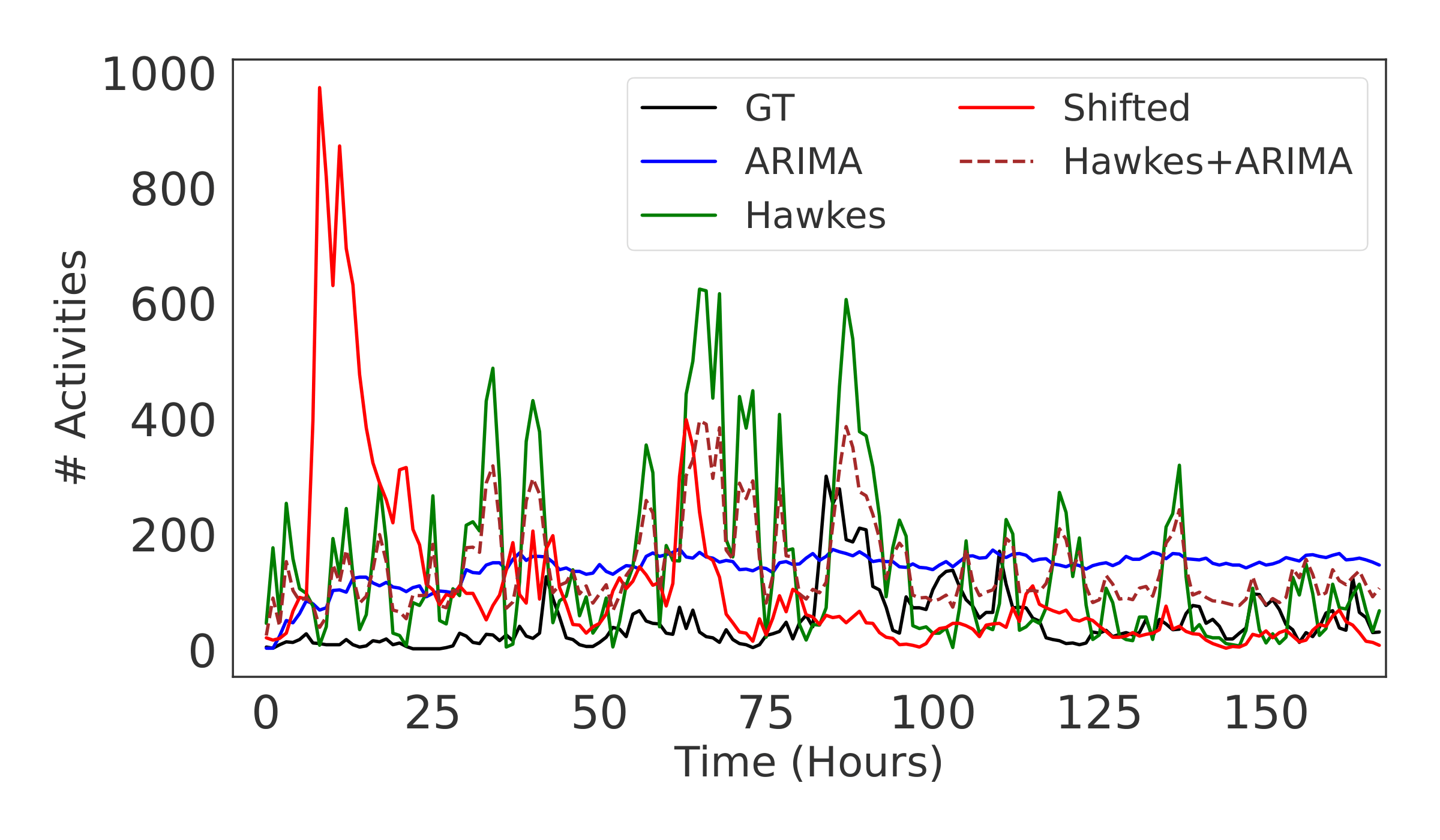}
			\label{fig:travel_tw_h}
	}
	&
	\subfloat[mistreatment (BRIA)]{
		\includegraphics[width=0.32\linewidth]{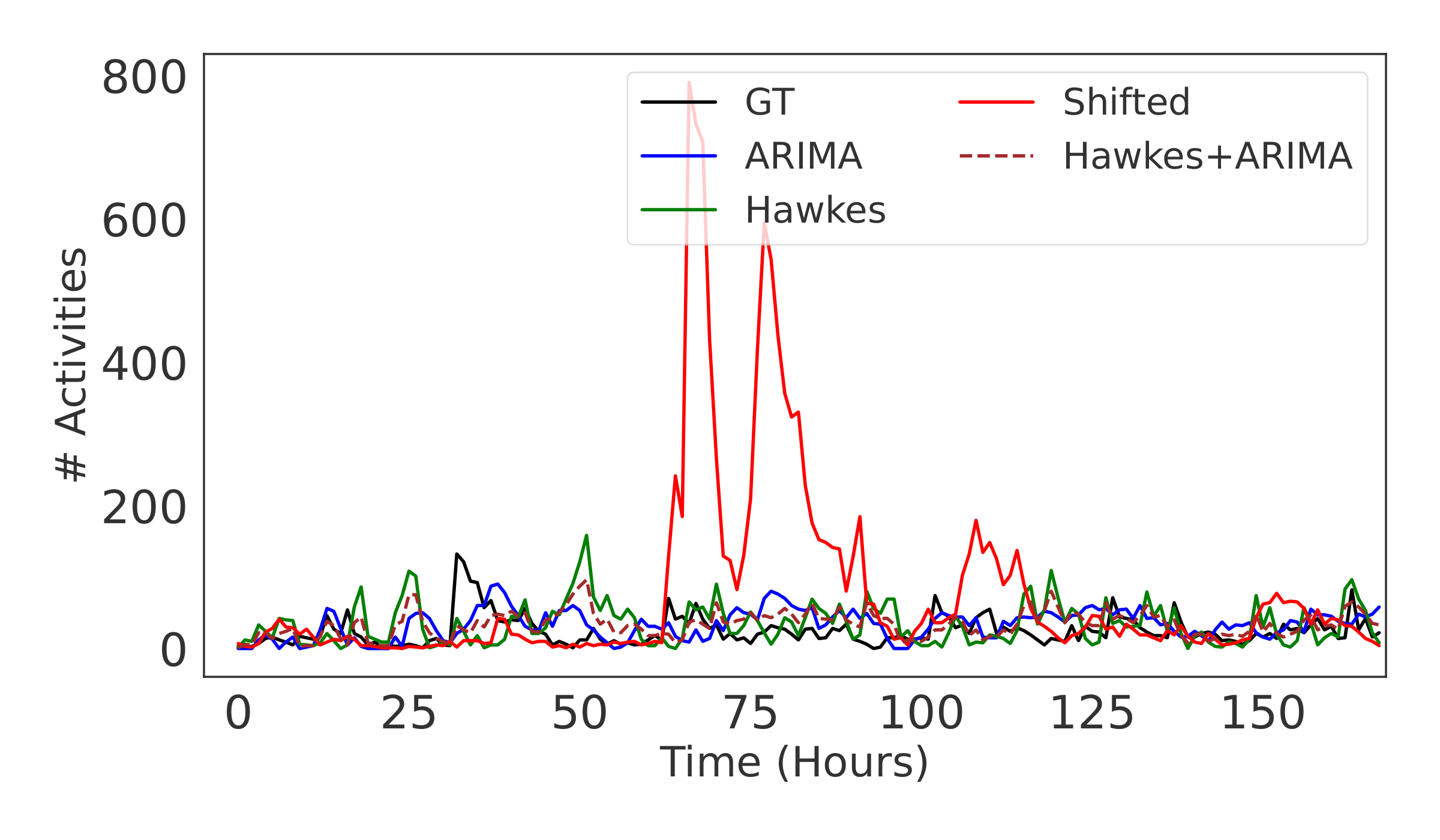}
			\label{fig:mistreatment_tw_h}
	}
\\
\end{tabular}
}
	 \caption{Hourly time series visualizations for Twitter on the top-3 most active topics for each context. We predict one week of activity (168 hours/time steps)}
	 \label{fig:time_series_tw_h}
\end{figure*}

\begin{figure*}[htbp]
\centering
\scalebox{0.85}{
\begin{tabular}{ccc}
	\subfloat[maduro/dictator (Vz19)]{
		\includegraphics[width=0.32\linewidth]{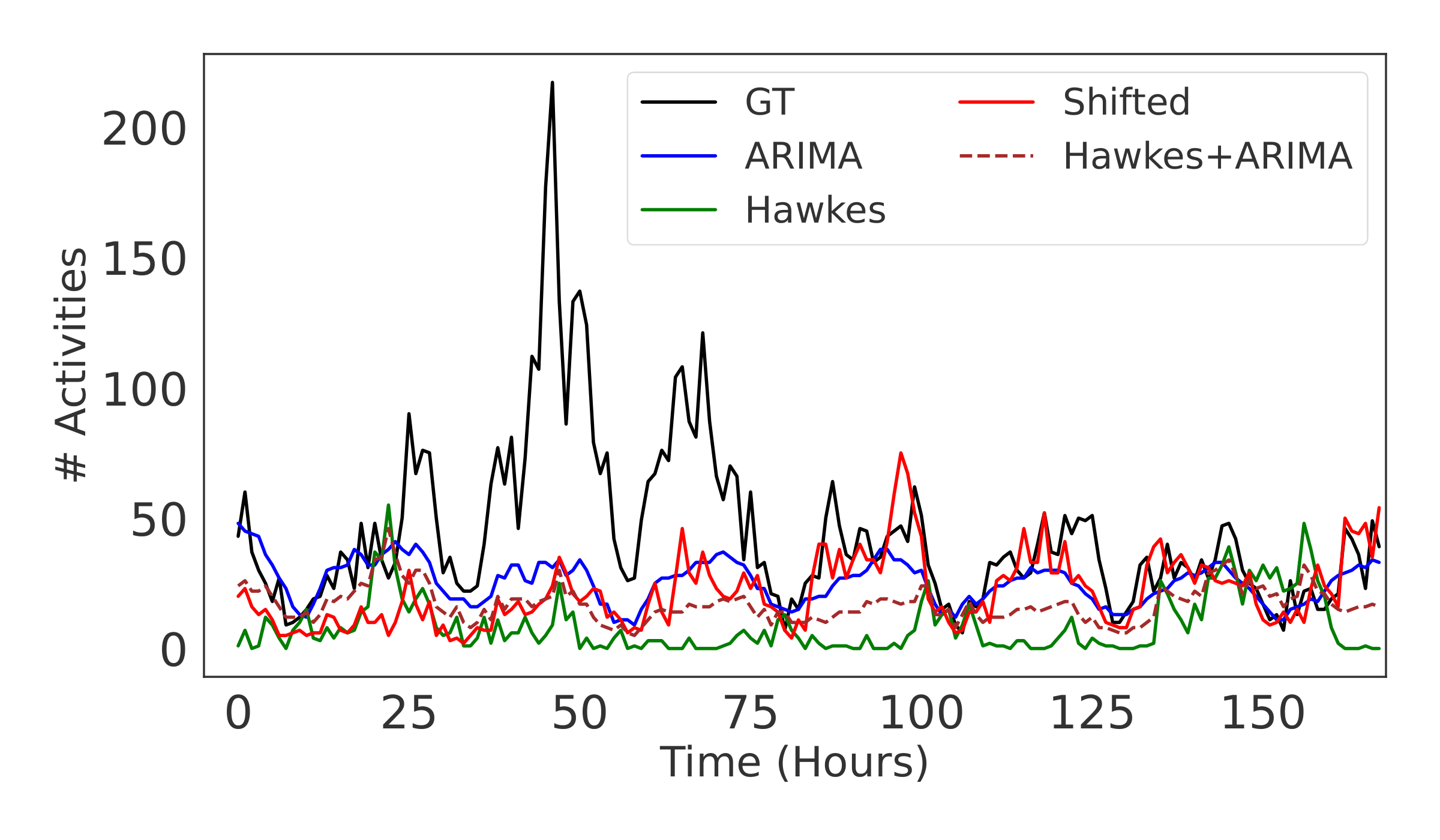}
			\label{fig:maduro_dictator_yt_h}
	}
	&
	\subfloat[international/aid (Vz19)]{
		\includegraphics[width=0.32\linewidth]{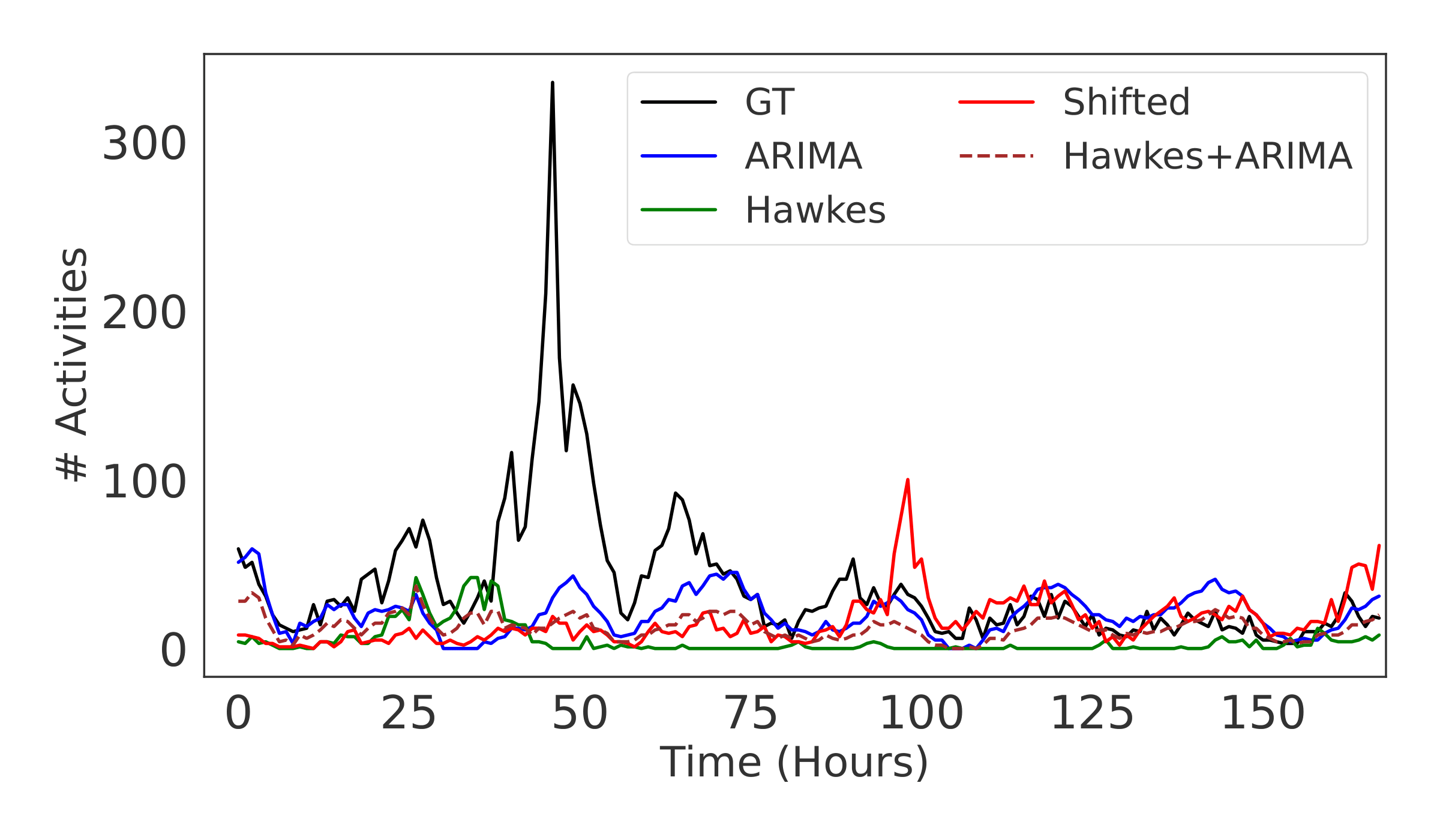}
			\label{fig:international_aid_yt_h}
	}
&
	\subfloat[other/ch\'{a}vez (Vz19)]{
		\includegraphics[width=0.32\linewidth]{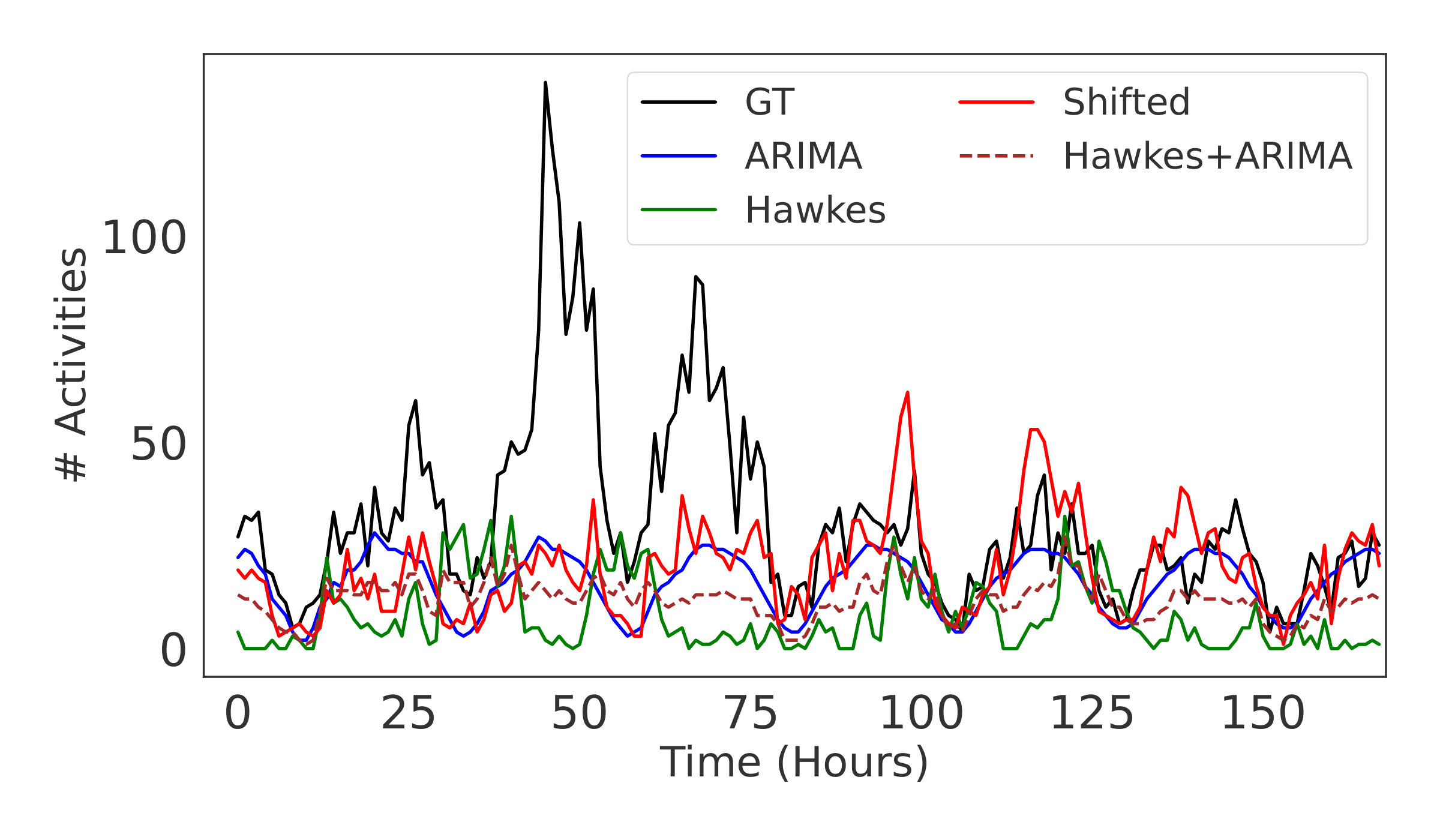}
			\label{fig:other_chavez_yt_h}
	}
\\
	\subfloat[controversies/china/border (CPEC)]{
		\includegraphics[width=0.32\linewidth]{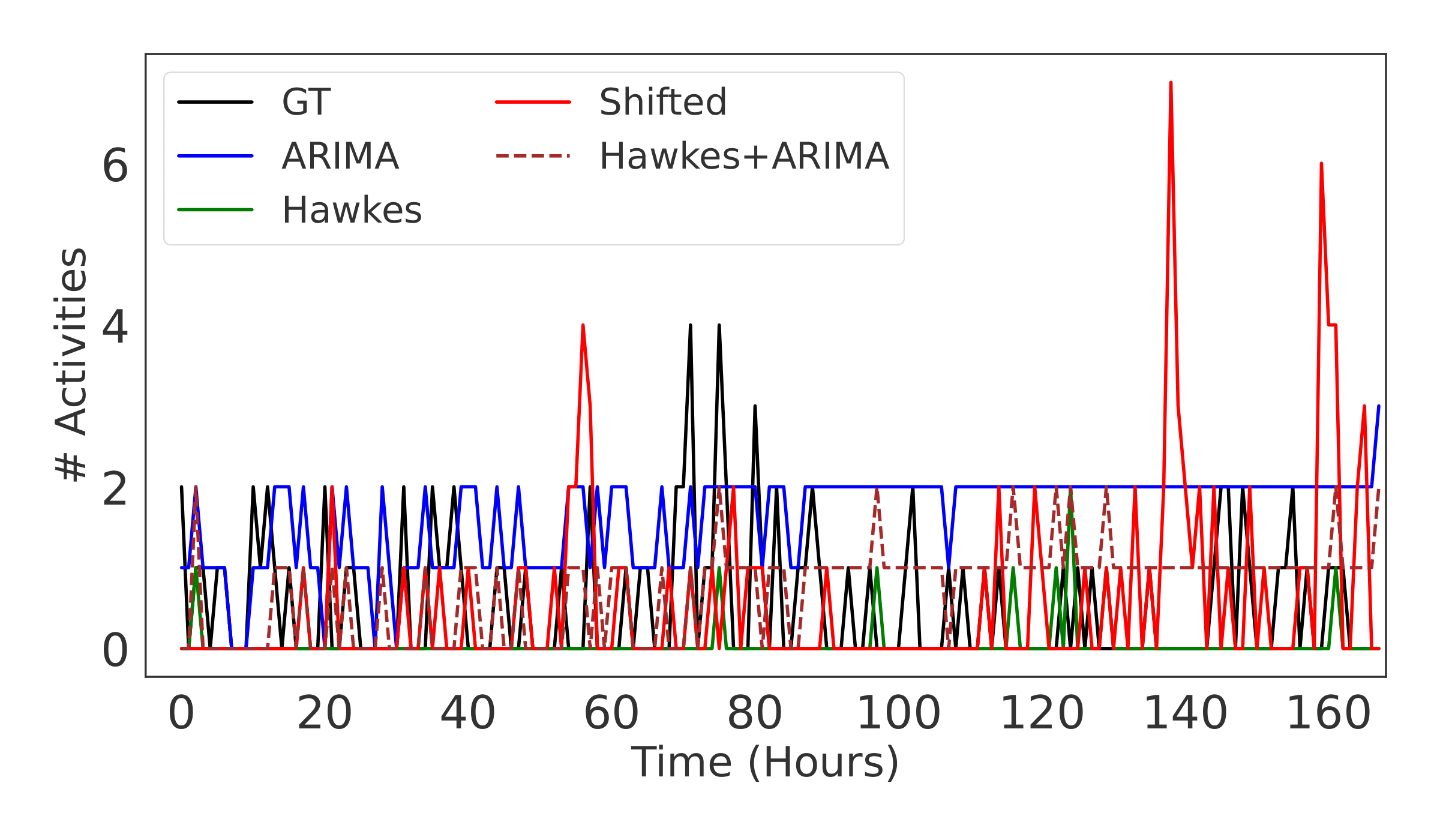}
			\label{fig:china_border_yt_h}
	}
	&
	\subfloat[leadership/sharif (CPEC)]{
		\includegraphics[width=0.32\linewidth]{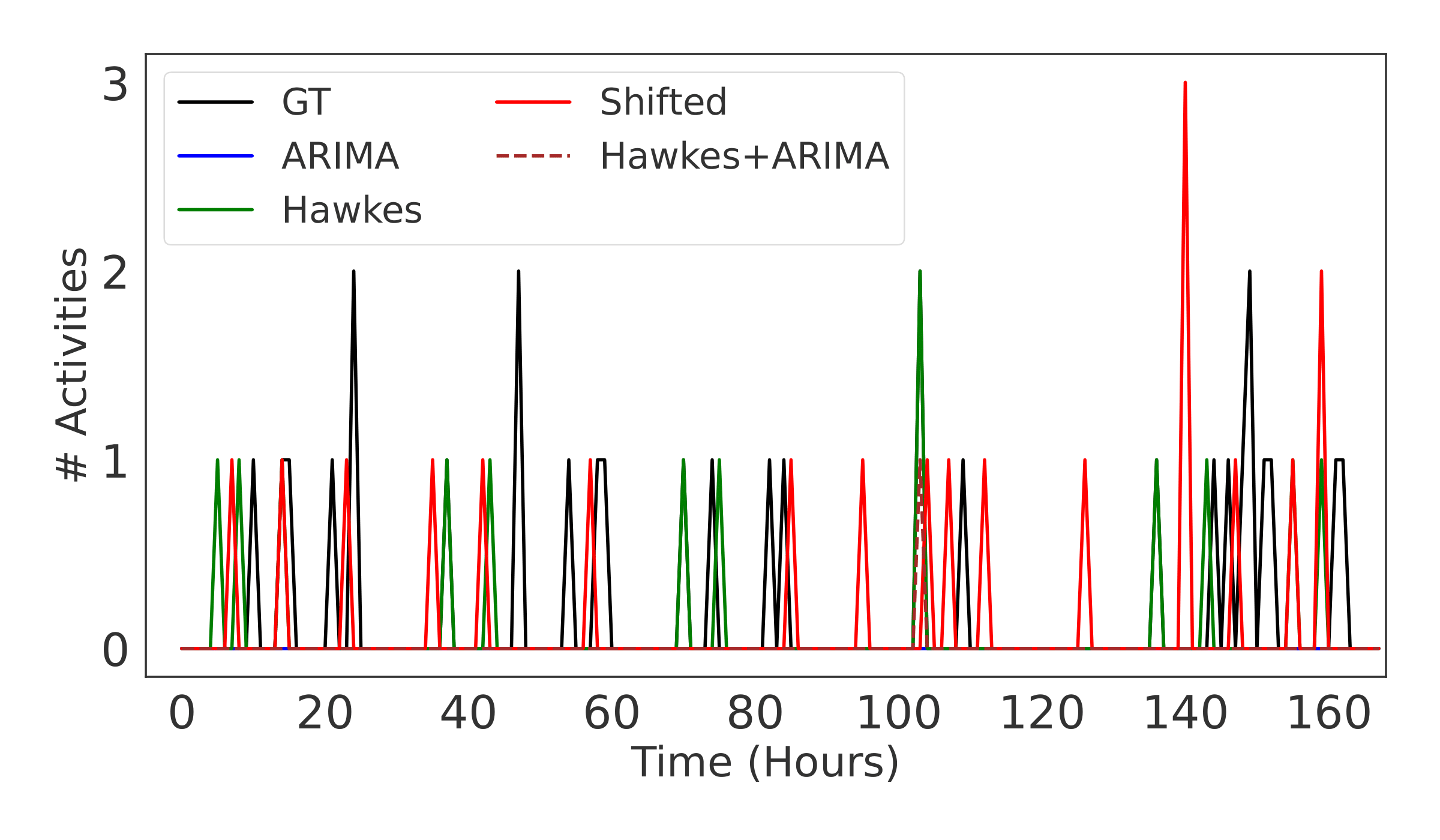}
			\label{fig:leadership_sharif_yt_h}
	}
	&
	\subfloat[benefits/development/roads (CPEC)]{
		\includegraphics[width=0.32\linewidth]{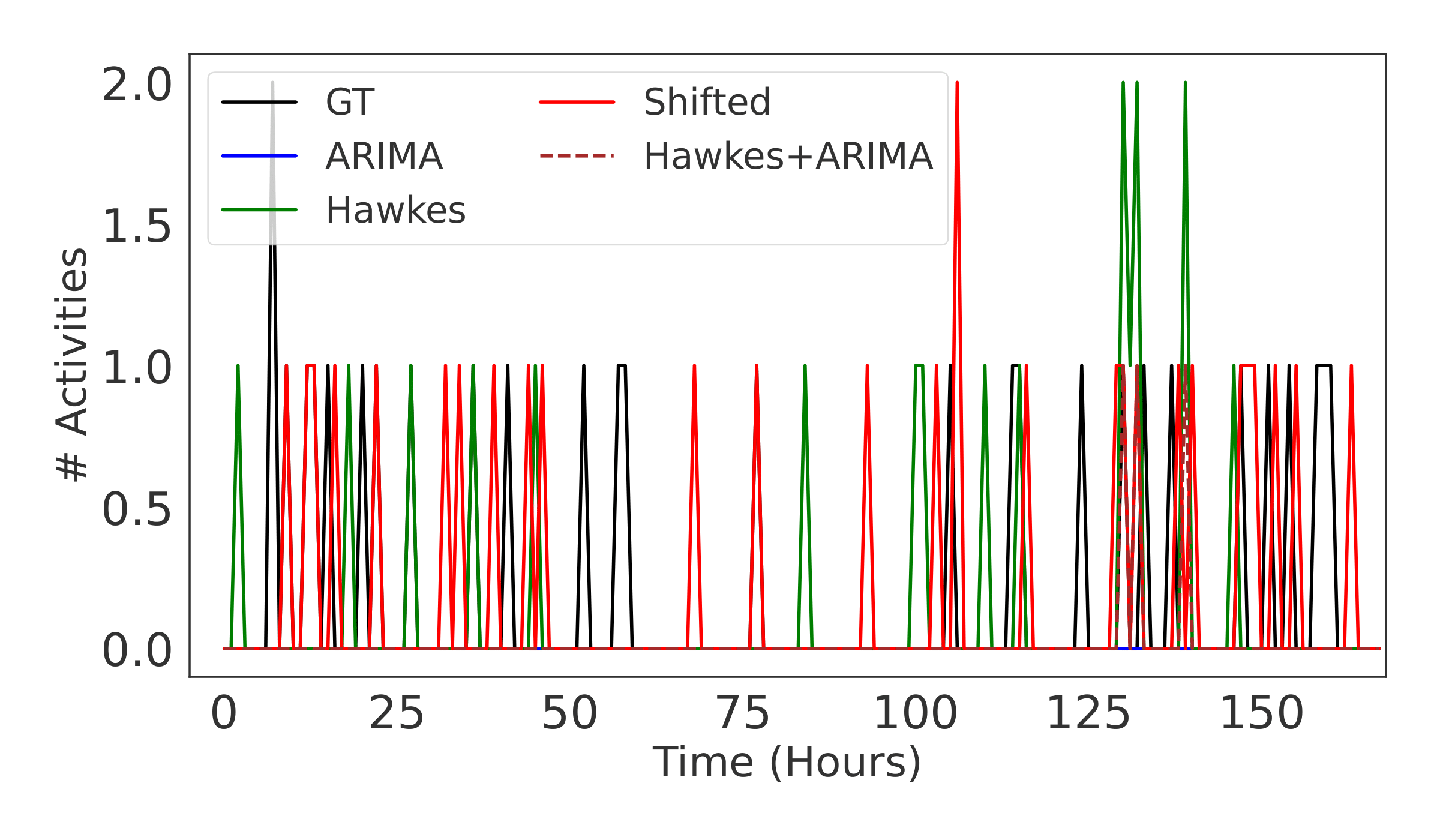}
			\label{fig:benefits_roads_yt_h}
	}
\\
	\subfloat[covid (BRIA)]{
		\includegraphics[width=0.32\linewidth]{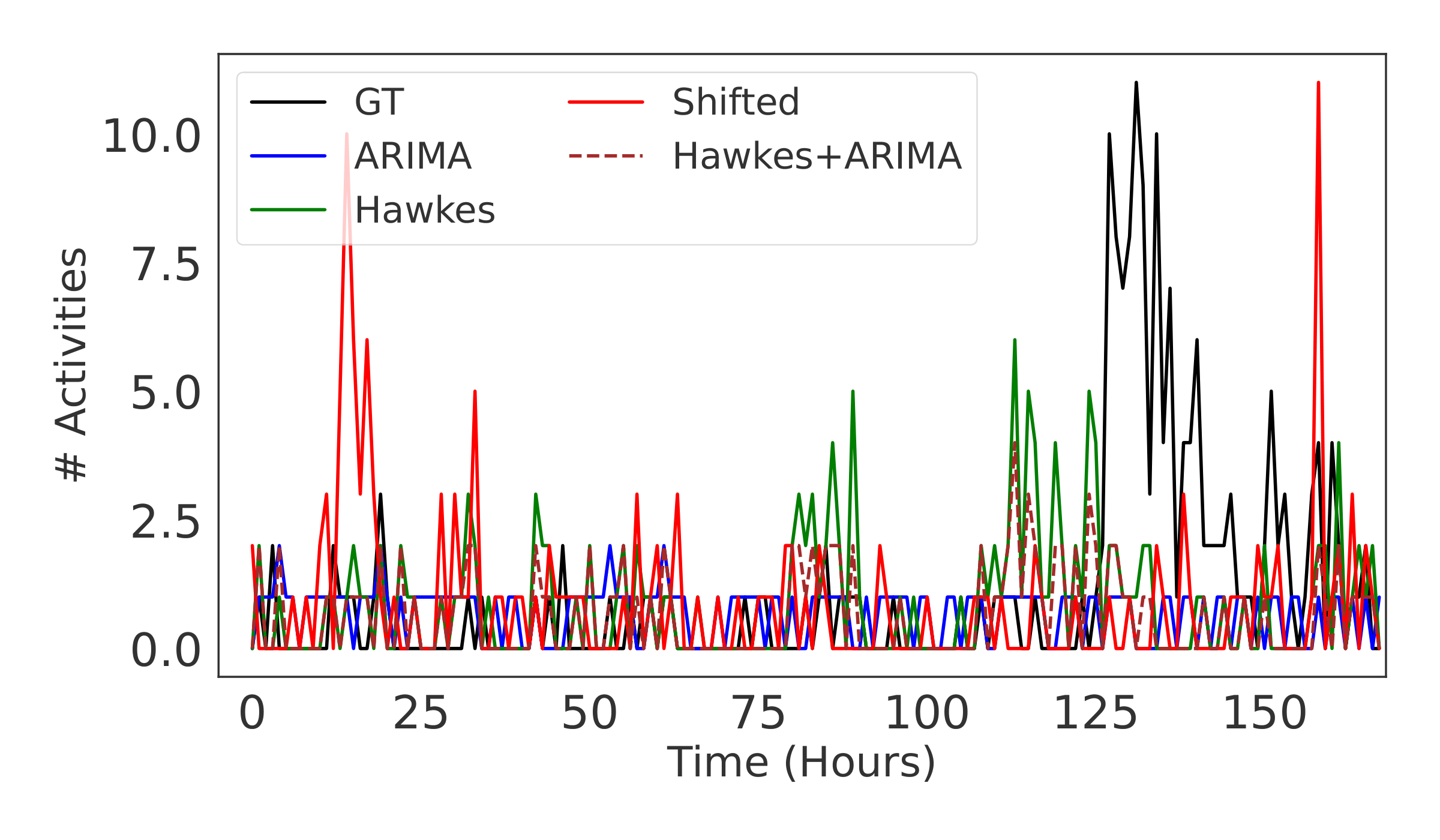}
			\label{fig:covid_yt_h}
	}
	&
	\subfloat[travel (BRIA)]{
		\includegraphics[width=0.32\linewidth]{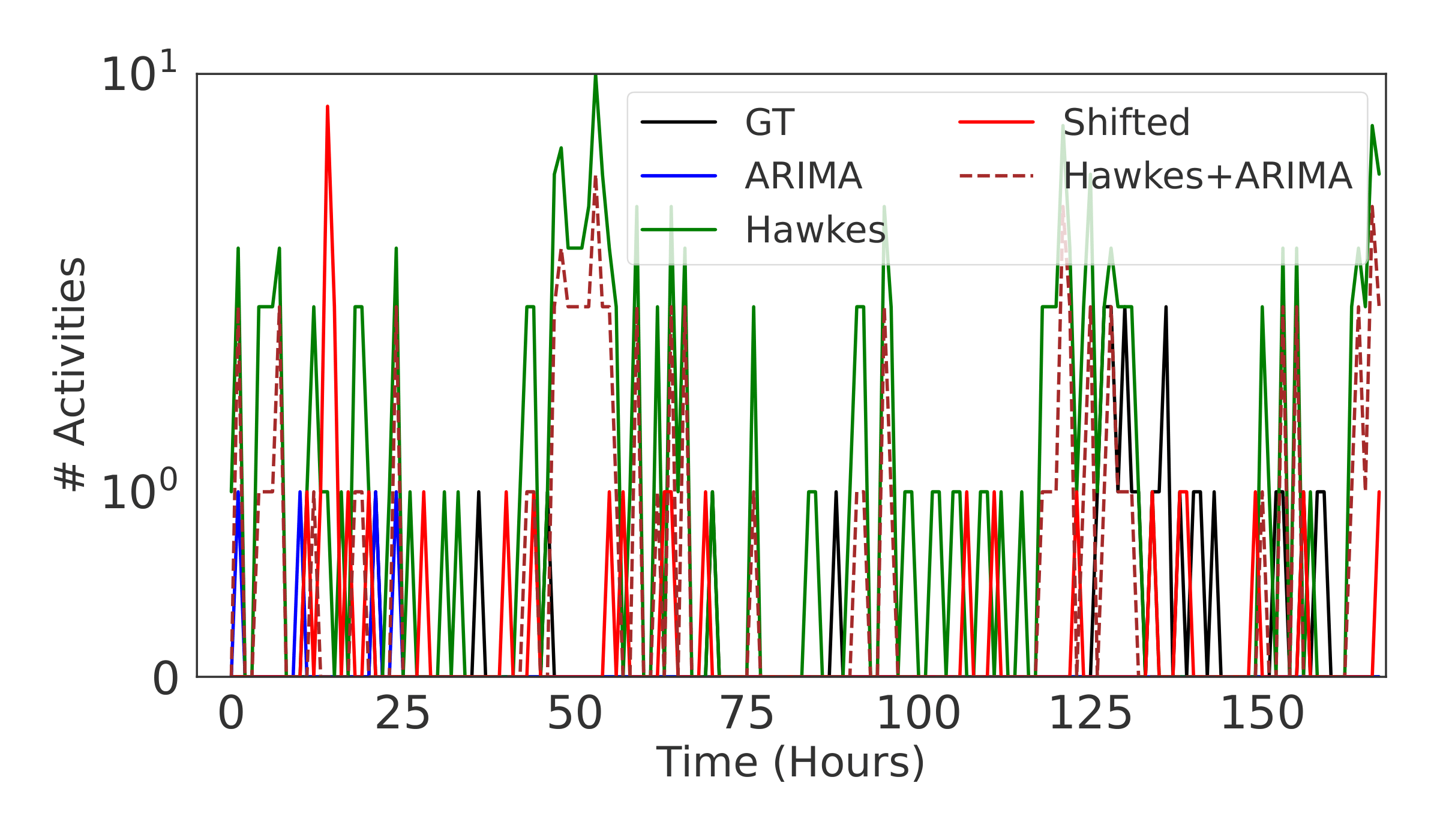}
			\label{fig:travel_yt_h}
	}
	&
	\subfloat[infrastructure (BRIA)]{
		\includegraphics[width=0.32\linewidth]{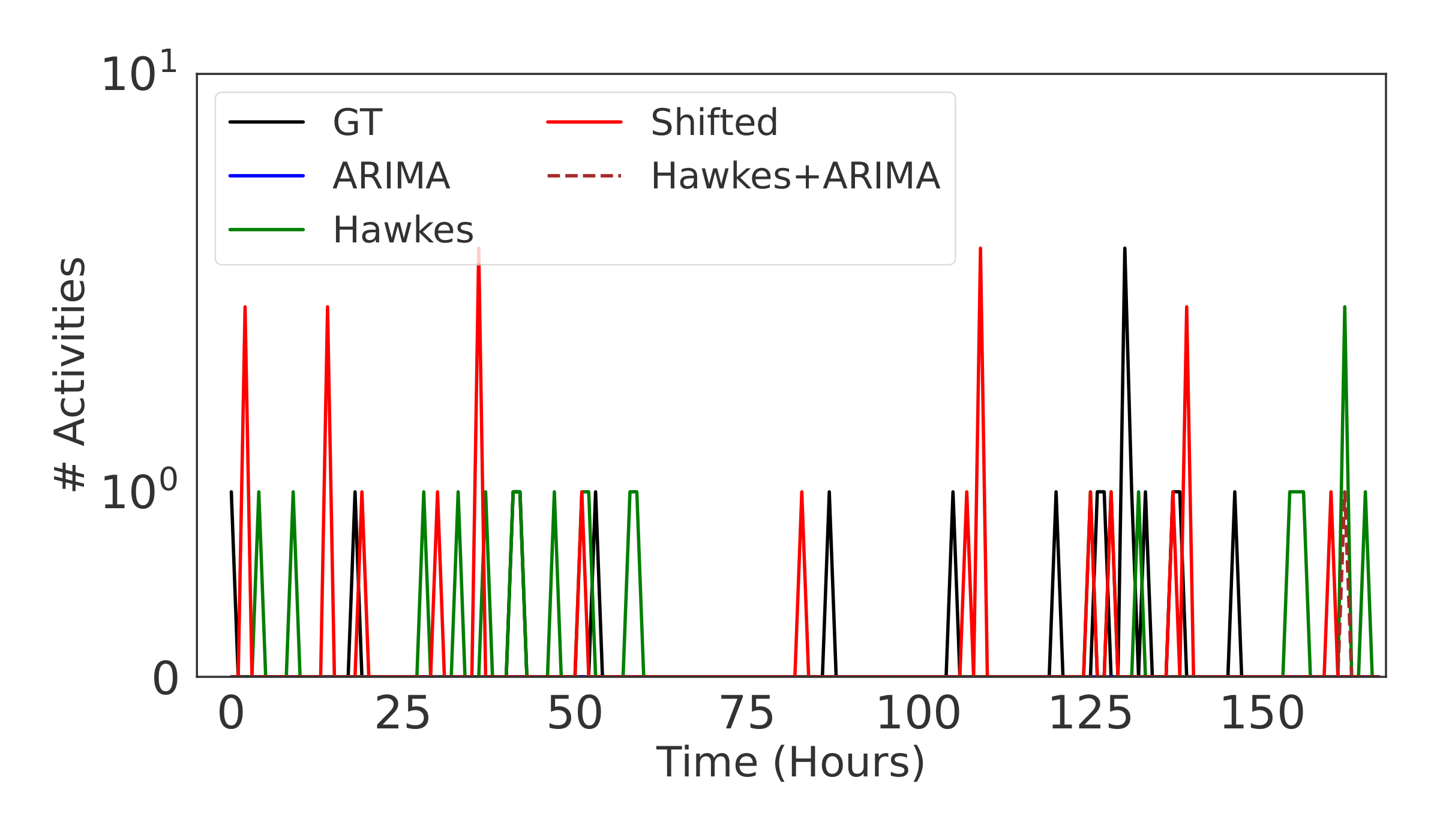}
			\label{fig:infrastruture_yt_h}
	}
\\
\end{tabular}
}
	 \caption{Hourly time series visualizations for YouTube on the top-3 most active topics for each context. We predict one week of activity (168 hours/time steps)}
	 \label{fig:time_series_yt_h}
\end{figure*}

Visual inspection of the time series in Figures~\ref{fig:time_series_tw_h} and~\ref{fig:time_series_yt_h} show that in many instances ARIMA produces a highly unrealistic pattern of variation that diminishes the seeming advantage in performance discussed above. 
While in some cases ARIMA captures well the periodicity in the ground truth data at hour granularity (as in, for example, Figures~\ref{fig:international_aid_tw_h},~\ref{fig:military_tw_h}~and~\ref{fig:covid_tw_h} for Twitter or Figures~\ref{fig:maduro_dictator_yt_h},~\ref{fig:international_aid_yt_h}~and~\ref{fig:other_chavez_yt_h} for YouTube), it either misses the amplitude of the signal by an order of magnitude (as shown in Figure~\ref{fig:leadership_sharif_tw_h}), or creates a highly regular (and unrealistic) pattern (Figure~\ref{fig:international_aid_tw_h}). 
In other cases, however, it predicts a linear time series that is highly unrealistic, as in Figure~\ref{fig:china_border_tw_h}. 

\begin{figure*}[htbp]
\centering
\begin{tabular}{ccc}
	\subfloat[Vz19 (Twitter)]{
		\includegraphics[width=0.4\linewidth]{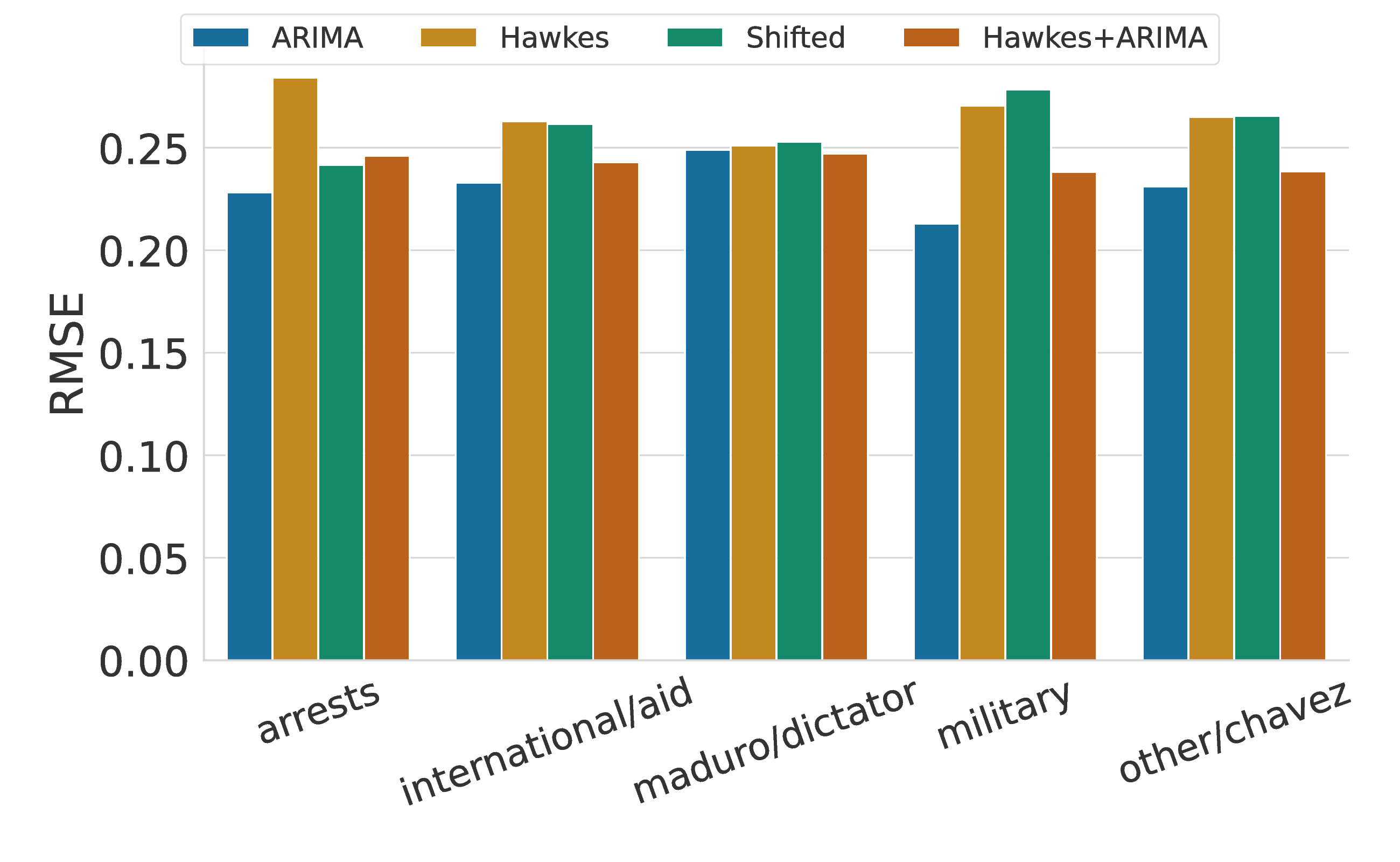}
			\label{fig:cp4_tw_rmse_h}
	}
	&
		\subfloat[Vz19 (YouTube)]{
		\includegraphics[width=0.4\linewidth]{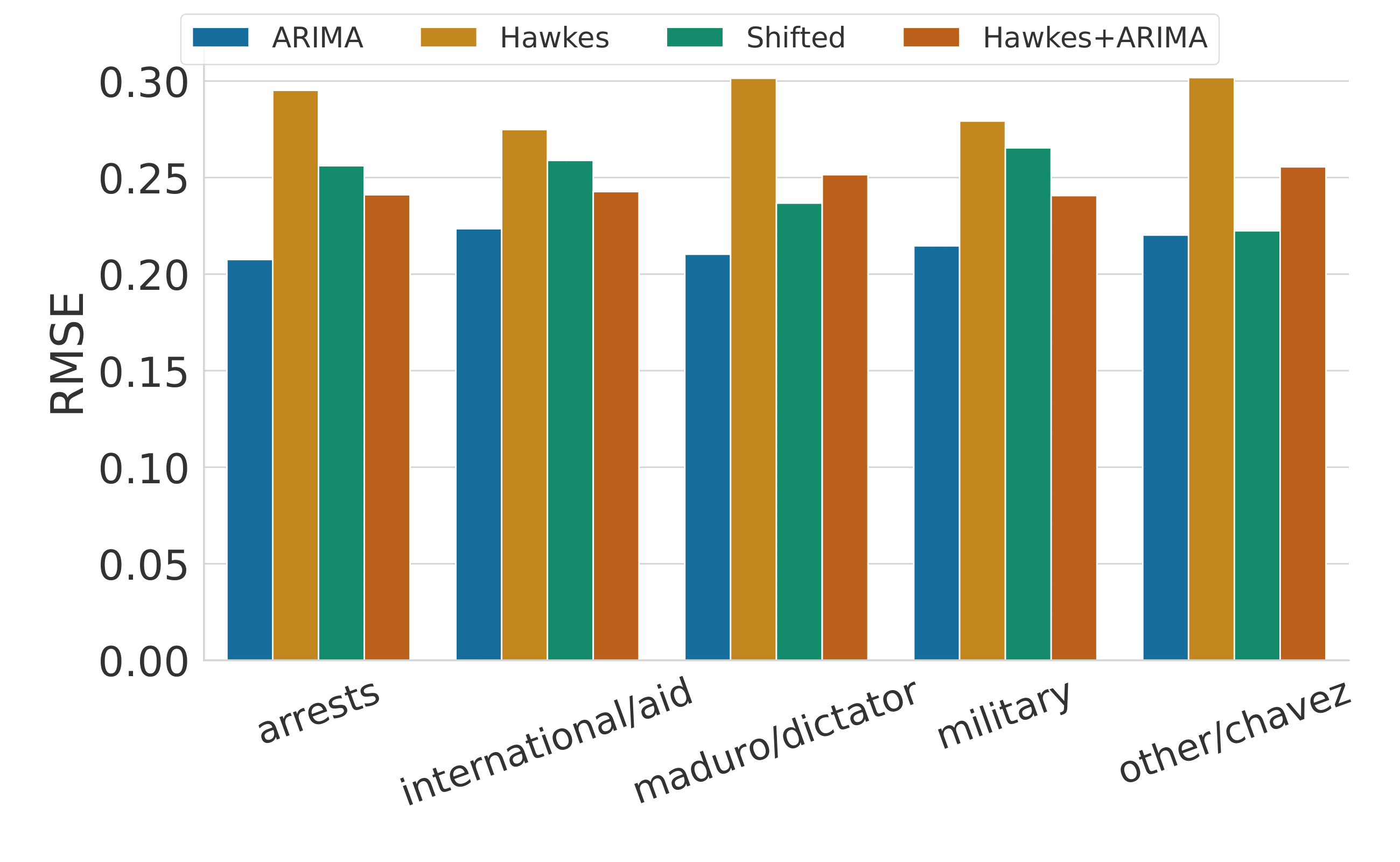}
			\label{fig:cp4_yt_rmse_h}
	}
	\\
	\subfloat[CPEC (Twitter)]{
		\includegraphics[width=0.4\linewidth]{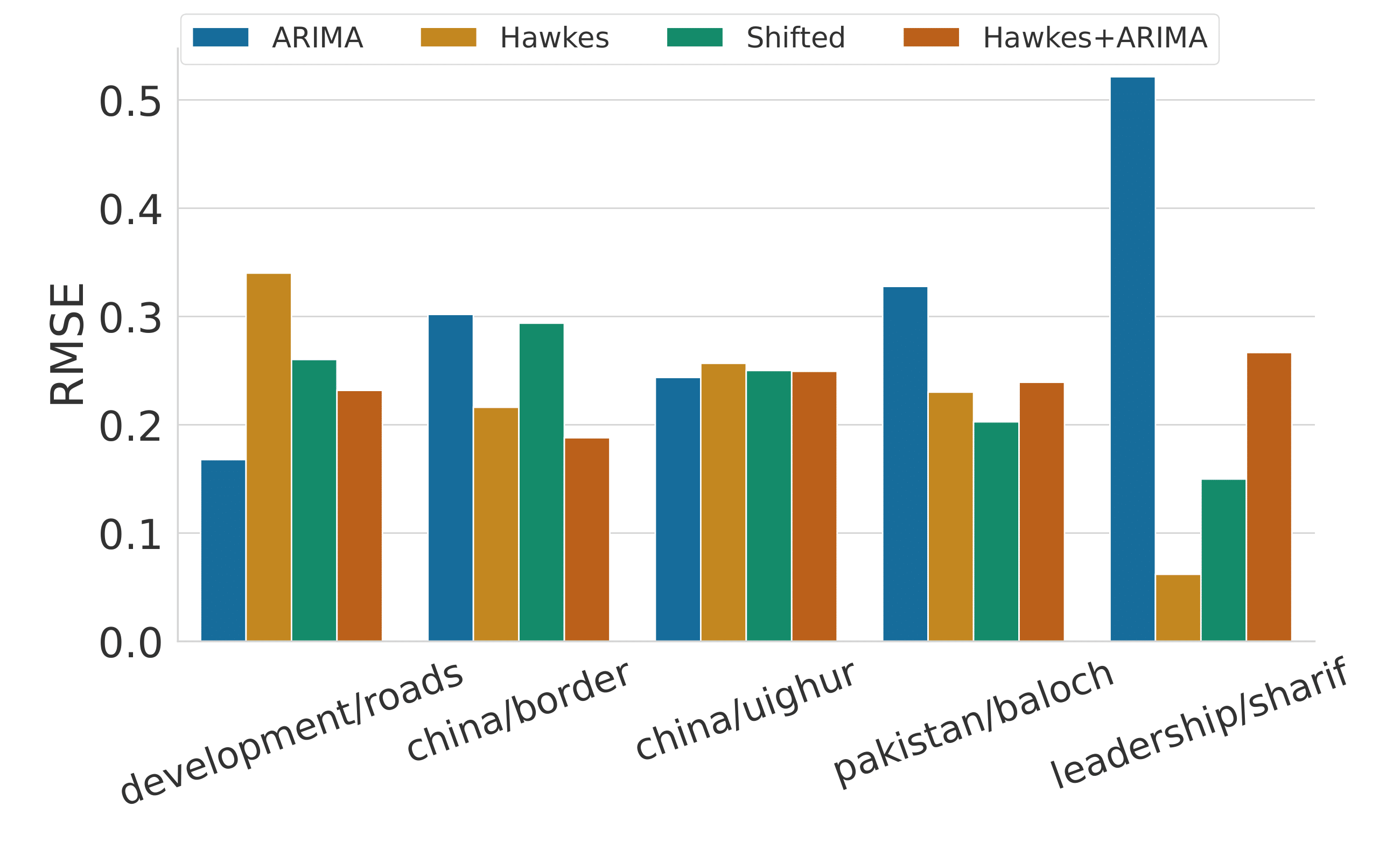}
			\label{fig:cp5_tw_rmse_h}
	}
    &
	\subfloat[CPEC (YouTube)]{
		\includegraphics[width=0.4\linewidth]{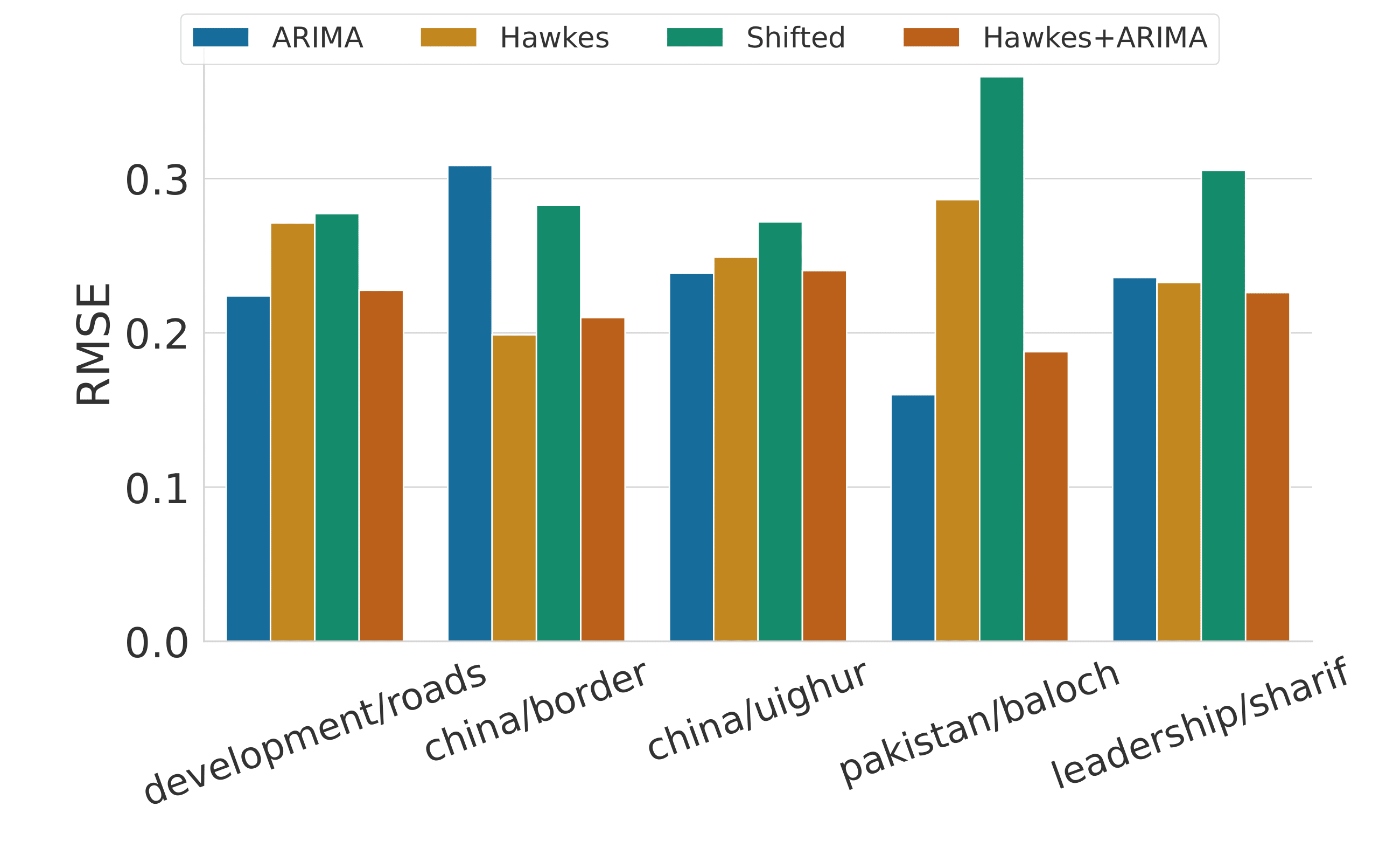}
			\label{fig:cp5_yt_rmse_h}
	}
	\\
	\subfloat[BRIA (Twitter)]{
		\includegraphics[width=0.4\linewidth]{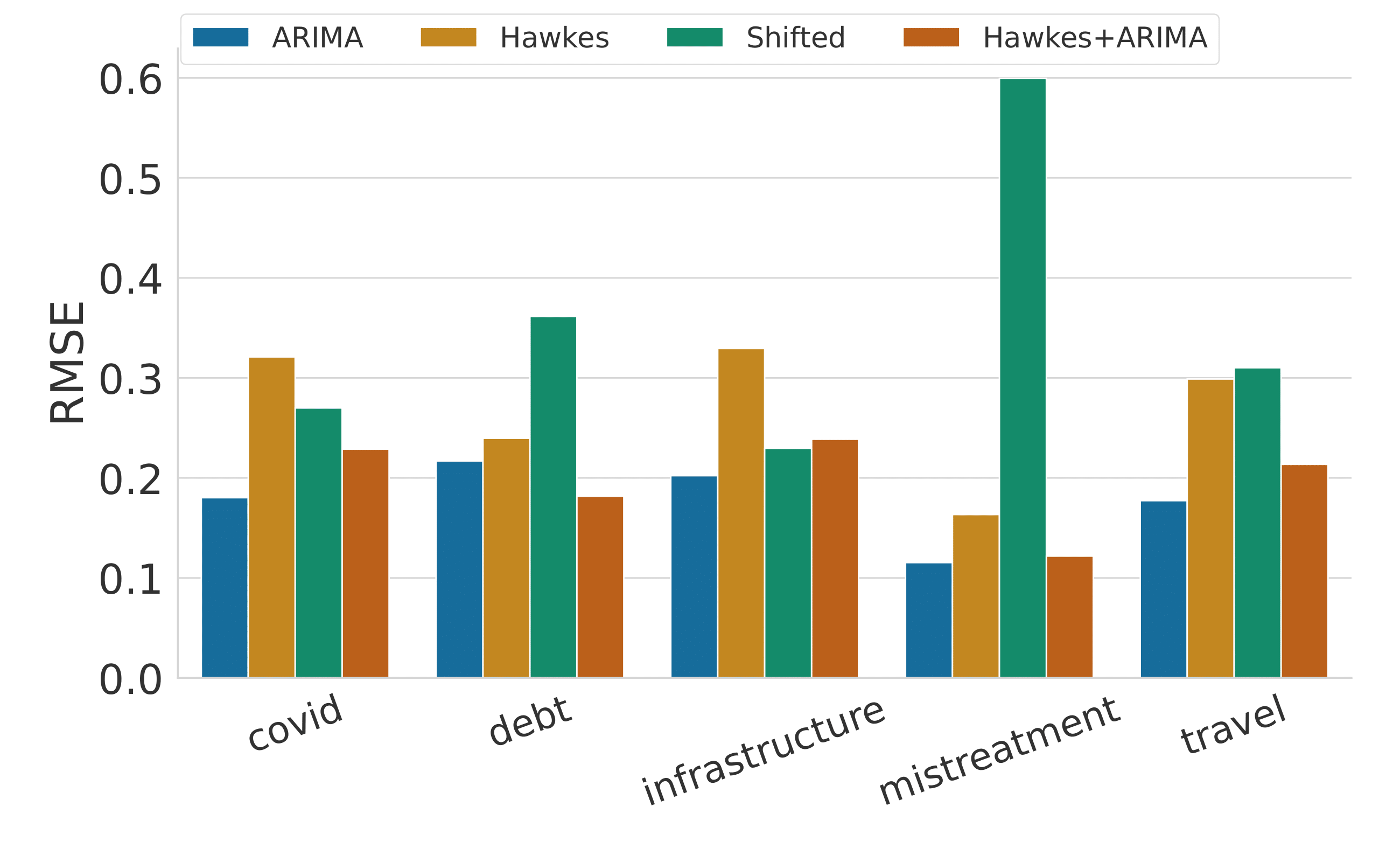}
			\label{fig:cp6_tw_rmse_h}
	}
	&
	\subfloat[BRIA (YouTube)]{
		\includegraphics[width=0.4\linewidth]{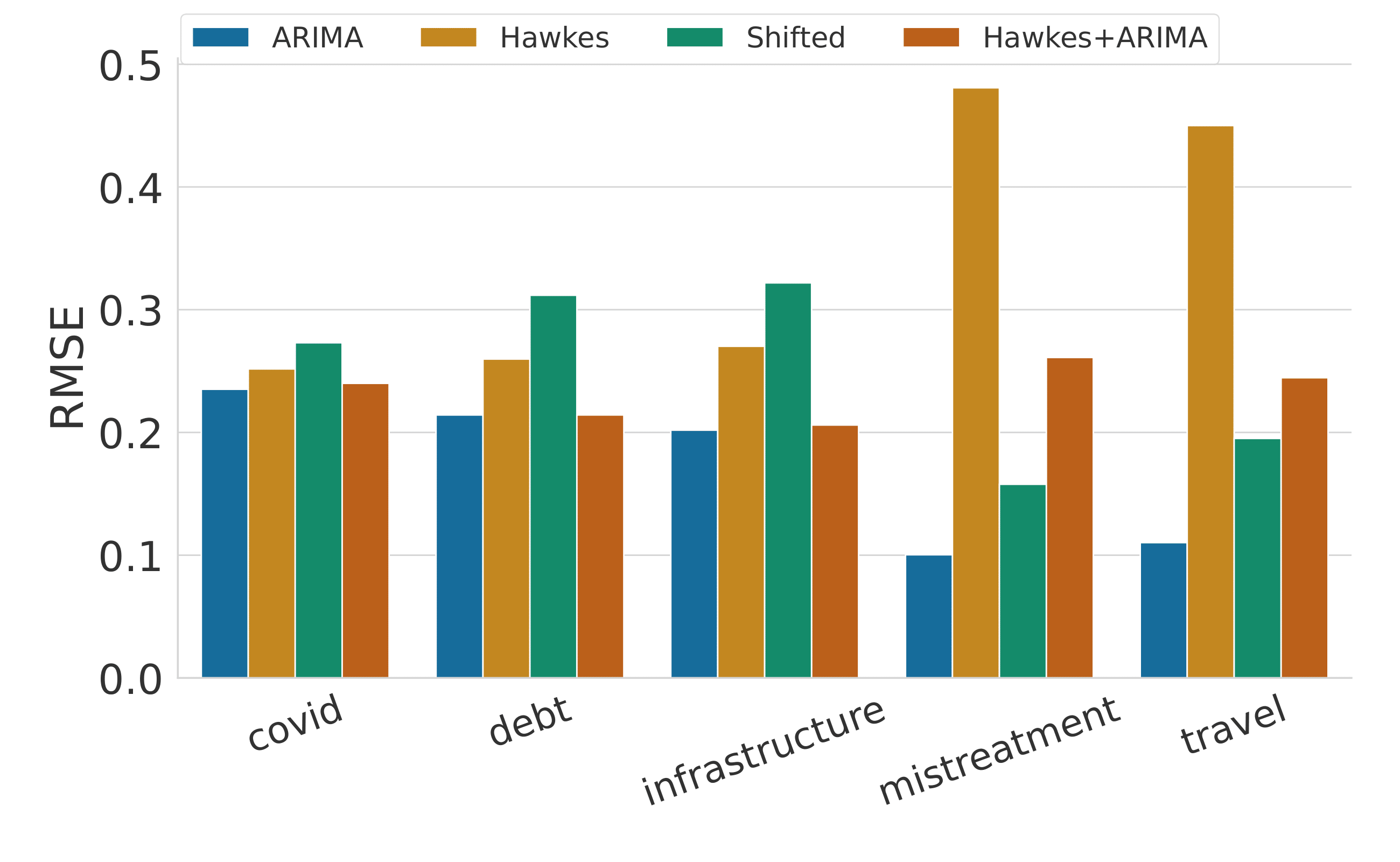}
			\label{fig:cp6_yt_rmse_h}
	}
\\
\end{tabular}
	 \caption{RMSE hourly performance across topics for each context and over two platforms (lower is better). RMSE scores for each topic are normalized between 0 to 1 relative to the sum of the baselines' errors. The results are for one week predictions at hourly granularity.}
	 \label{fig:rmse_topic_performance_h}
\end{figure*}

\begin{figure*}[htbp]
\centering
\begin{tabular}{cc}
	\subfloat[Vz19 (Twitter)]{
		\includegraphics[width=0.4\linewidth]{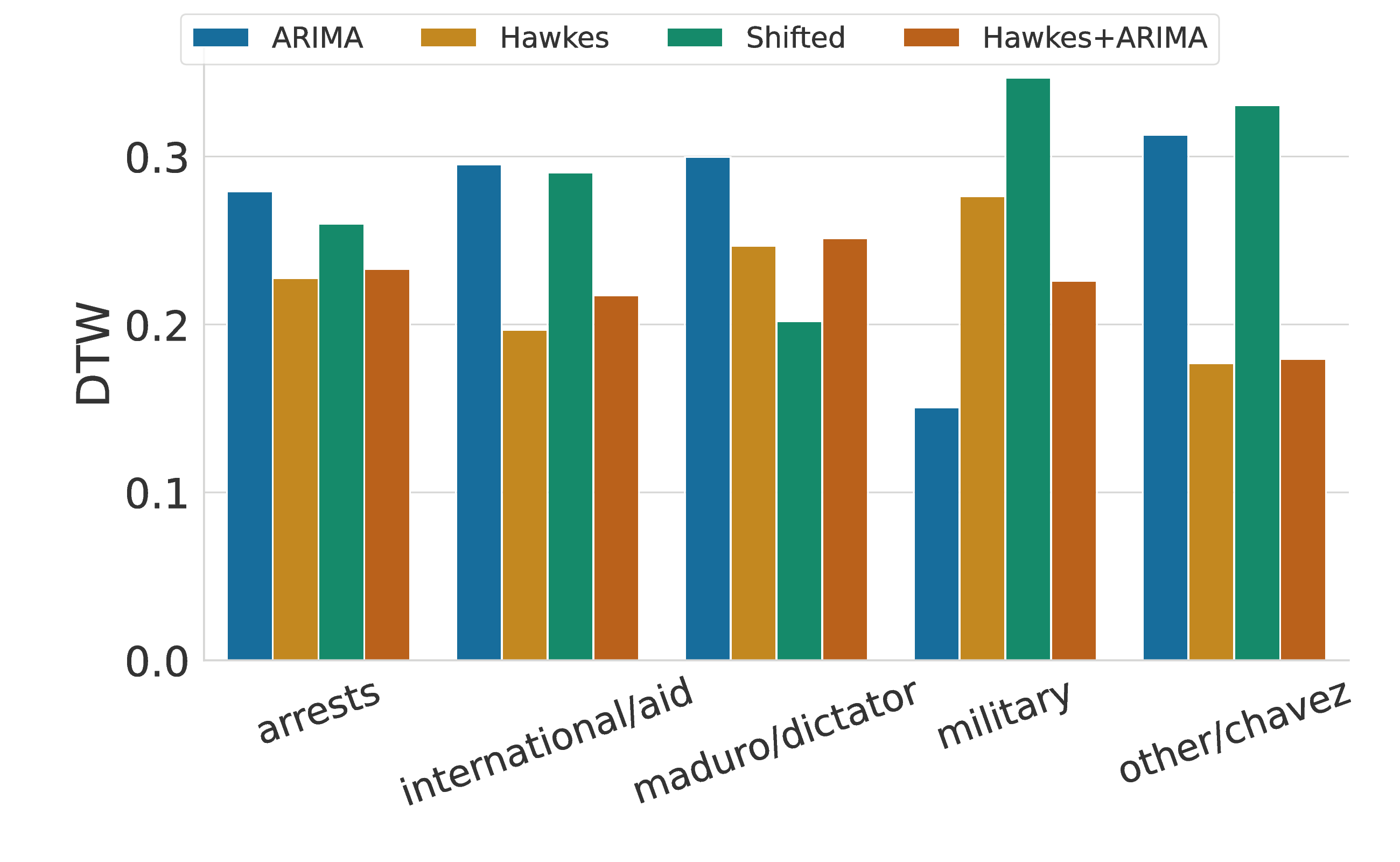}
			\label{fig:cp4_tw_dtw_h}
	}
	&
	\subfloat[Vz19 (YouTube)]{
		\includegraphics[width=0.4\linewidth]{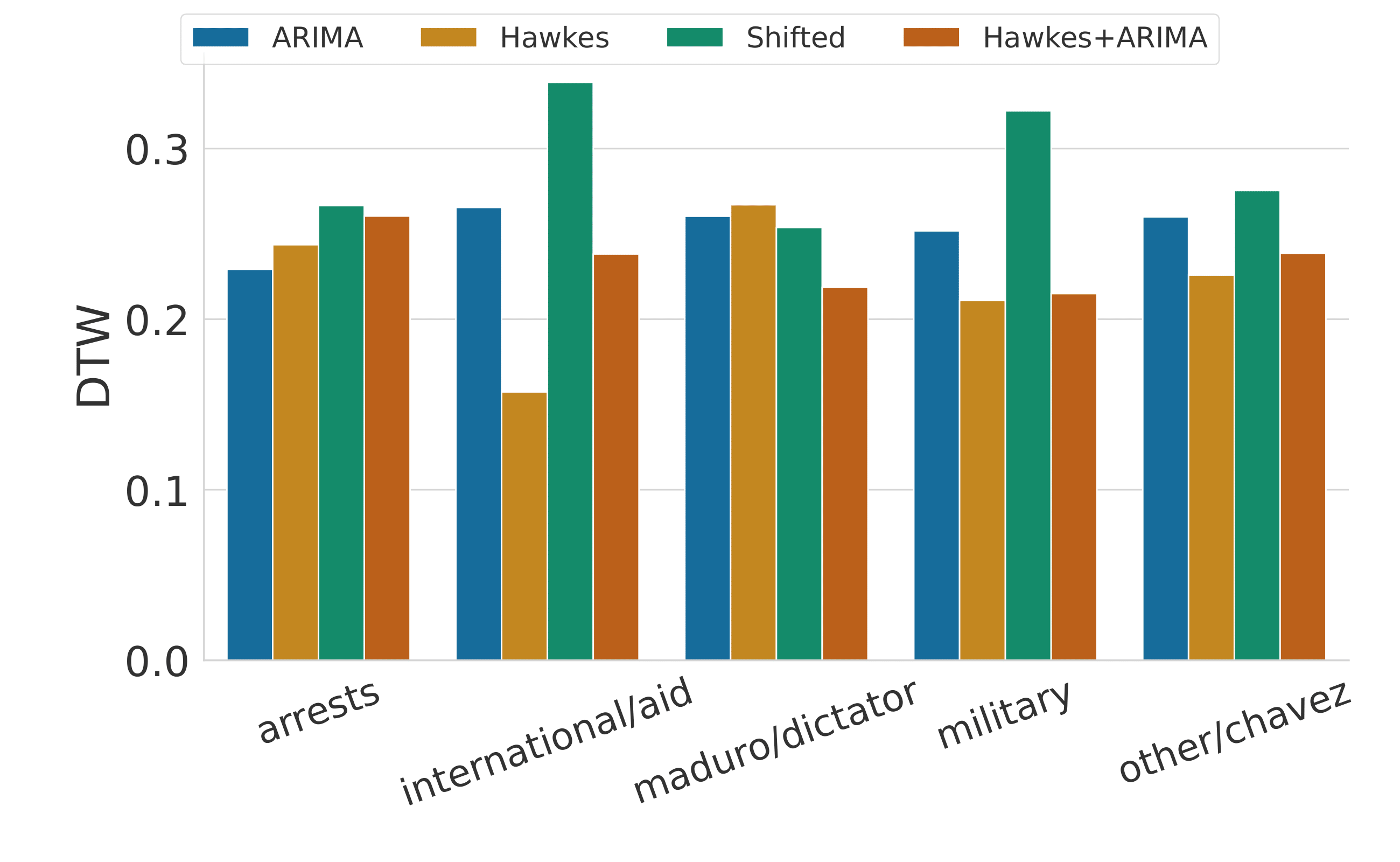}
			\label{fig:cp4_yt_dtw_h}
	}
	\\
	\subfloat[CPEC (Twitter)]{
		\includegraphics[width=0.4\linewidth]{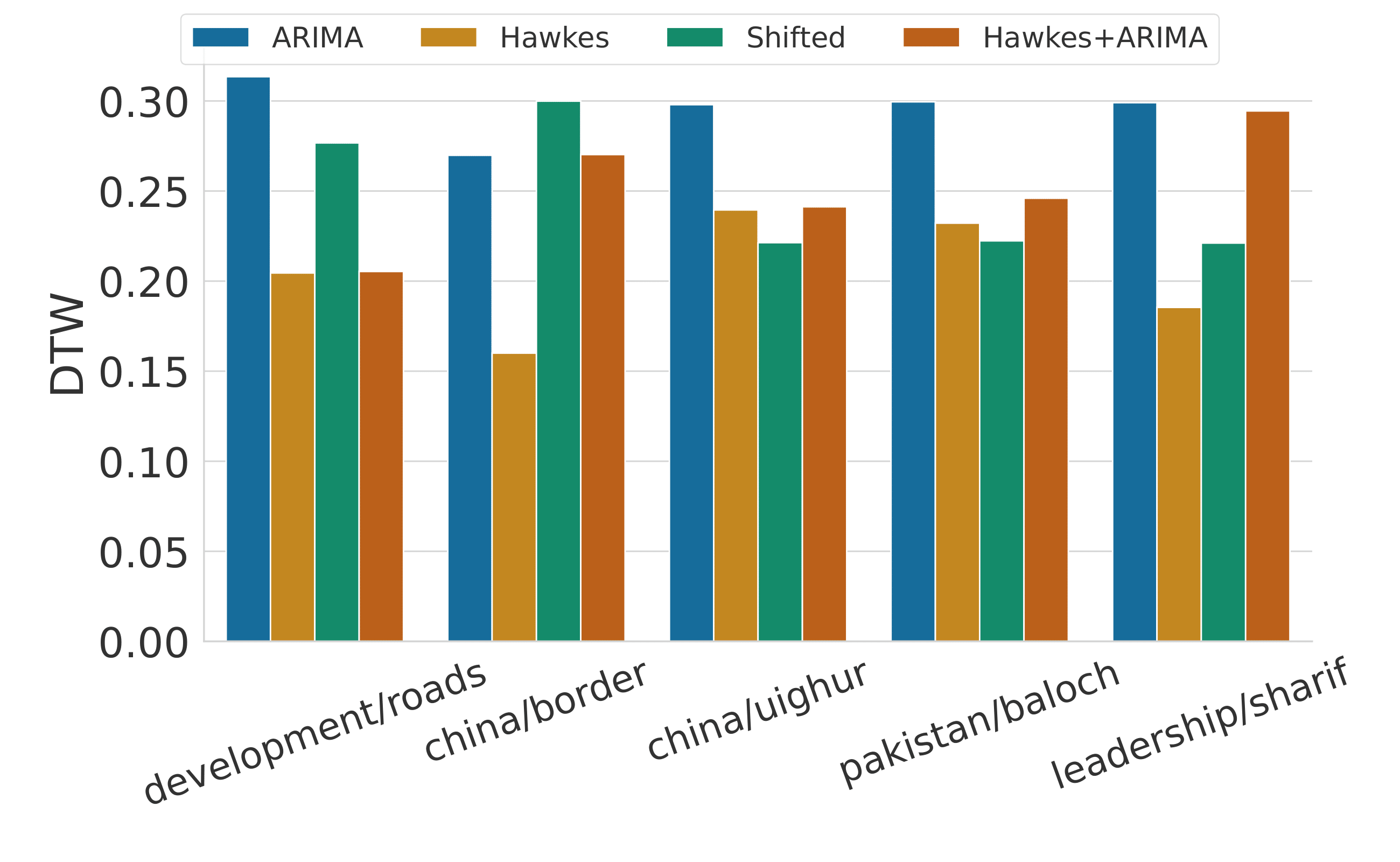}
			\label{fig:cp5_tw_dtw_h}
	}
	&
	\subfloat[CPEC (YouTube)]{
		\includegraphics[width=0.4\linewidth]{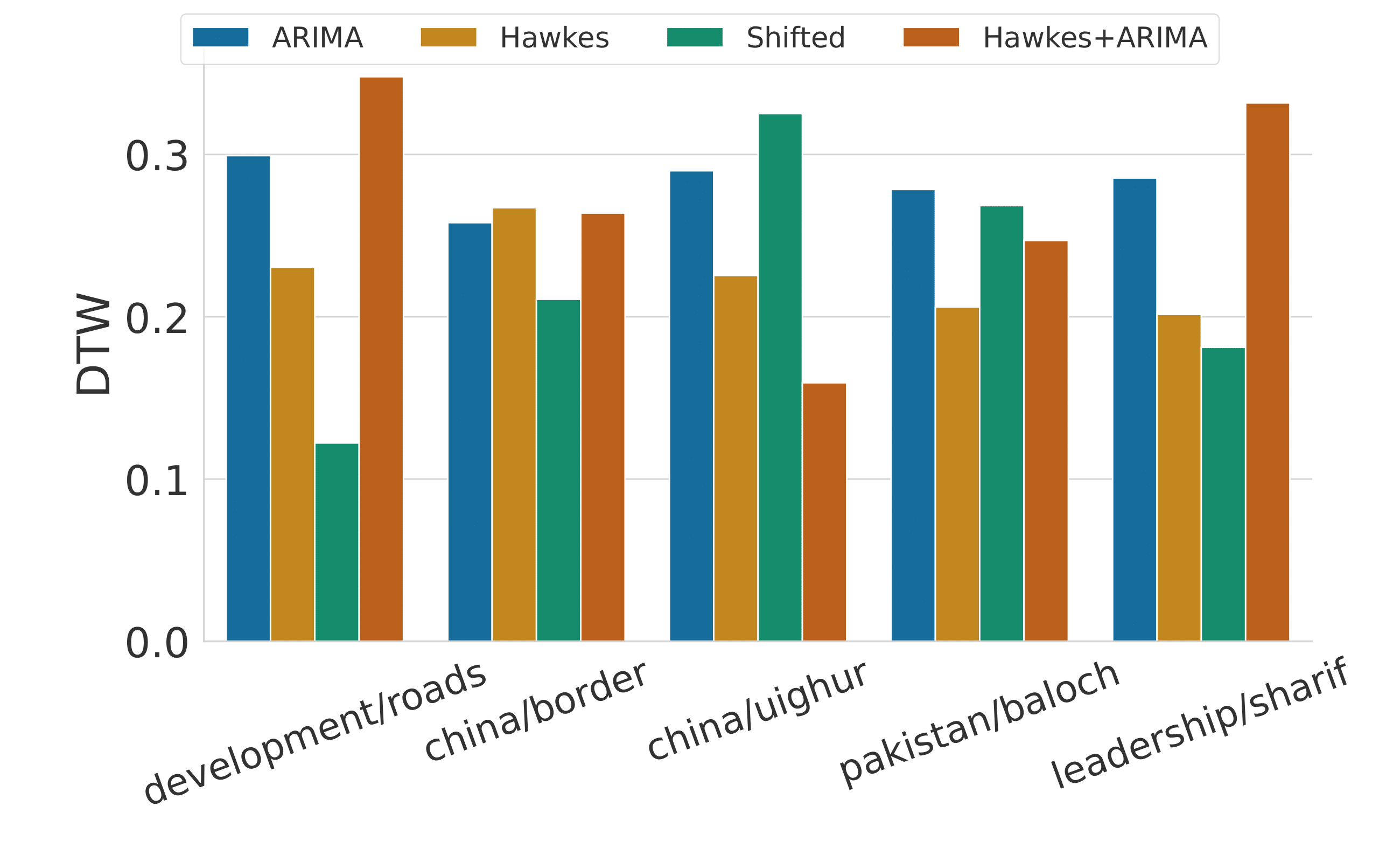}
			\label{fig:cp5_yt_dtw_h}
	}
	\\
	\subfloat[BRIA (Twitter)]{
		\includegraphics[width=0.4\linewidth]{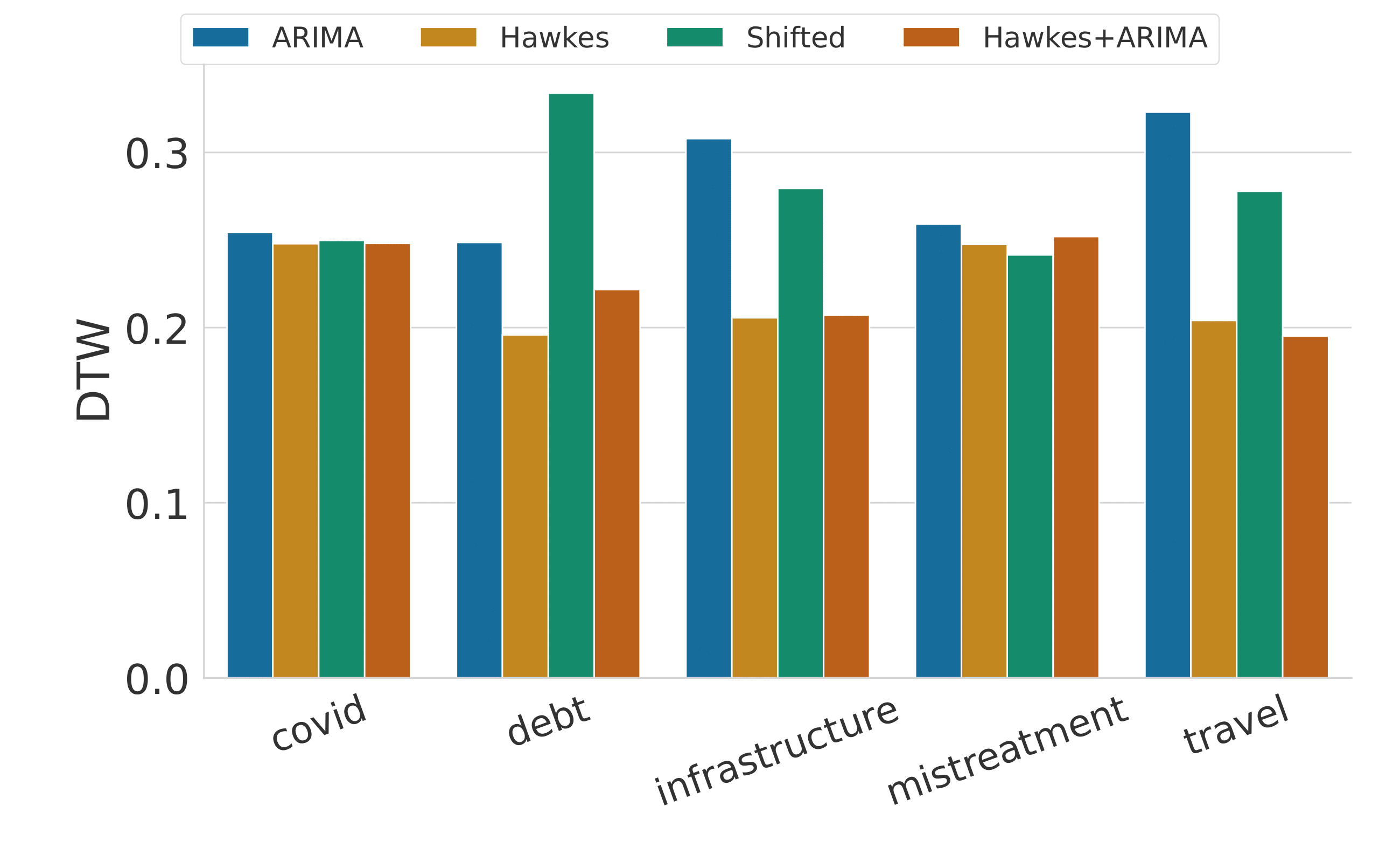}
			\label{fig:cp6_tw_dtw_h}
	}
	&
	\subfloat[BRIA (YouTube)]{
		\includegraphics[width=0.4\linewidth]{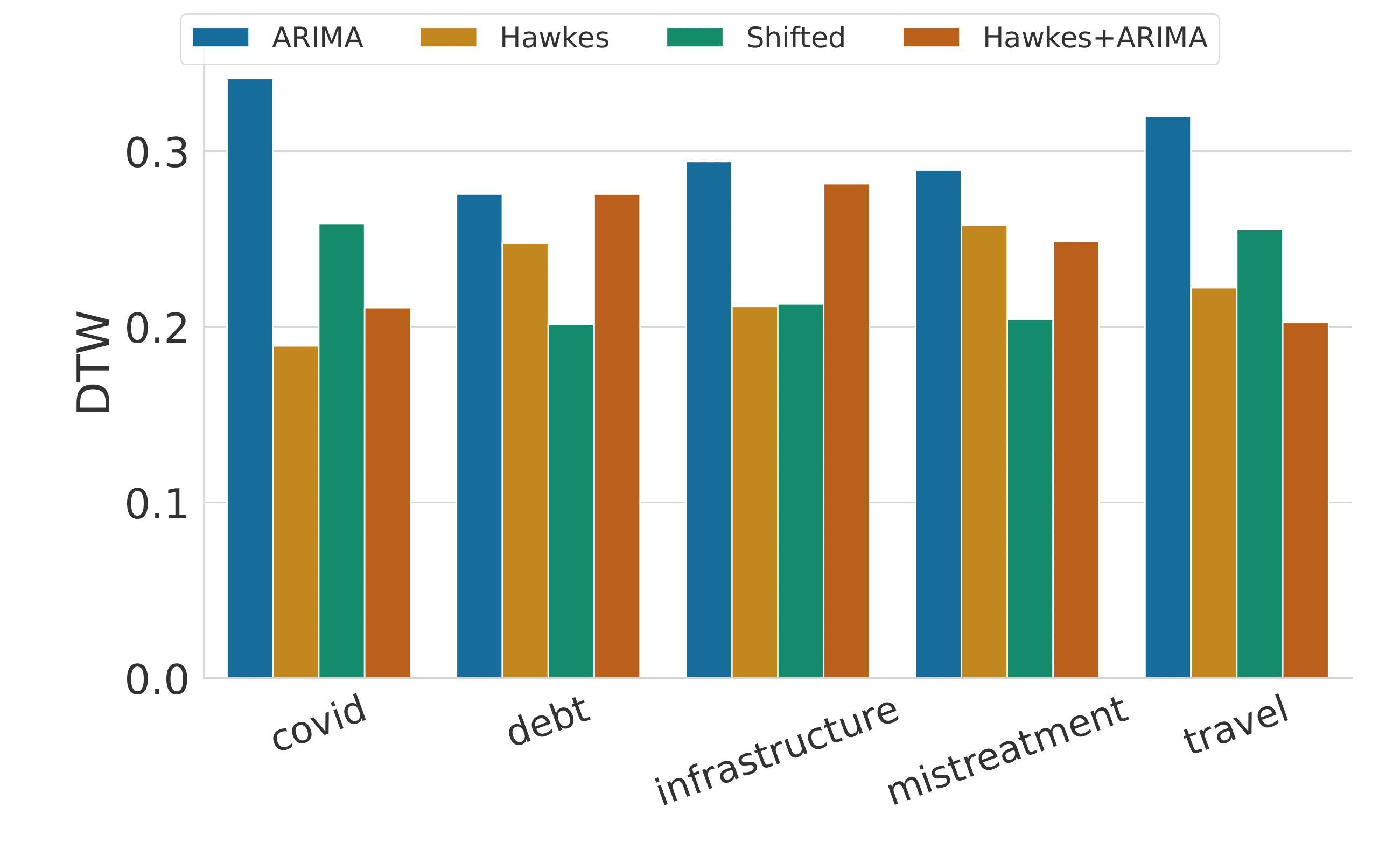}
			\label{fig:cp6_yt_dtw_h}
	}
\\
\end{tabular}
	 \caption{DTW hourly performance across topics for each context and over two platforms (lower is better). DTW scores for each topic are normalized between 0 to 1 relative to the sum of the baselines' errors. The results are for one week predictions at hourly granularity.}
	 \label{fig:dtw_topic_performance_h}
\end{figure*}

Yet despite these visible limitations, traditional time series error metrics such as RMSE show a more favorable picture of ARIMA when compared to both Hawkes and the shifted baselines.
For example, in Table~\ref{tab:aggregate_counts_performance}, which shows the weighted performance over all topics in a context/plaform, ARIMA appears to be the best performing baseline on RMSE (5 out of 6 possible instances).
This observation also holds when evaluated per topic, where ARIMA outperforms other baselines in 23 out of 30 instances (as shown in Figure~\ref{fig:rmse_topic_performance_h}).
This contradiction emphasizes the importance of model performance evaluation across additional dimensions. 
For instance, if we desire to accurately capture patterns over time, then metrics that measure similarity between temporal sequences (e.g., DTW) might be a more appropriate alternative. 
From Figure~\ref{fig:dtw_topic_performance_h}, we observe that ARIMA often fails to outperform other baselines, especially Hawkes, with regards to DTW as in many cases it produces unrealistic temporal sequences.
This observation excludes ARIMA as a meaningful methodology for creating representative social media time series patterns for scientific purposes. 

\subsubsection{Skewness and Volatility}

\begin{figure*}[htbp]
\centering
\begin{tabular}{ccc}
	\subfloat[Vz19 (Twitter)]{
		\includegraphics[width=0.4\linewidth]{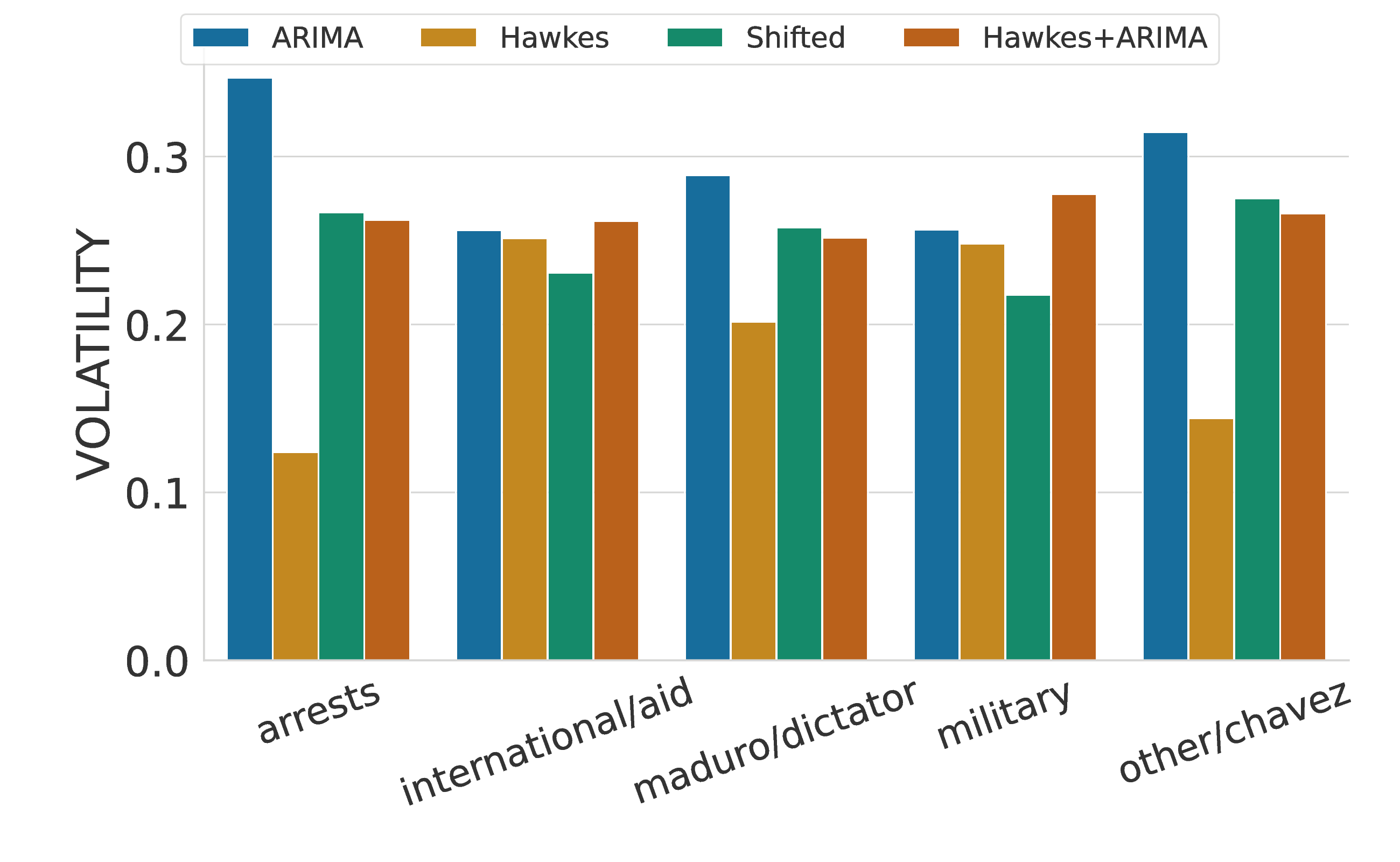}
			\label{fig:cp4_tw_volatility_h}
	}
	&
	\subfloat[Vz19 (YouTube)]{
		\includegraphics[width=0.4\linewidth]{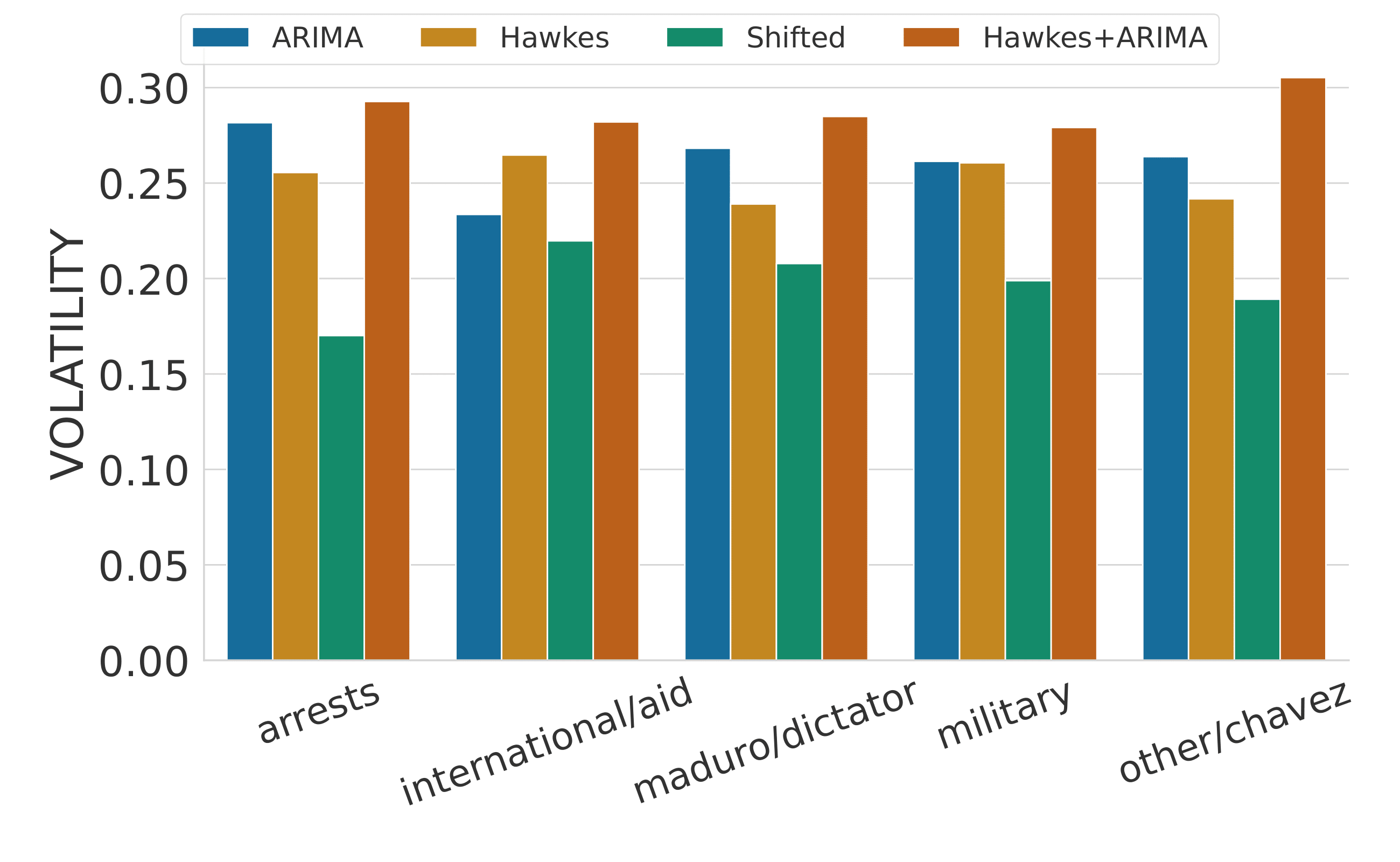}
			\label{fig:cp4_yt_volatility_h}
	}
	\\
	\subfloat[CPEC (Twitter)]{
		\includegraphics[width=0.4\linewidth]{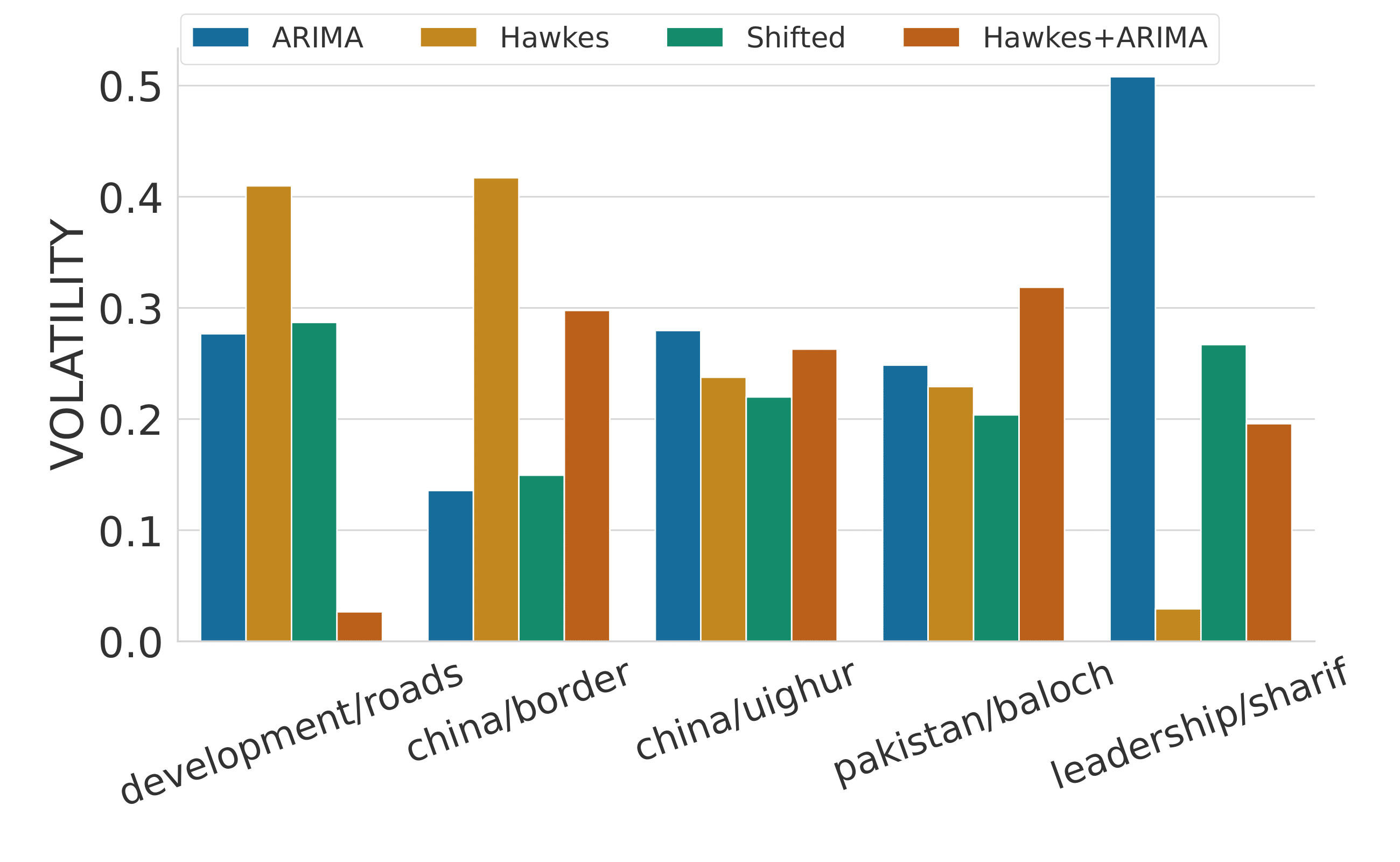}
			\label{fig:cp5_tw_volatility_h}
	}
	&
	\subfloat[CPEC (YouTube)]{
		\includegraphics[width=0.4\linewidth]{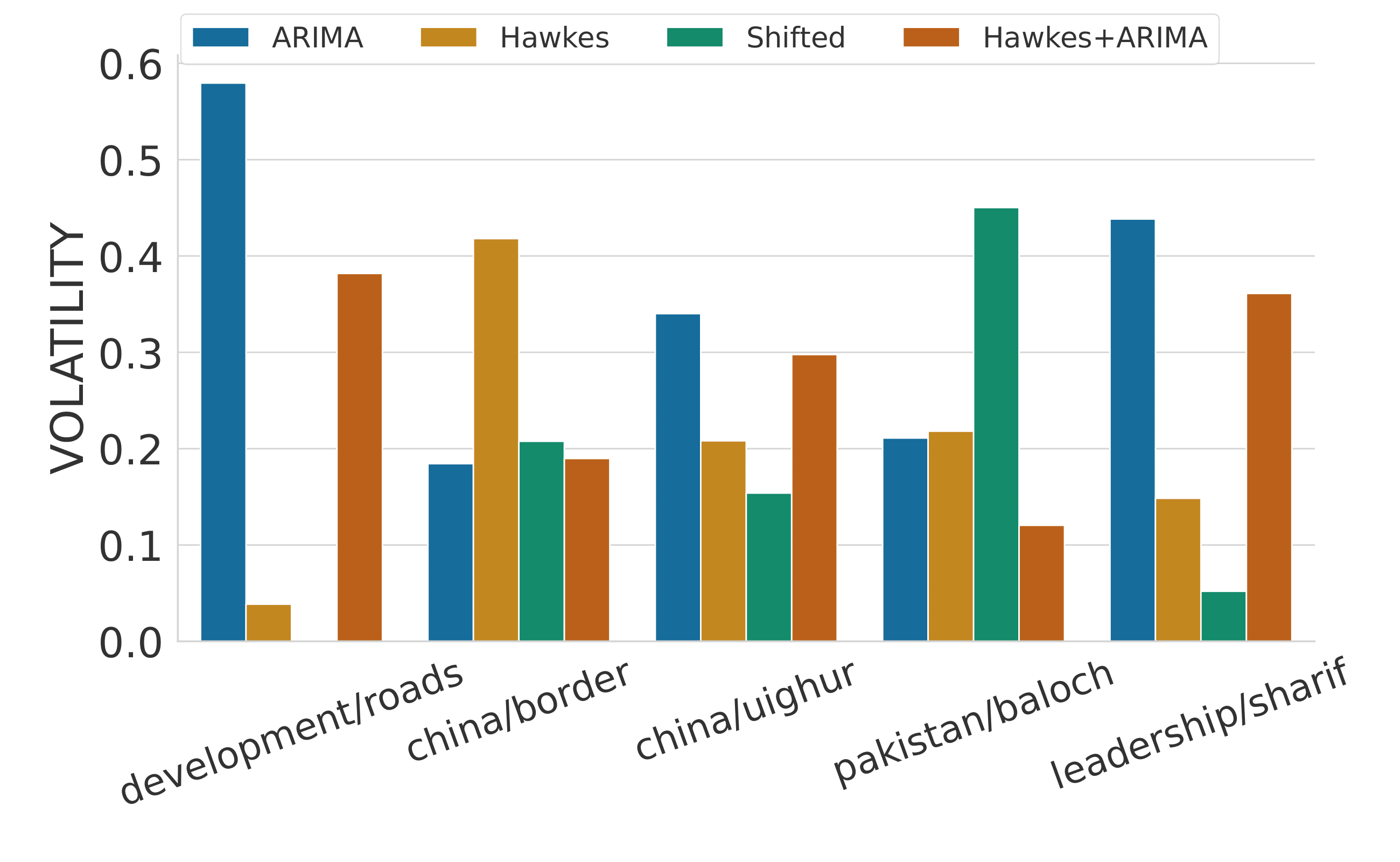}
			\label{fig:cp5_yt_volatility_h}
	}
	\\
	\subfloat[BRIA (Twitter)]{
		\includegraphics[width=0.4\linewidth]{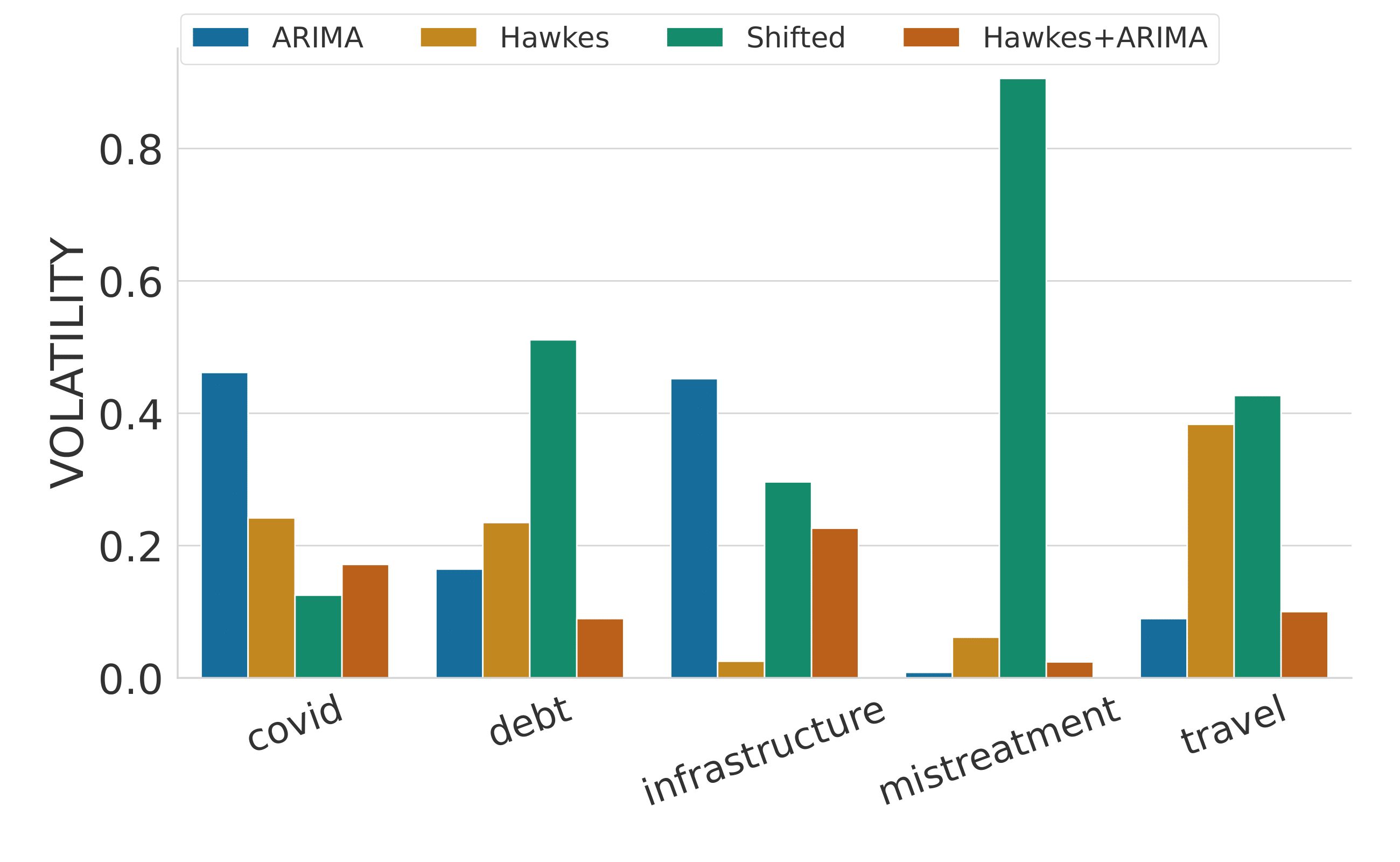}
			\label{fig:cp6_tw_volatility_h}
	}
	&
	\subfloat[BRIA (YouTube)]{
		\includegraphics[width=0.4\linewidth]{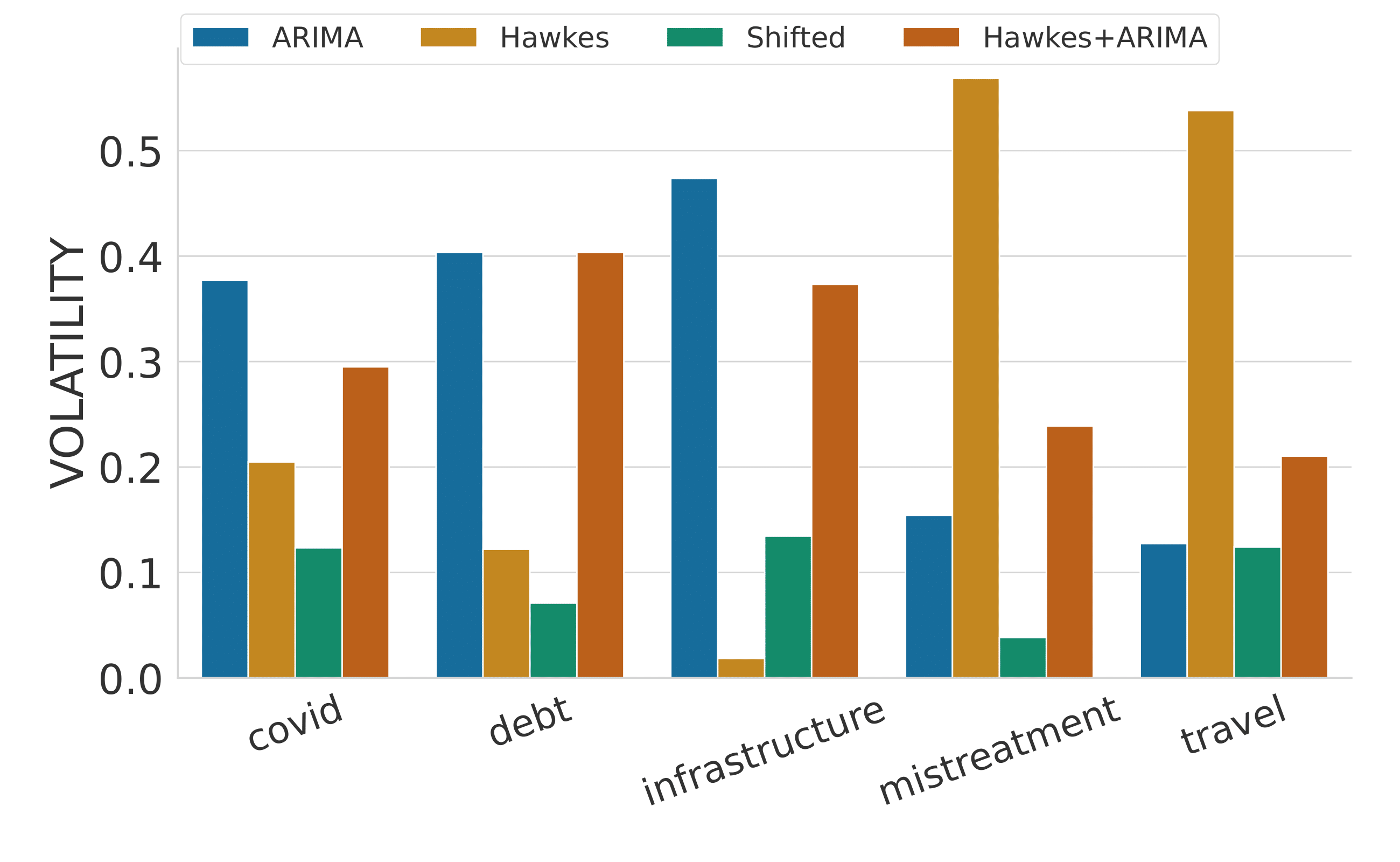}
			\label{fig:cp6_yt_volatility_h}
	}
\\
\end{tabular}
	 \caption{Volatility hourly performance across topics for each context and over two platforms (lower is better). Volatility scores for each topic are normalized between 0 to 1 relative to the sum of the baselines' errors. The results are for one week predictions at hourly granularity.}
	 \label{fig:volatility_topic_performance_h}
\end{figure*}

\begin{figure*}[htbp]
\centering
\begin{tabular}{ccc}
	\subfloat[Vz19 (Twitter)]{
		\includegraphics[width=0.4\linewidth]{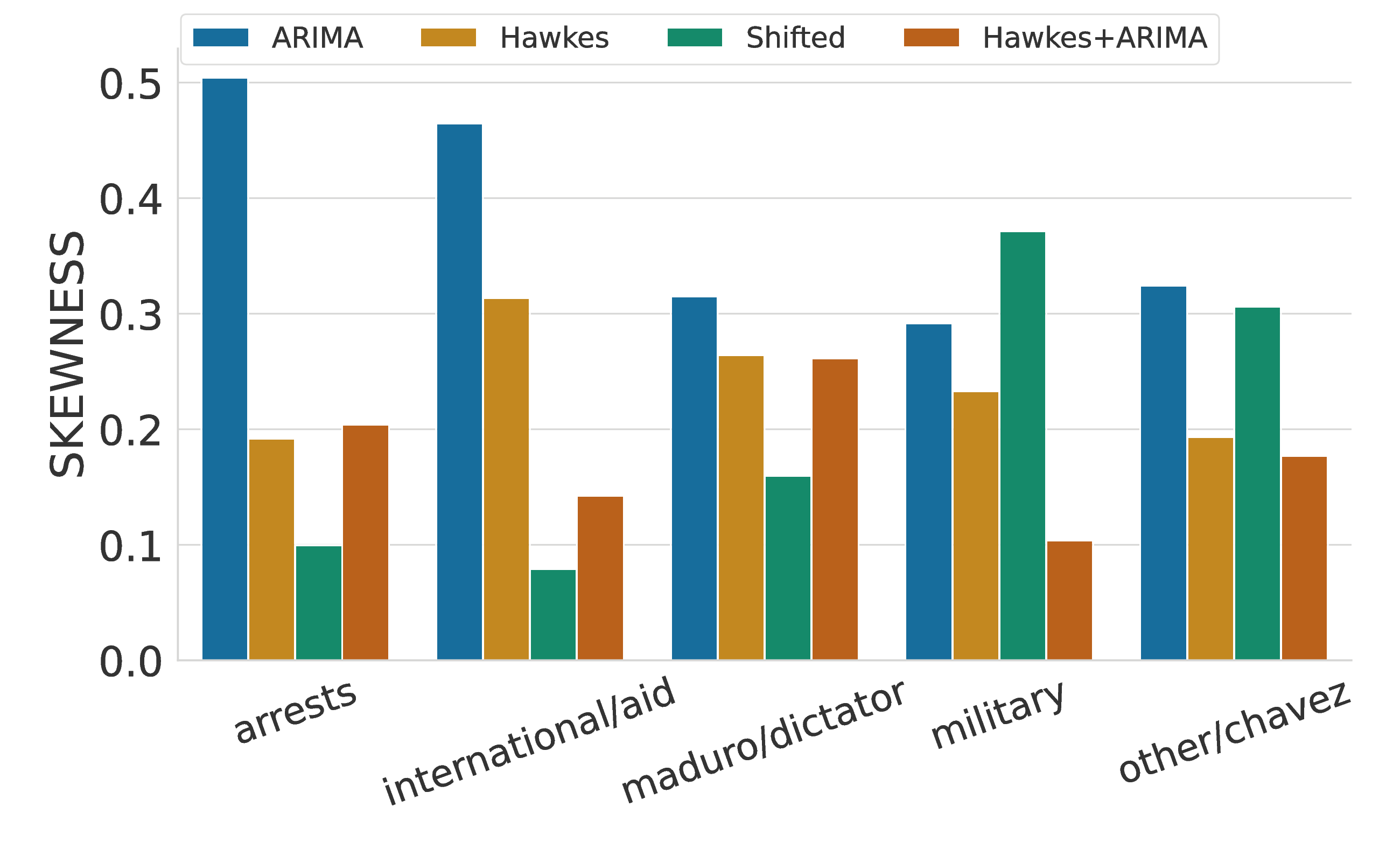}
			\label{fig:cp4_tw_skewness_h}
	}
	&
	\subfloat[Vz19 (YouTube)]{
		\includegraphics[width=0.4\linewidth]{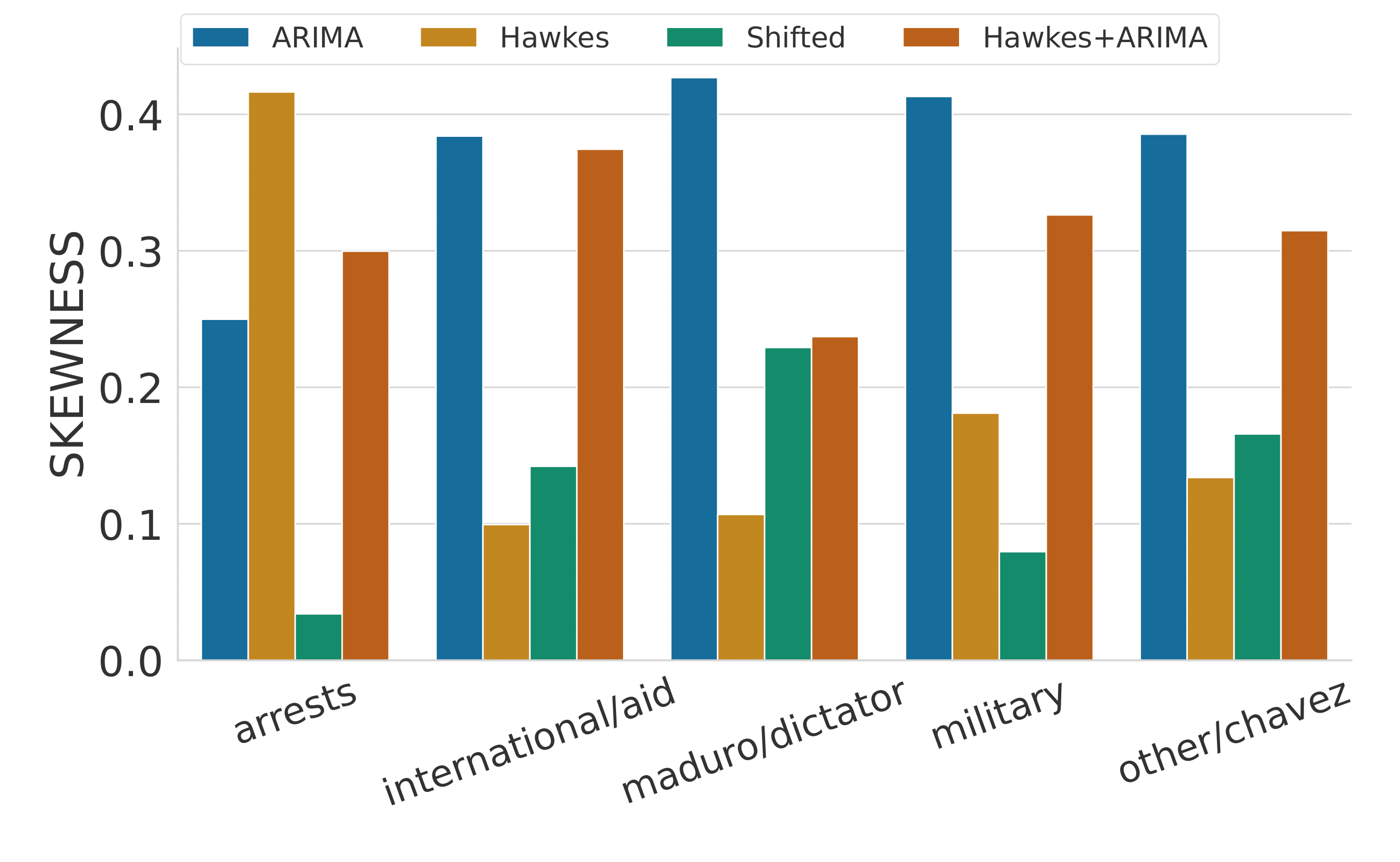}
			\label{fig:cp4_yt_skewness_h}
	}
	\\
	
	\subfloat[CPEC (Twitter)]{
		\includegraphics[width=0.4\linewidth]{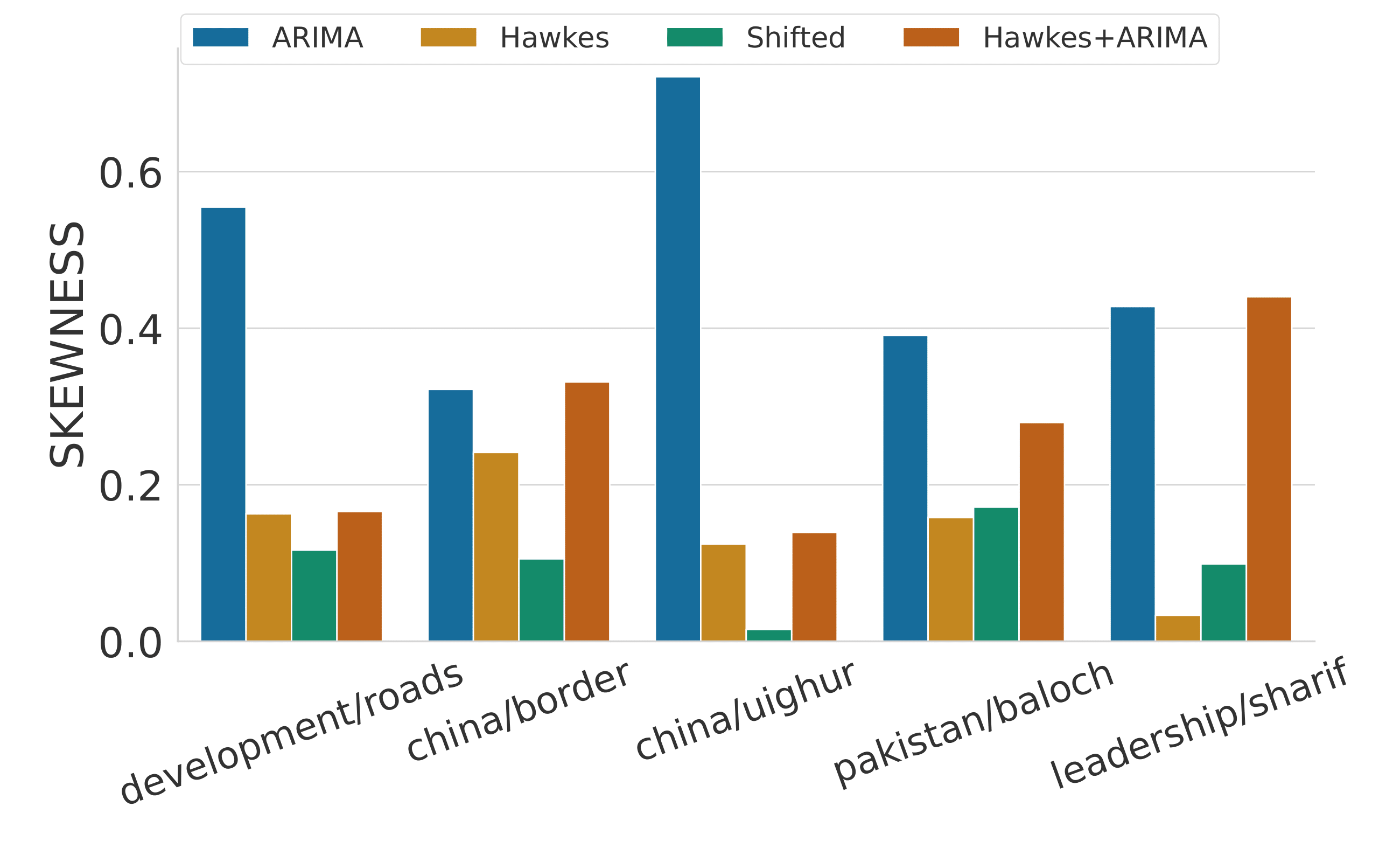}
			\label{fig:cp5_tw_skewness_h}
	}
    &
	\subfloat[CPEC (YouTube)]{
		\includegraphics[width=0.4\linewidth]{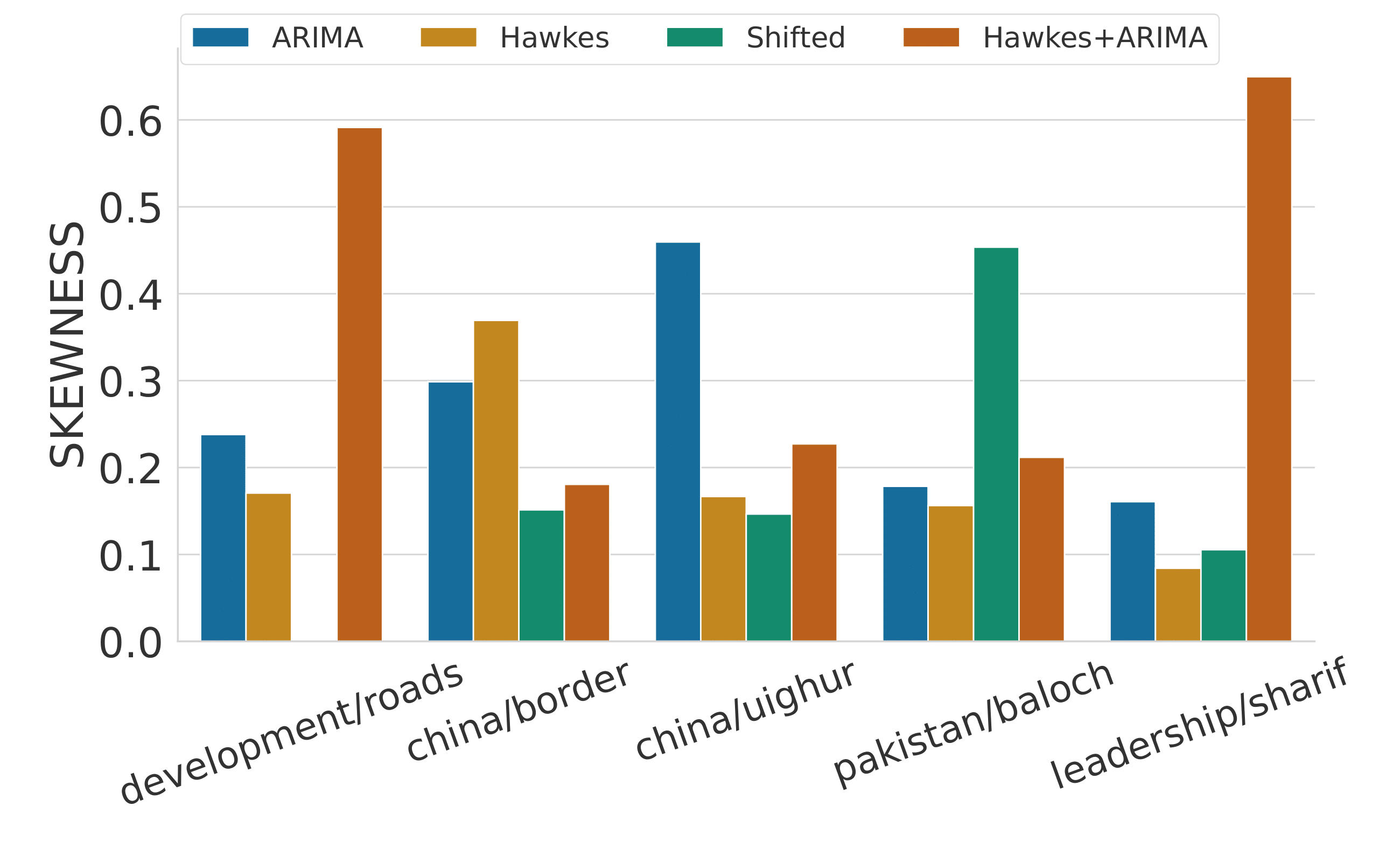}
			\label{fig:cp5_yt_skewness_h}
	}
	\\
	\subfloat[BRIA (Twitter)]{
		\includegraphics[width=0.4\linewidth]{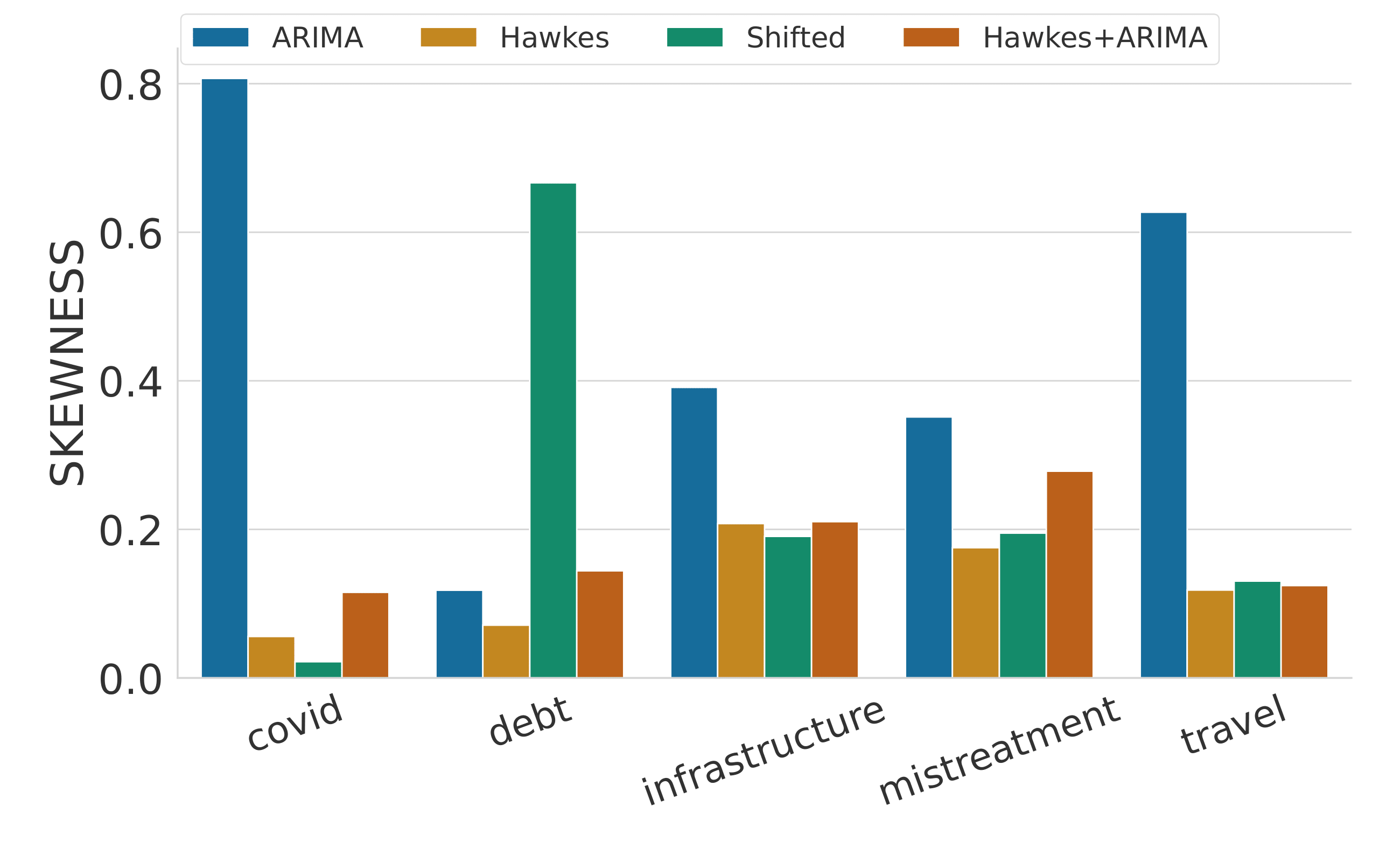}
			\label{fig:cp6_tw_skewness_h}
	}
	&
	\subfloat[BRIA (YouTube)]{
		\includegraphics[width=0.4\linewidth]{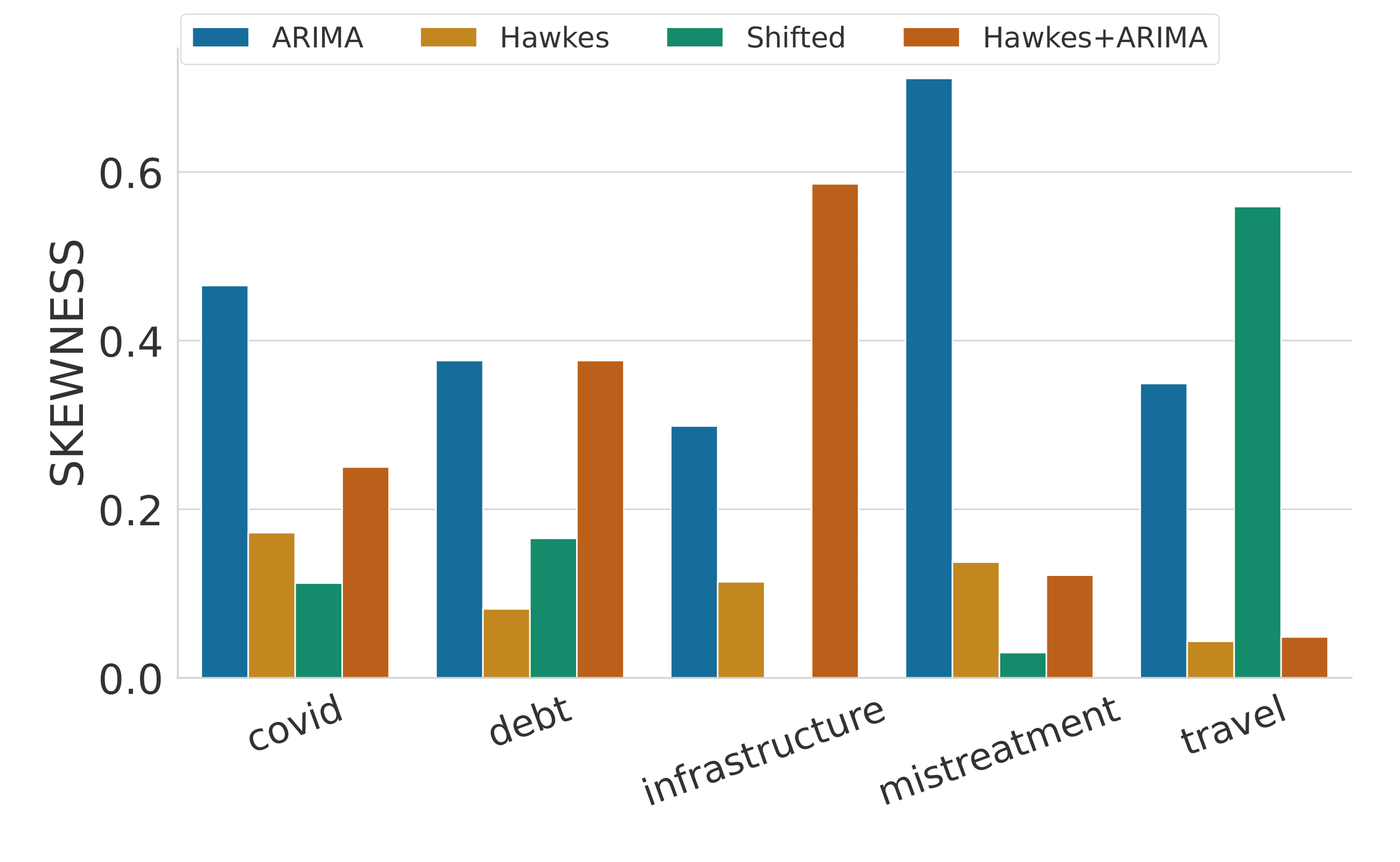}
			\label{fig:cp6_yt_skewness_h}
	}
\\
\end{tabular}
	 \caption{Skewness hourly performance across topics for each context and over two platforms (lower is better). Skewness scores for each topic are normalized between 0 to 1 relative to the sum of the baselines' errors. The results are for one week predictions at hourly granularity.}
	 \label{fig:skewness_topic_performance_h}
\end{figure*}

Despite ARIMA's unrealistic predictions, we noticed that traditional time series forecasting metrics fail to penalize it enough.
Relying only on these particular metrics to make decisions on which baseline to select can be misleading.
In fact, a representative baseline for social media time series is one that not only produces realistic activity levels over time but also captures well key properties related to the distribution of data.
Figures~\ref{fig:volatility_topic_performance_h}~and~\ref{fig:skewness_topic_performance_h} show baselines performance on capturing two basic characteristics of a distribution: spread (in terms of volatility error) and shape (in terms of skewness).

While in some topics ARIMA captures volatility well (e.g., ``china/border'' in CPEC or ``travel'' in BRIA as shown in Figures~\ref{fig:cp5_tw_volatility_h}~and~\ref{fig:cp6_tw_volatility_h}), on the majority of topics it is the worst across different contexts and platforms.
From time series visualizations, it is evident that ARIMA completely misses the variations observed in the ground truth time series data.
For example, in highly volatile topics such as ``controversies/china/border'' (Figure~\ref{fig:china_border_tw_h}) or ``leadership/sharif'' (Figure~\ref{fig:leadership_sharif_tw_h}), ARIMA predicts smooth time series with no signs of volatility. 
On the other hand, Hawkes and Shifted baselines do a better job at capturing rapid changes despite sometimes exhibiting irregular up-and-down trends.
Similarly, ARIMA does poorly on capturing the shape of the data distribution based on its symmetry (Figure~\ref{fig:skewness_topic_performance_h}). 
These patterns become clearer when looking at the aggregated results on volatility and skewness across topics in Table~\ref{tab:aggregate_counts_performance}, where ARIMA's limitations on distribution-based metrics are more evident. 
Finally, we might want to select relevant baselines as informed by the utility of time series predictions: should we compare a particular time series against different baselines for different performance metrics? 
For example, consider ARIMA as a relevant baseline for volume, but another baseline for capturing representative patterns over time.
To explore this question, we experimented with an ensemble approach (Hawkes+ARIMA), which takes advantage of both ARIMA's volume predictions and Hawkes' temporal patterns to possibly produce more accurate predictions.
Overall the ensemble approach exhibits  well-rounded performance across different metrics (Table~\ref{tab:aggregate_counts_performance}), and proves promising for improving in some topics.

\subsubsection{Overall Normalized Metric Error Analysis}
We sought to have a more clear understanding of which models perform the best, so to that end, we utilized an aggregate metric called the \textit{Overall Normalized Metric Error}, previously used in \citep{vam-venez-xplore}. It combines the 6 metrics shown in Table \ref{tab:aggregate_counts_performance} into 1 metric per model. Intuitively, it shows how each model performed relative to the other models.

Tables \ref{tab:onme-full-results} and \ref{tab:onme-summary} contain the ONME results for each \textit{(domain, platform)} pairing for each model. Table \ref{tab:onme-full-results} contains both the ONME results as well as the values (called normalized errors) used to calculate them. Table \ref{tab:onme-summary} contains only the ONME values from \ref{tab:onme-full-results} by themselves, for easier viewing and interpretation of results.

For a given \textit{(domain, platform)} pairing, the ONME was calculated in the following way.

\begin{enumerate}
    \item For each model, we calculated a \textit{normalized error} score. This was done by taking the metric result for a particular model and metric, and then dividing that metric result by the sum of all the other models’ metric results for the current \textit{(domain, platform)} pair of interest. We indicate the normalized error for each metric using the prefix n (e.g. \textit{nAPE}). These normalized errors are shown in Table \ref{tab:onme-full-results}.

    For example, for \textit{(Vz19, Twitter)}, we calculated ARIMA’s normalized APE score by taking its raw APE from Table \ref{tab:aggregate_counts_performance} (35.54), and then dividing it by the sum of all the models’ APE scores for \textit{(Vz19, Twitter)} from Table \ref{tab:aggregate_counts_performance}. So, the sum of all APEs would be 35.54+74.61+67.88+55.07, which is 233.1 We then divided 35.54 by this number, which yielded ARIMA’s normalized APE score (nAPE) for \textit{(Vz19, Twitter)}, which was about 0.1525.  We repeated this process for each model and metric. The normalized error results for each model can be seen in Table \ref{tab:onme-full-results}.

    \item By repeating this process for each model and metric, this resulted in 6 normalized metric values per model, each between 0 and 1, for a given \textit{(domain, platform)} pair. Intuitively, each normalized metric score indicates the relative performance of a model on a particular metric, in comparison to the other models. Note that since each error value is in the range of 0 to 1, the sum of each row in Table \ref{tab:onme-full-results} adds up to 1.

    \item Finally, for each model, we calculated the ONME score by taking the average of the 6 normalized metric error values for that particular model and \textit{(domain, topic)} pairing. For example, for the ARIMA model in the \textit{(Vz19, Twitter)} category, its nAPE, nRMSE, nSMAPE, nDTW, nSkewness, and nVolatility results were 0.1525, 0.2235, 0.2325, 0.2038, 0.3971, and 0.2548; respectively. The average of these 6 values is 0.244, which is the ONME value shown in Table \ref{tab:onme-full-results}.
\end{enumerate}

\begin{table}[htpb]
\centering
\caption{Normalized error results for each metric result along with the Overall Normalized Metric Error (ONME).}
\begin{tabular}{|c|l|r|r|r|c|}
\hline
 Platform                 & \multicolumn{1}{c|}{Metric} & \multicolumn{1}{c|}{ARIMA} & \multicolumn{1}{c|}{Hawkes} & \multicolumn{1}{c|}{Shifted} & \multicolumn{1}{c|}{Hawkes+ARIMA} \\ \hline
\multirow{7}{*}{Twitter (Vz19)} & nAPE                        & \textbf{0.1525}            & 0.3201                      & 0.2912                       & 0.2363                            \\
                                                & nRMSE                       & \textbf{0.2235}            & 0.2655                      & 0.2688                       & 0.2421                            \\
                                                 & nSMAPE                      & 0.2325                     & 0.2679                      & 0.281                        & \textbf{0.2186}                   \\
                                                 & nDTW                        & \textbf{0.2038}            & 0.2577                      & 0.3192                       & 0.2192                            \\
                                                 & nSkewness                   & 0.3971                     & \textbf{0.0145}             & 0.3275                       & 0.2609                            \\
                                                 & nVolatility                 & 0.2548                     & 0.2588                      & \textbf{0.2208}              & 0.2656                            \\
                                                 & ONME                        & 0.244                      & \textbf{0.2307}             & 0.2848                       & 0.2405                            \\ \hline
                        \multirow{7}{*}{YouTube (Vz19)} & nAPE                        & \textbf{0.1732}            & 0.3449                      & 0.2229                       & 0.259                             \\
                                                 & nRMSE                       & \textbf{0.2142}            & 0.2895                      & 0.2487                       & 0.2476                            \\
                                                 & nSMAPE                      & \textbf{0.1477}            & 0.3919                      & 0.2418                       & 0.2186                            \\
                                                 & nDTW                        & 0.2594                     & \textbf{0.1887}             & 0.3208                       & 0.2311                            \\
                                                 & nSkewness                   & 0.3253                     & 0.1981                      & \textbf{0.1741}              & 0.3025                            \\
                                                 & nVolatility                 & 0.2435                     & 0.2735                      & \textbf{0.2032}              & 0.2798                            \\
                                                 & ONME                        & \textbf{0.2272}            & 0.2811                      & 0.2352                       & 0.2565                            \\ \hline
\multirow{7}{*}{Twitter (CPEC)} & nAPE                        & 0.6087                     & 0.1709                      & \textbf{0.0014}              & 0.2189                            \\
                                                 & nRMSE                       & 0.422                      & \textbf{0.1555}             & 0.2138                       & 0.2087                            \\
                                                 & nSMAPE                      & 0.2973                     & 0.2614                      & \textbf{0.2043}              & 0.237                             \\
                                                 & nDTW                        & 0.3046                     & \textbf{0.149}              & 0.245                        & 0.3013                            \\
                                                 & nSkewness                   & 0.4228                     & \textbf{0.0403}             & 0.1043                       & 0.4325                            \\
                                                 & nVolatility                 & \textbf{0.1265}            & 0.4236                      & 0.2526                       & 0.1974                            \\
                                                 & ONME                        & 0.3637                     & 0.2001                      & \textbf{0.1702}              & 0.266                             \\  \hline 
                        \multirow{7}{*}{YouTube (CPEC)} & nAPE                        & 0.3236                     & 0.4291                      & \textbf{0.04}                & 0.2073                            \\
                                                 & nRMSE                       & 0.2278                     & 0.2322                      & 0.3151                       & \textbf{0.2249}                   \\
                                                 & nSMAPE                      & \textbf{0.231}             & 0.2535                      & 0.2601                       & 0.2554                            \\
                                                 & nDTW                        & 0.2694                     & \textbf{0.2251}             & 0.2325                       & 0.2731                            \\
                                                 & nSkewness                   & 0.4576                     & \textbf{0.1381}             & 0.2496                       & 0.1547                            \\
                                                 & nVolatility                 & 0.3426                     & 0.2176                      & \textbf{0.1528}              & 0.287                             \\
                                                 & ONME                        & 0.3087                     & 0.2493                      & \textbf{0.2083}              & 0.2337                            \\ \hline
 \multirow{7}{*}{Twitter (BRIA)} & nAPE                        & \textbf{0.2179}            & 0.2575                      & 0.2869                       & 0.2377                            \\
                                                 & nRMSE                       & \textbf{0.1772}            & 0.302                       & 0.3012                       & 0.2197                            \\
                                                 & nSMAPE                      & 0.2455                     & 0.2832                      & \textbf{0.221}               & 0.2503                            \\
                                                 & nDTW                        & 0.2637                     & \textbf{0.2473}             & 0.2418                       & \textbf{0.2473}                   \\
                                                 & nSkewness                   & 0.8351                     & 0.0351                      & \textbf{0.0216}              & 0.1081                            \\
                                                 & nVolatility                 & 0.3819                     & 0.1849                      & 0.2769                       & \textbf{0.1563}                   \\
                                                 & ONME                        & 0.3536                     & 0.2183                      & 0.2249                       & \textbf{0.2032}                   \\ \hline 
                        \multirow{7}{*}{YouTube (BRIA)} & nAPE                        & 0.3506                     & 0.5111                      & \textbf{0.0667}              & 0.0716                            \\
                                                 & nRMSE                       & 0.2256                     & 0.2756                      & 0.2861                       & \textbf{0.2127}                   \\
                                                 & nSMAPE                      & 0.268                      & 0.2451                      & 0.2479                       & \textbf{0.239}                    \\
                                                 & nDTW                        & 0.3537                     & \textbf{0.1789}             & 0.2683                       & 0.1992                            \\
                                                & nSkewness                   & 0.3971                     & \textbf{0.1862}             & 0.2292                       & 0.1875                            \\
                                                 & nVolatility                 & 0.6145                     & 0.0512                      & \textbf{0.0422}              & 0.2922                            \\
                                                 & ONME                        & 0.3682                     & 0.2413                      & \textbf{0.19}                & 0.2004                            \\ \hline
\end{tabular}
\label{tab:onme-full-results}
\end{table}

As previously mentioned, Table \ref{tab:onme-summary} contains only the ONME results from Table \ref{tab:onme-full-results} by themselves, for easier viewing and interpretation. There is also a new row, that shows the average ONME for each model across all domains and platforms. 

As one can see in the table, the Shifted Baseline has the best overall performance. Out of the 6 \textit{(domain, platform)} pairings, it had the lowest ONME three times. In contrast, the Hawkes, ARIMA,  and Hawkes+ARIMA models won in this regard only once each.

Furthermore, the Shifted model had the lowest average ONME across all domains and platforms. Its score was 0.2189. In the second, third, and fourth places were the Hawkes+ARIMA, Hawkes, and ARIMA models; respectively. Their ONMEs were 0.2334, 0.2368, and 0.3109; respectively. It is notable that ARIMA performed much worse than the other models in this regard, as it is such a commonly used time series baseline.

\begin{table}[htpb]
\centering
\caption{Overall Normalized Metric Error results for each domain and platform per model.}
\begin{tabular}{|c|c|c|c|c|c|}
\hline
Domain                & Platform      & ARIMA           & Hawkes          & Shifted         & Hawkes+ARIMA          \\ \hline
\multirow{2}{*}{Vz19} & Twitter       & 0.244           & \textbf{0.2307} & 0.2848          & 0.2405                \\  
                      & YouTube       & \textbf{0.2272} & 0.2811          & 0.2352          & 0.2565                \\ \hline
\multirow{2}{*}{CPEC} & Twitter       & 0.3637          & 0.2001          & \textbf{0.1702} & 0.266               \\
                      & YouTube       & 0.3087          & 0.2493          & \textbf{0.2083} & 0.2337               \\ \hline
\multirow{2}{*}{BRIA} & Twitter       & 0.3536          & 0.2183          & 0.2249          & \textbf{0.2032}  \\ 
                      & YouTube       & 0.3682          & 0.2413          & \textbf{0.19}   & 0.2004               \\ \hline
All-Domains           & All-Platforms & 0.3109          & 0.2368          & \textbf{0.2189} & 0.2334               \\ \hline
\end{tabular}
\label{tab:onme-summary}
\end{table}

\subsubsection{Forecast Duration vs. Baseline Performance}
One of our objectives is to understand how the baselines perform under different prediction window durations. 
Towards this, we developed a new set of models that predict just one day of activity at the hourly granularity (thus, 24 values, each corresponding to one hour of the day) to contrast with the 7 day predictions just discussed. 
We refer to these new models as \emph{Short-Term}, while the previous models are called \emph{Long-Term}. 

For the Short-Term configuration, the Shifted model predictions were created by shifting the previous 24 hours of ground truth into the next day. This was done for all seven days in the test period. 

All of the Short-Term ARIMA models predicted 24 hours out into the future at a time, for all 7 test day periods. Their \textit{p, d, } and \textit{q} hyperparameters were determined using a validation set, in a similar fashion to the Long-Term ARIMA model hyperparameters in Table \ref{tab:arima_params}. Table \ref{tab:st-arima-combos} contains the Short-Term ARIMA hyperparameter combinations.

\begin{table}[htpb]
\centering
\caption{The Short-Term ARIMA model best parameter combinations. Similar to the Long-Term configurations, a validation set was used to determine the combinations for each topic and platform.}
\begin{tabular}{|c|l|cc|}
\hline
\multicolumn{1}{|l|}{\multirow{2}{*}{Domain}} & \multicolumn{1}{c|}{\multirow{2}{*}{Topic}} & \multicolumn{2}{c|}{(p, d, q)}                 \\  
\multicolumn{1}{|l|}{}                        & \multicolumn{1}{c|}{}                       & \multicolumn{1}{c}{Twitter}     & YouTube     \\ \hline
\multirow{5}{*}{Vz19}                         & arrests                                     & \multicolumn{1}{c|}{(24, 1, 24)} & (48, 2, 72) \\ 
                                              & international/aid                           & \multicolumn{1}{c|}{(24, 1, 48)} & (72, 1, 24) \\  
                                              & maduro/dictator                             & \multicolumn{1}{c|}{(24, 2, 72)} & (72, 2, 72) \\  
                                              & military                                    & \multicolumn{1}{c|}{(24, 1, 24)} & (24, 2, 72) \\ 
                                              & other/chavez                                & \multicolumn{1}{c|}{(48, 2, 72)} & (24, 1, 96) \\ \hline
\multirow{5}{*}{CPEC}                         & benefits/development/roads                  & \multicolumn{1}{c|}{(96, 1, 96)} & (48, 2, 24) \\ 
                                              & controversies/china/border                  & \multicolumn{1}{c|}{(96, 1, 96)} & (24, 1, 24) \\ 
                                              & controversies/china/uighur                  & \multicolumn{1}{c|}{(96, 2, 48)} & (48, 2, 96) \\  
                                              & controversies/pakistan/baloch               & \multicolumn{1}{c|}{(24, 2, 48)} & (24, 1, 24) \\ 
                                              & leadership/sharif                           & \multicolumn{1}{c|}{(24, 1, 72)} & (96, 1, 96) \\ \hline
\multirow{5}{*}{BRIA}                         & covid                                       & \multicolumn{1}{c|}{(24, 2, 24)} & (48, 1, 72) \\ 
                                              & debt                                        & \multicolumn{1}{c|}{(48, 2, 24)} & (72, 1, 48) \\  
                                              & infrastructure                              & \multicolumn{1}{c|}{(48, 2, 24)} & (48, 2, 96) \\ 
                                              & mistreatment                                & \multicolumn{1}{c|}{(24, 2, 96)} & (24, 2, 24) \\ 
                                              & travel                                      & \multicolumn{1}{c|}{(24, 2, 24)} & (48, 2, 24) \\ \hline
\end{tabular}
\label{tab:st-arima-combos}
\end{table}

As a reminder, the Long-Term models predict the seven days of activity following the training period at hour granularity.
In the Long-Term configuration, the Shifted model predictions were created by shifting a full 168-hour block of past ground truth into the 168-hour test period and the ARIMA models predicted 168 hours of values without any ground truth after the start time as input to the model. 
The predictions of the previous 24-hour blocks were fed into the model as inputs for the next 24-hour block's prediction. We focus on the Shifted model because it had the strongest Overall Normalized Metric Error, and on the ARIMA model because of its high popularity in the literature \citep{zhang1998forecasting, box2015time, hipel1994time, siami2018comparison, prasha-naam}.

Note that we use the terms \textit{Short-Term} and \textit{Long-Term} in a relative sense, and not an absolute one. 
The purpose of the following analysis is to better understand how extending the prediction windows of the Shifted and ARIMA models would affect their relative performance in comparison to one another. 
The reader may have their own preference for prediction window size, however we just want to demonstrate how extending or shortening it may alter relative model performance.

\begin{figure}[htbp]
	\centering
\begin{tabular}{c}
		\subfloat[Twitter (Short-Term)]{
         \includegraphics[scale=0.4]{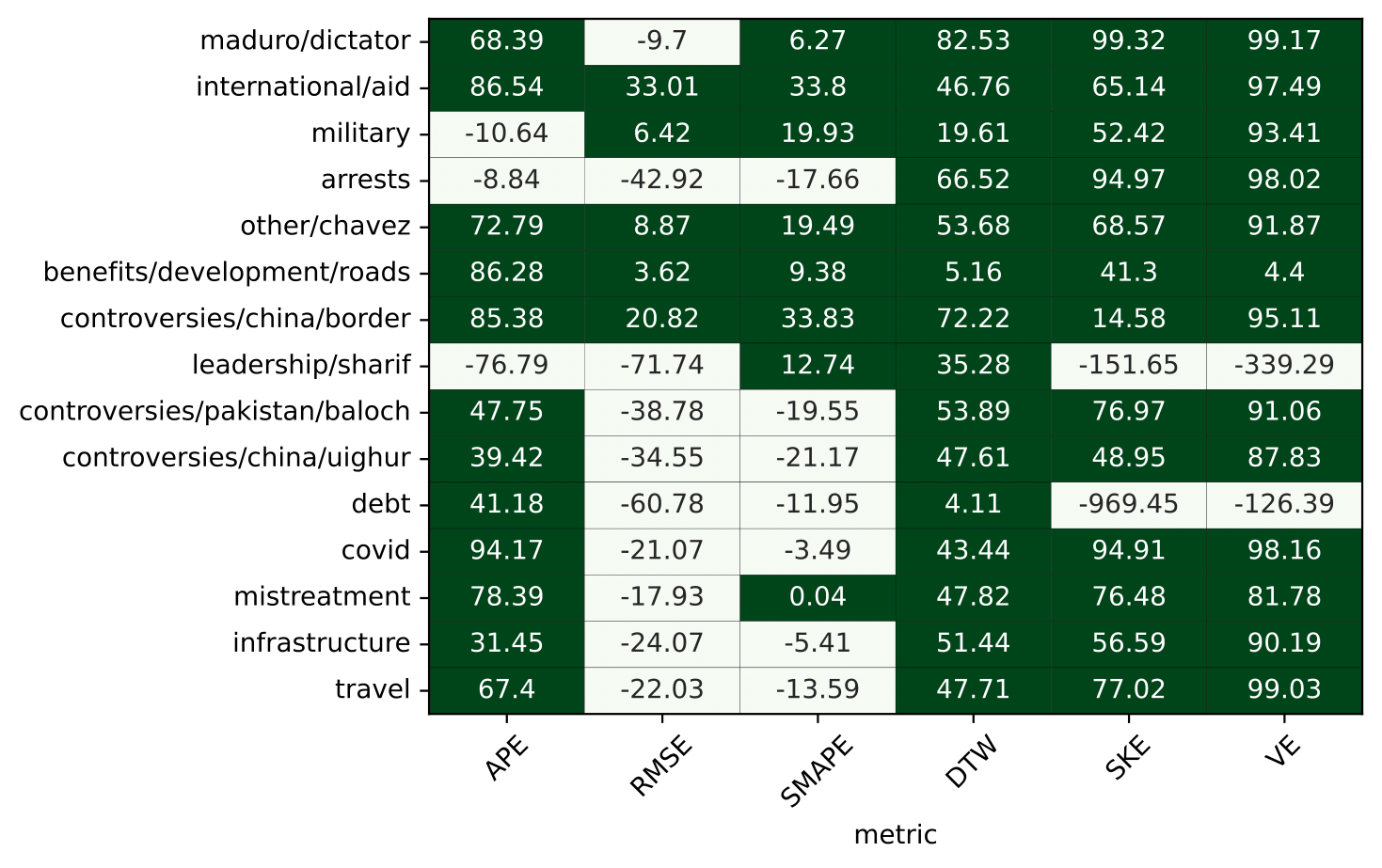} 
		    \label{fig:st-twitter-hourly}
	}
	
	\\
	
			\subfloat[YouTube (Short-Term)]{
         \includegraphics[scale=0.4]{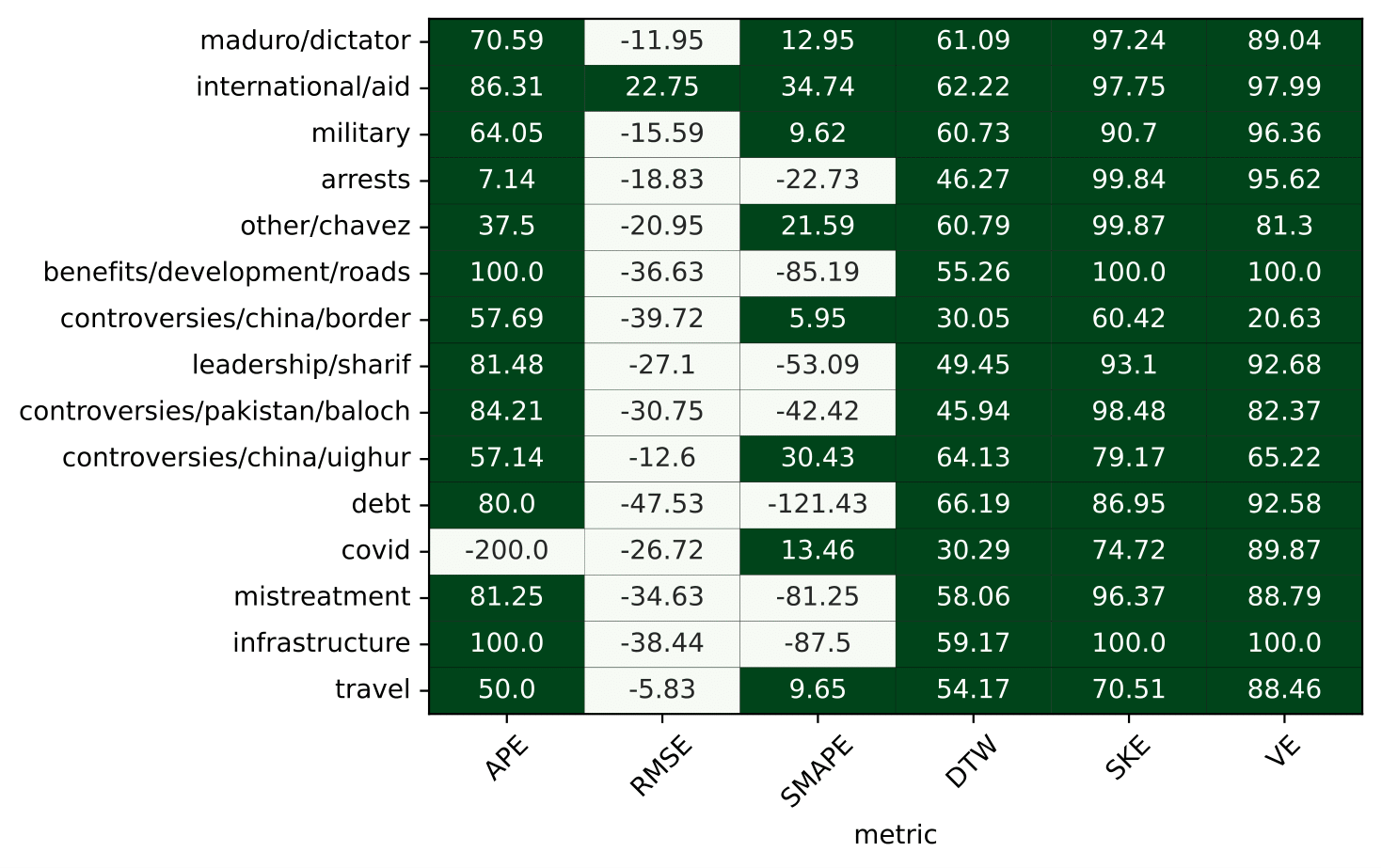} 
		    \label{fig:st-youtube-hourly}
	}

\end{tabular}
\caption{Heatmaps comparing the performance of the Shifted and ARIMA models in the short-term model setting. A cell is green if the Shifted model outperformed the ARIMA model on the particular topic-metric pair.}
\label{fig:short-term-comps-hourly}
\end{figure}

\begin{figure}[htbp]
	\centering
\begin{tabular}{c}

		\subfloat[Twitter (Long-Term)]{
         \includegraphics[scale=0.4]{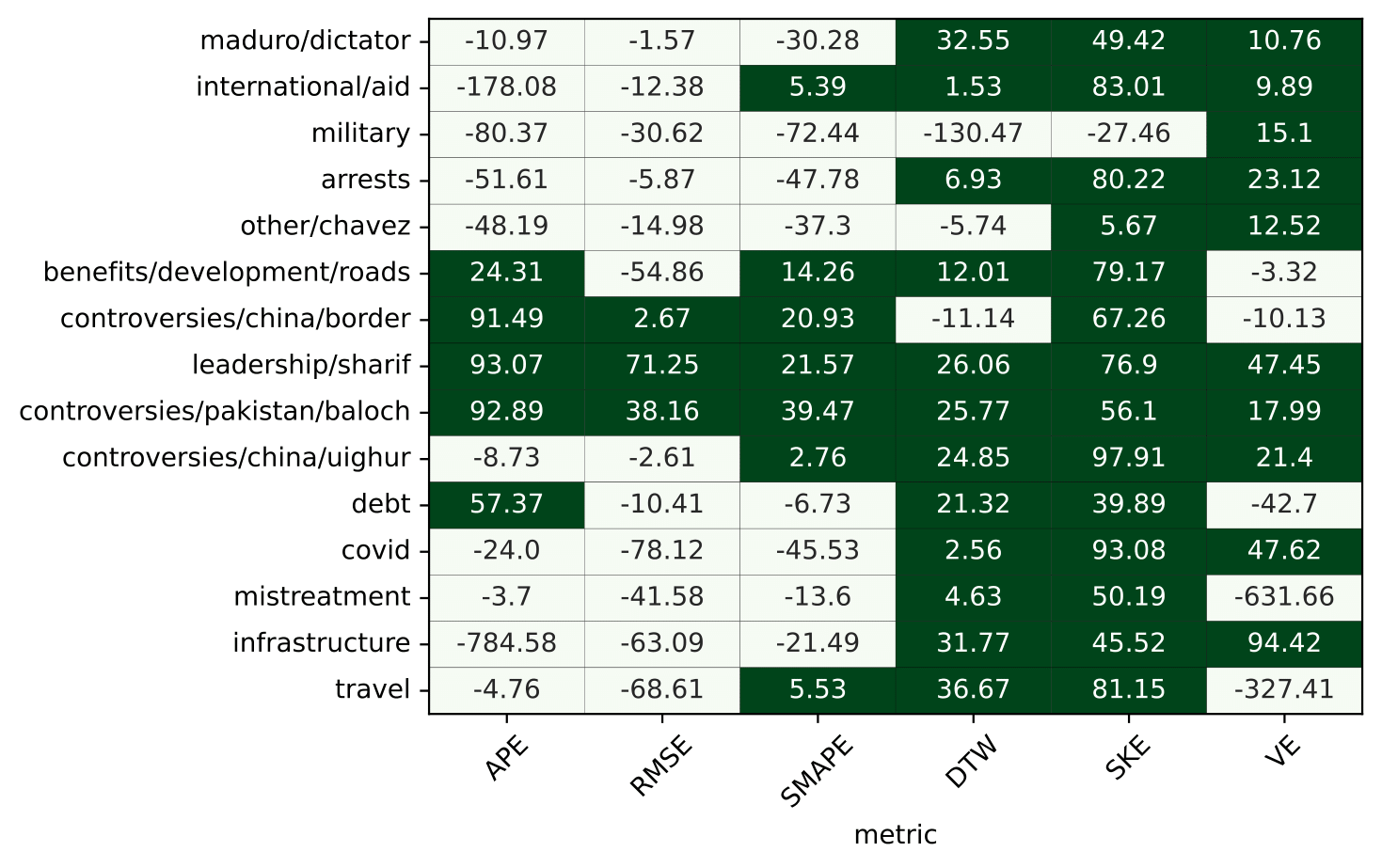} 
		    \label{fig:lt-twitter-hourly}
		    }

	\\

		\subfloat[YouTube (Long-Term)]{
         \includegraphics[scale=0.4]{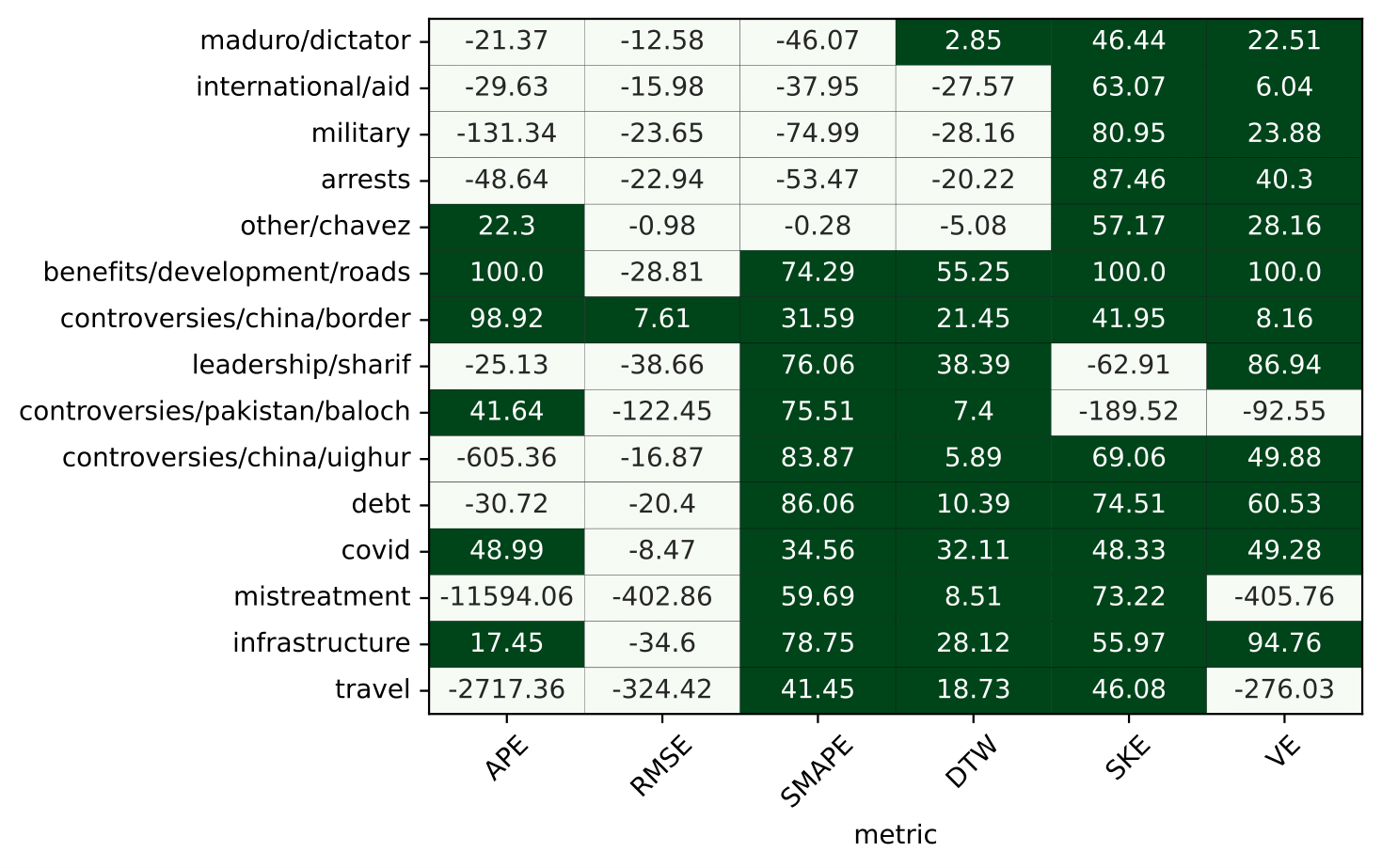} 
		    \label{fig:lt-youtube-hourly}
	}

\end{tabular}
\caption{Heatmaps comparing the performance of the Shifted and ARIMA models in the long-term model setting. A cell is green if the Shifted model outperformed the ARIMA model on the particular topic-metric pair.}
\label{fig:long-term-comps-hourly}
\end{figure}

The Figure~\ref{fig:short-term-comps-hourly} heatmap shows the Short-Term prediction results and the Figure~\ref{fig:long-term-comps-hourly} heatmap shows the Long-Term prediction results of the Shifted and ARIMA models.
In these heatmaps, if the Shifted Model outperforms the ARIMA model, a cell is green; otherwise, the cell is white. 
The values in each cell are the percent improvement scores of the Shifted models over the ARIMA models (thus, the larger the value, the better Shifted performs compared to ARIMA in the metric shown on the x axis and on the topic shown on the y axis).

For short-term (one day) predictions (Figure~\ref{fig:short-term-comps-hourly}), the Shifted model outperforms ARIMA on most topics and most metrics on both platforms.
For the Short-Term Twitter predictions (Figure~\ref{fig:st-twitter-hourly}), the Shifted model wins 66 out of 90 trials, or 73.33\% of the time. 
Similarly, for YouTube, the Shifted model wins 68 out of 90, or 75.56\% of the time (Figure~\ref{fig:st-youtube-hourly}).  
APE is the challenge for ARIMA in this configuration, a result we have not seen in the long-term prediction results presented before.

However, the Shifted model loses its relative advantages over ARIMA in the Long-Term configuration, as shown in Figure~\ref{fig:long-term-comps-hourly}. 
In the Long-Term setup on Twitter data (\ref{fig:lt-twitter-hourly}), the Shifted model wins only slightly more than 50\%. 
For YouTube (Figure~\ref{fig:lt-youtube-hourly}), the Shifted model outperforms ARIMA 59\% of the time. 

Moreover, the advantage that ARIMA gained in the long term predictions are mainly in the volume-related metric (APE), and the exacting-timing-volume metrics (RMSE and sMAPE); Shifted maintains advantages in the approximate-timing-volume metric (dynamic time warping), as well as in the volatility-metrics (Skewness Error and Volatility Error). 

To conclude, for short-term predictions, the Shifted baseline is reliable (and in our experience, very challenging to outperform~\citep{ng2021forecasting}) while ARIMA misses on most metrics in all contexts and both platforms. 
For long-term predictions, however, the two models perform similarly when aggregated over all metrics we analyzed, but with clear advantages in the volume-related and exact-timing-volume-related  metrics for ARIMA and the approximate-timing-volume and volatility-related metrics for Shifted.

\section{Summary and Recommendations}

This study was motivated by the desire to understand and document the relative performance benefits of various time series forecasting methodologies for social media activity. 
For this, we defined the objective of forecasting the number of social media activities per hour within a discussion context for the next 168 hours without using the ground truth information during the forecasting interval. 
This objective is from a typical forecasting problem, and was the task during the scientific challenges which were a part of the DARPA SocialSim project output evaluation. 
Forecasting social media activity has numerous practical applications, from generating realistic simulations and/or synthetic datasets for scientific experiments to testing intervention techniques meant to control processes on social media. 

We chose three representative techniques for forecasting time series: Shifted, that simply replays the past, Hawkes processes, and ARIMA. 
We used three different conversations from different geo-political contexts, each manifested simultaneously on two social media platforms (Twitter and YouTube), and each comprised of five topics. 
Based on a combination of manual and automated annotation, each message was annotated with/assigned to one of the topics.

Using these rich datasets, we focused on answering the following question: What is a competitive baseline for forecasting social media activity? 
We discovered that the answer is nuanced: the Shifted Baseline performed the best in terms of Overall Normalized Metric Error, however it was still not universally the best. Rather, a representative baseline should be chosen based on what the predicted time series will be used for. 
How should a competitive baseline be chosen for a particular scenario? 
It varies based on 1)~the duration of the forecast window; 2)~the metrics that are most important for the forecast task (operational scenario, perhaps).

Specifically, we discovered that ARIMA best estimates the volume of activity when measured in RMSE, but it fails to accurately model temporal patterns. 
The Shifted baseline works well for short prediction windows in most measurements, but loses power for longer prediction windows. 
Hawkes processes create very agitated time series and never dominate in all measurements. 

Overall, it appears that the Shifted baseline is the winner: replaying the recent past is a cheap and highly reliable estimation of the near future in many scenarios and performance metrics. 
Where the Shifted baseline will likely fail is when a topic on social media is heavily influenced by real life events, such as political unrest, protests and armed conflicts. 
In such cases, ARIMA provides a better estimation of the overall volume of social media posts, although it fails at capturing the variation over time. 

The choice of baselines depends on the particular characteristics that the problem attempts to capture/model. 
For example, if modeling the final volume of information cascades, or predicting next day activity (assuming previous day ground truth is always available), then a baseline like ARIMA could be a good choice. 
However, if the focus is to capture spiky behavior to identify anomalous periods in the future, for instance, then comparing against baselines such as Shifted or Hawkes might be more meaningful.
        
An ensemble of baselines could result in more competitive models to compare against. 
For example, when combining Hawkes and ARIMA by averaging, we observed an improvement for some topics in both volume of activity and temporal patterns. 

We note that experiments were done with different granularity: in addition to predicting hourly activity, we also predicted daily activity for the same seven-day interval (not included here). 
Given how the Shifted and Hawkes processes work (that is, they predict the exact time when an event will occur, totally independent of the time granularity chosen), the only difference between day vs hour granularities are in the aggregation of the performance results over time.
Thus, comparing these two methods at different time granularities tells more about the performance metrics chosen to evaluate their quality rather than about the methods themselves. 
ARIMA did not show significant differences in terms of volume prediction.
Because many of the other metrics are not normalized, a direct comparison is not meaningful. 

A final observation from this study is that different performance metrics tell incomplete and different stories---what is the minimum optimal set of performance metrics for estimating time series forecasting especially in the context of social media activity remains an open question. 

Future work would include analysis of burstiness as described in \citep{karsai-bursty}. Also of interest would be an analysis of the features of the time series across domains and platforms. This may yield some better understanding as to why certain models perform better or worse on certain time series. Lastly, we would try incorporating the Shifted model into the ARIMA+Hawkes ensemble model to see if better performance could be obtained.

\backmatter


\bibliography{main}


\end{document}